\documentclass[11pt]{article}
\pdfoutput=1
\usepackage[square,comma,numbers,sort&compress]{natbib}
\usepackage{bbold}
\usepackage{amsmath}
\usepackage{amsfonts}
\usepackage{amssymb}
\usepackage{graphicx, rotating}
\usepackage{epstopdf}
\usepackage{epsfig}
\usepackage{empheq}
\usepackage{latexsym}
\usepackage{multirow}
\usepackage{stmaryrd}
\usepackage{tikz}
\usepackage{color}
\usepackage{slashed}
\usepackage[colorlinks=true,linkcolor=black,anchorcolor=black,citecolor=blue,filecolor=cyan,menucolor=red,runcolor=filecolor,urlcolor=blue,bookmarks=true,bookmarksnumbered=true]{hyperref}%
\usepackage{color}
\definecolor{rosso}{cmyk}{0,1,1,0.4}
\definecolor{rossos}{cmyk}{0,1,1,0.55}
\definecolor{rossoc}{cmyk}{0,1,1,0.2}
\definecolor{blu}{cmyk}{1,1,0,0.3}
\definecolor{blus}{cmyk}{1,1,0,0.6}
\definecolor{bluc}{cmyk}{1,1,0,0.1}
\definecolor{verde}{cmyk}{0.92,0,0.59,0.25}
\definecolor{verdec}{cmyk}{0.92,0,0.59,0.15}
\definecolor{verdes}{cmyk}{0.92,0,0.59,0.4}
\definecolor{grigio}{cmyk}{0,0,0,0.07}
\definecolor{rosa}{cmyk}{0,0.1,0.1,0.02}
\definecolor{rosino}{cmyk}{0,0.05,0.05,0.02}
\definecolor{rosas}{cmyk}{0,0.3,0.25,0.05}
\definecolor{celeste}{cmyk}{0.1,0,0,0.02}
\definecolor{giallino}{cmyk}{0,0,0.4,0.02}
\definecolor{rosso}{cmyk}{0,1,1,0.4}
\definecolor{rossos}{cmyk}{0,1,1,0.55}
\definecolor{rossoc}{cmyk}{0,1,1,0.2}
\definecolor{blu}{cmyk}{1,1,0,0.3}
\definecolor{bluc}{cmyk}{1,1,0,0.1}
\definecolor{blucc}{cmyk}{0.7,0.5,0,0}
\definecolor{viola}{cmyk}{0,1,0,0.6}
\definecolor{viola2}{cmyk}{0,1,0.2,0.6}
\definecolor{verde}{cmyk}{0.92,0,0.59,0.25}
\definecolor{verdec}{cmyk}{0.92,0,0.59,0.15}
\definecolor{verdes}{cmyk}{0.92,0,0.59,0.4}
\definecolor{verdino}{cmyk}{0.12,0,0.09,0.05}
\definecolor{giallo}{cmyk}{0,0,1,0}
\definecolor{gialloverde}{cmyk}{0.44,0,0.74,0}
\definecolor{grey}{rgb}{0.6,0.6,0.6}
\definecolor{fuchsia}{rgb}{1,0,1}

\newcommand{\grey}[1]{{\color{grey}\color{grey}#1\color{grey}}}
\usepackage{chngcntr}
\counterwithin{figure}{section}
\counterwithin{table}{section}
\counterwithin{equation}{section}

\def\beq{\begin{equation}}
\def\eeq{\end{equation}}
\def\bes{\begin{subequations}}
\def\ees{\end{subequations}}
\def\bea{\begin{eqnarray}}
\def\eea{\end{eqnarray}}
\def\bry{\begin{array}}
\def\ery{\end{array}}
\def\bit{\begin{itemize}}
\def\eit{\end{itemize}}
\def\nn{\nonumber}
\def\mc{\mathcal}

\def\dst{\displaystyle}
\def\f{\frac}
\def\gst{g_{V}}
\def\eq{Eq.~\eqref}
\newcommand{\eqs}[2]{Eqs.~\eqref{#1} and \eqref{#2}}

\title{
\vspace{-1.5cm}
\normalsize{\hspace{9.9cm}DFPD-2014/TH/01,LPN14-050}
\vspace{1cm}
\vspace{0.0 cm}
{\huge
Heavy Vector Triplets:
\\ \vspace*{5pt} 
Bridging Theory and Data 
}}
\vspace{0.5cm}
\author{{\Large\text{Duccio Pappadopulo$^{1}$\,}\footnote{\href{pappadopulo@berkeley.edu}{pappadopulo@berkeley.edu}}\text{\,\,, Andrea Thamm$^{2}$\,}\footnote{\href{andrea.thamm@epfl.ch}{andrea.thamm@epfl.ch}}\text{\,\,, }}\vspace{1mm}\\{\Large\text{Riccardo Torre$^{3,4}$\,}\footnote{\href{riccardo.torre@pd.infn.it}{riccardo.torre@pd.infn.it}}\text{\,\, and Andrea Wulzer$^{3}$\,}\footnote{\href{andrea.wulzer@pd.infn.it}{andrea.wulzer@pd.infn.it}}} \vspace{4mm}\\ 
{\small\emph{$^{1}$Department of Physics, University of California, Berkeley, CA 94720, USA and}}\\ \vspace{-15pt} \\
{\small\emph{Theoretical Physics Group, Lawrence Berkeley National Laboratory, Berkeley, CA 94720, USA}} \\
{\small\emph{$^{2}$Institut de Th\'eorie des Ph\'enom\`enes Physiques, EPFL, CHÐ1015 Lausanne, Switzerland}} \\
{\small\emph{$^{3}$Dipartimento di Fisica e Astronomia, Universit\`a di Padova and}}\\ \vspace{-15pt} \\
{\small\emph{INFN, Sezione di Padova, via Marzolo 8, I-35131 Padova, Italy}} \\
{\small\emph{$^{4}$SISSA, Via Bonomea 265, I-34136 Trieste, Italy}}
}

\date{}

\begin{document}
\baselineskip=16pt

\maketitle \thispagestyle{empty}
\begin{center}
{\Large Abstract}
\end{center}
We introduce a model-independent strategy to study narrow resonances which we apply to a heavy vector triplet of the Standard Model (SM) group for illustration. The method is based on a simplified phenomenological Lagrangian which reproduces a large class of explicit models. Firstly, this allows us to derive robust model-independent phenomenological features and, conversely, to identify the peculiarities of different explicit realizations. Secondly, limits on $\sigma\times \text{BR}$ can be converted into bounds on a few relevant parameters in a fully analytic way, allowing for an interpretation in any given explicit model. Based on the available 8 TeV LHC analyses, we derive current limits and interpret them for vector triplets arising in weakly coupled (gauge) and strongly coupled (composite) extensions of the SM. We point out that a model-independent limit setting procedure must be based on purely on-shell quantities, like $\sigma\times \text{BR}$. Finite width effects altering the limits can be considerably reduced by focusing on the on-shell signal region. We illustrate this aspect with a study of the invariant mass distribution in di-lepton searches and the transverse mass distribution in lepton-neutrino final states. In addition to this paper we provide a set of online tools available at a dedicated webpage \cite{projectwebpage}.

\parbox[c]{12cm}{

\medskip
\noindent

}

\newpage
\tableofcontents
\section{Introduction}

Ensuring proper communication among theory and experiment is an important and stimulating task, particularly in the context of hypothetical TeV-scale extensions of the Standard Model (SM) which have to be compared with LHC data. The central aspect is that the theory is not developed enough to provide sharp predictions of the experimental observables. Indeed, as of today, no single explicit complete model of New Physics, by which precise predictions could be made, has emerged as a particularly motivated or compelling possibility. Instead, we have interesting and motivated generic ``frameworks'' which are defined as a set of broad assumptions on the New Physics and can not be translated into a single concrete model. Robust qualitative predictions, like the existence of a given set of particles, can be made within each framework but a quantitative comparison with the data requires some explicit implementations of the general idea. Several models can be constructed within each framework and since we have no idea how to choose one we would need all of them to be compared with the data. Obviously, this program can not be completed directly by the experimental collaborations because it would require tens of different models for each New Physics analysis and a separate presentation of the results for each of them. Moreover even if we knew the ``true'' New Physics theory, it would typically depend on so many free parameters that a direct comparison with data, obtained by scanning the multi-dimensional parameter space with numerical simulations, would be impossible. The typical example of this situation is the Minimal Supersymmetric Standard Model (MSSM) which, in spite of its well-known limitations, is still sometimes regarded as a plausible benchmark model of low-energy supersymmetry. A full scan over its parameters is not feasible, and one is forced to one of its restricted versions. 

While the problem of data/theory comparison is probably too hard to be tackled in full generality, progress can be made if we restrict our attention to direct experimental manifestations of New Physics which consist of the production of reasonably narrow new particles. In this case one can conveniently adopt the so-called ``Simplified Model'' strategy \cite{Alves:2011dz} which has by now become a standard method in supersymmetry searches and starts to be developed also in non-supersymmetric frameworks \cite{Bauer:2009p1295,Barbieri:2011p2759,Han:2010p2696,Accomando:2010p2249,Schmaltz:2010p2610,Grojean:2011vu,Chiang:2011kq,Torre:2011vn,Torre:2012ub,deBlas:2012tc,DeSimone:2012ul,Buchkremer:2013uj,AguilarSaavedra:2013hg,Lizana:2013vz}.
The idea is extremely simple and after all just the standard strategy adopted  in hadron spectroscopy where, exactly like in the present case, no complete predictive model is available. The point is that resonant searches are typically not sensitive to all the details and the free parameters of the underlying model, but only to those parameters or combinations of parameters that control the mass of the resonance and the interactions involved in its production and decay. Therefore one can employ a simplified description of the resonance defined by a phenomenological Lagrangian where only the relevant couplings and mass parameters are retained. Aside from symmetry constraints, the Simplified Model Lagrangian does not need to fulfill any particular theoretical requirement. Its sole goal is to provide a phenomenological parametrization of a broad enough set of explicit models and should thus contain all and only those terms which are present in the explicit constructions. The experimental results should be presented in the parameter space of the phenomenological Lagrangian, expressed by confidence level curves or, if possible, in terms of a likelihood function. In this way they could be easily translated into any specific model where the phenomenological parameters can be computed explicitly. The advantage of this two-step approach is that the phenomenological parameters can always be expressed analytically in terms of those of the ``fundamental'' theory. No matter how complicated the model is, the comparison with the data will always be performed analytically rather than with numerical simulations, in a way that furthermore does not require any knowledge of the experimental details of the analysis.

\begin{figure}[t]
\begin{center}
\includegraphics[scale=0.8]{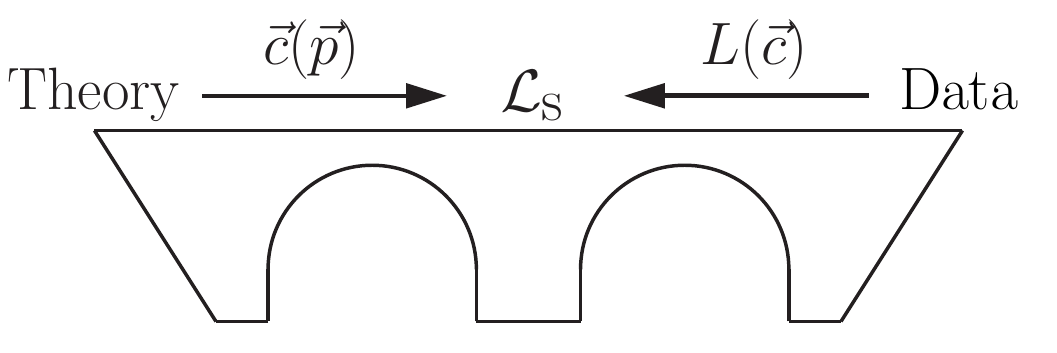}
\caption{\small Pictorial view of the Bridge Method.}\label{fig:uno}
\end{center}
\end{figure}

The procedure is conveniently depicted as a two-span bridge, shown in Figure~\ref{fig:uno}, where the Simplified Model constitutes the central pillar and the two spans represent the fundamental/phenomenological parameter relations and the comparison of the Simplified Model with the data respectively. In the Figure, we denote collectively as $\vec{c}$ the parameters of the phenomenological Lagrangian and as $L(\vec{c})$ the likelihood function, or the CL curves, as extracted from the experimental data. Notice that $L(\vec{c})$ could very well be the result of a combination of different analyses, which can be preformed directly on the Simplified Model parameter space. Once the likelihood or the CL limits are known, the experimental information is immediately translated into the free parameters $\vec{p}$ of any explicit model by computing the phenomenological/explicit parameter relations $\vec{c}(\vec{p})$.

When comparing the Simplified Model with the data, some care is required. The crucial point is that the Simplified Model, differently for instance from the SM or the MSSM, is not supposed to be a complete theory and attention must be paid not to use it outside its realm of validity. Namely, the Simplified Model is constructed to describe only the \emph{on-shell} resonance production and decay. A good experimental search should thus be only sensitive to the on-shell process and insensitive to the off-shell effects. The simplest example of this situation, which we will discuss in detail, is the Drell-Yan (DY) process where the invariant mass distribution of the final state is studied. Aside from the resonant peak, the distribution is characterized by a low mass tail which can become prominent, because of the rapidly-falling parton distribution functions, when the resonance approaches the kinematical production threshold or when a large interference with the SM background is present. Many different New Physics effects, not included in the Simplified Model, might contribute to the tail and radically change the Simplified Model prediction. This could come, for instance, from extra contact interactions or from heavier resonances produced in the same channel. Around the peak, and only in this region, these effects are negligible and the Simplified Model prediction is trustable. Indeed the peak shape is well described, through the Breit-Wigner (BW) formula, in terms of purely on-shell quantities such as the production rate times the Branching Ratio (BR) to the relevant final state, \mbox{$\sigma\times$BR}, and by the resonance total decay width. Experimental searches should focus on the peak and avoid contamination from the other regions as much as possible. More in general, any resonance search relies on the measurement of a given observable, either the number of events or a distribution, restricted by suitable identification and selection cuts. Only ``on-shell'' observables, which are exclusively sensitive to the resonance formation and decay, should be employed in Simplified Model searches. Notice that whether an experimental observable is on-shell or not can crucially depend on the cuts and must be checked case by case. 

Aside from addressing the conceptual issues previously outlined, the usage of on-shell observables is also an important practical simplification. Because of factorization of the production cross-section and the decay BR, on-shell observables are ``easy'' to predict within the Simplified Model since they do not depend on all the parameters of the phenomenological Lagrangian in a complicated way but only on few combinations that describe the on-shell re-sonance. In the example of the invariant mass distribution, a search performed at the peak can be turned into limits on \mbox{$\sigma\times$BR} as a function of the resonance mass and possibly of its width. The width and BRs are simple analytical functions of the Simplified Model parameters and also the total production rate can be expressed semi-analytically in terms of the parton luminosities at each mass point. The mass- and width- dependent limits can thus be mapped \emph{analytically} into the phenomenological parameter space. Obviously, taking the experimental efficiencies properly into account is essential. This is typically rather easy because, as in the examples discussed in the following, the efficiencies only depend on the resonance mass and can be extracted from a few benchmark simulations. The tail of the invariant mass distribution, instead, has a more complicated dependence on the model parameters and can not be predicted analytically. Therefore a search which is sensitive to the tail can not be cast into a limit on \mbox{$\sigma\times$BR} and it can be interpreted within the model only by scanning the parameter space with long and demanding simulations.

The aim of the present paper is to illustrate these general concepts in detail by focusing on the simple but well-motivated example of electroweak-charged spin one resonances which are a common prediction of many New Physics scenarios. The latter can be weakly coupled, like for instance $Z'$ \cite{1980PhRvD..22..727B,Hewett:1989dr,Cvetic:1995vc,Rizzo:2006wq,Langacker:2008yv,Accomando:2010p2249,Salvioni:2009mt,Salvioni:2010p2769,Salvioni:2010p1209,Chanowitz:2011ew,Salvioni:2012gya,Accomando:2013ve} or $W'$ \cite{Langacker:1989p2578,Sullivan:2002p2617,Rizzo:2007bk,Grojean:2011vu,Torre:2011vn,Torre:2012ub,Schmaltz:2010p2610,Frank:2010p2250,Accomando:2011up} models, or strongly coupled constructions such as Composite Higgs models \cite{Agashe:2007hh,Agashe:2009bj,Agashe:2009ve,Contino:2011np,Bellazzini:2012tv,Accomando:2012us,Hernandez:2013wd} and some variants of Technicolor \cite{Chanowitz:1993fc,Barbieri:2008p1580,Barbieri:2010p144,Barbieri:2010p1577,CarcamoHernandez:2010p1578,Hernandez:2010qp,Cata:2009ka,Accomando:2011gt,Falkowski:2011ua}. The experimental searches for these particles, performed by ATLAS \cite{ATLASCollaboration:2012ex,ATLASCollaboration:2012gi,ATLAS-CONF-2012-130,ATLAS-CONF-2012-150,ATLAS-CONF-2013-050,ATLAS-CONF-2013-052,ATLAS-CONF-2013-066,ATLAS-CONF-2014-017,Aad:2014cka,Aad:2014pha,Aad:2014aqa} and CMS \cite{CMS-PAS-EXO-12-021,CMS-PAS-EXO-12-022,CMS-PAS-EXO-12-023,CMS-PAS-EXO-12-024,CMS-PAS-EXO-12-059,CMS-PAS-EXO-12-060,CMS-PAS-EXO-12-061,CMS-PAS-B2G-12-005,CMS-PAS-B2G-12-010,Khachatryan:2014xja}, provide theoretical interpretations of the results in terms of an extremely small subset of the possible models and moreover restrict to limited benchmark regions of the parameter space. This strategy does not provide a sufficient coverage of the theoretical possibilities and furthermore it precludes reinterpretation in other models. In this paper we will show that a great improvement can be achieved with the Bridge method. 

The paper is organised as follows. In Section~\ref{2} we introduce the Simplified Model Lagrangian and discuss some basic aspects of its phenomenology. We also show how the resonance production cross-section in the two relevant channels, DY and Vector Boson Fusion (VFB), can be parametrized semi-analytically in a way that, as previously described, allows for an efficient comparison of the model with the experimental results. We restrict, for definiteness, to the case of an $SU(2)_L$ iso-triplet of resonances. The extension to other representations should be straightforward and is left to future work. Section~\ref{3} is a survey of the present experimental situation where, based on the present experimental limits, we derive $95\%$~CL exclusion bounds in the Simplified Model parameter space. This is done by taking the experimental results at face value, {\it{i.e.}} by assuming that the limits are properly set on \mbox{$\sigma\times$BR} as a function of the resonance mass as presented by the experimental collaborations. However this might not be completely correct, since important effects associated with the finite resonance width could affect the \mbox{$\sigma\times$BR} currently extracted by the experiments, which would result in an incorrect definition of the quantity on which the limit is set. In Section~3.3 we will illustrate these effects in detail by focusing on the examples of di-lepton and lepton-neutrino searches. In Section~\ref{4} we relate the Simplified Model to explicit constructions. Two examples are considered as representatives of weakly and strongly coupled theories, showing that the Simplified Model is general enough to describe both cases in different regions of the parameter space. The examples are the extension of the SM gauge group described in Ref.~\cite{1980PhRvD..22..727B} and the effective description of Composite Higgs models vectors of Ref.~\cite{Contino:2011np}. 
In Section~\ref{5} we present our Conclusions.
Our Simplified Model is implemented in a series of tools described in Appendix~\ref{AppC} and available on the webpage \cite{projectwebpage}.

\section{A Simple Simplified Model}
\label{2}

In addition to the SM fields and interactions we consider a real vector $V_\mu^a$, $a=1,2,3$, in the adjoint representation of $SU(2)_L$ and with vanishing hypercharge. It describes one charged and one neutral heavy spin-one particle with the charge eigenstate fields defined by the familiar relations 
\beq
\displaystyle
V_\mu^\pm=\frac{V_\mu^1\mp i V_\mu^{2}}{\sqrt{2}}\,,\;\;\;\;\; V_\mu^0=V_\mu^{3}\,.
\eeq
Similarly to Ref.~\cite{deBlas:2012tc}, we describe the dynamics of the new vector by a simple phenomenological Lagrangian
\beq\label{sml}
\bry{lll}
\dst{\mathcal{L}}_V &=&\dst-\frac14 D_{[\mu}V_{\nu ]}^a D^{[\mu}V^{\nu ]\;a}+\frac{m_V^{2}}2V_\mu^a V^{\mu\;a}\vspace{2mm}\\
&&\dst+\, i\,\gst  c_H V_\mu^a H^\dagger \tau^a {\overset{{}_{\leftrightarrow}}{D}}^\mu H+\frac{g^2}{\gst } c_F V_\mu^a J^{\mu\;a}_F\vspace{2mm}\\
&&\dst+\,\frac{\gst }2 c_{VVV}\, \epsilon_{abc}V_\mu^a V_\nu^b D^{[\mu}V^{\nu]\; c}+\gst ^{2} c_{VVHH} V_\mu^aV^{\mu\;a} H^\dagger H
-\frac{g}2 c_{VVW}\epsilon_{abc}W^{\mu\,\nu\; a} V_\mu^b V_\nu^c\,.
\ery
\eeq
The first line of the above equation contains the $V$ kinetic and mass term, plus trilinear and quadrilinear interactions with the vector bosons from the covariant derivatives
\beq\label{Vcovder}
D_{[\mu}V^a_{\nu ]}=D_{\mu}V^a_{\nu } -D_{\nu}V^a_{\mu }\,,\;\;\;\;\; D_\mu V_\nu^a = \partial_\mu V_\nu^a
+g\,\epsilon^{abc}W_\mu^bV_\nu^c\,,
\eeq
where $g$ denotes the $SU(2)_L$ gauge coupling. Notice that the $V_\mu^a$ fields are not mass eigenstates as they mix with the $W_\mu^a$ after EWSB and the mass parameter $m_V$ does not coincide with the physical mass of the resonances.\\
The second line contains direct interactions of $V$ with the Higgs current
\beq
i\,H^\dagger \tau^a {\overset{{}_{\leftrightarrow}}{D}}^\mu H=i\,H^\dagger \tau^a D^\mu H\,-\,i\,D^\mu H^\dagger \tau^a  H\,,
\eeq
and with the SM left-handed fermionic currents
\beq
\displaystyle
J_F^{\mu\;a}=\sum_f\overline{f}_L\gamma^\mu\tau^a f_L\,,
\eeq
where $\tau^a=\sigma^a/2$. The Higgs current term $c_H$ leads to vertices involving the physical Higgs field and the three unphysical Goldstone bosons. By the Equivalence Theorem \cite{Chanowitz:1985hj}, the Goldstones represent the longitudinally polarized SM vector bosons $W$ and $Z$ in the high-energy regime which is relevant for the resonance production and decay. Thus $c_H$ controls the $V$ interactions with the SM vectors and with the Higgs, and in particular its decays into bosonic channels. Similarly, $c_F$ describes the direct interaction with fermions, which is responsible for both the resonance production by DY and for its fermionic decays. In \eq{sml} we  reported, for shortness, a universal coupling of $V$ to fermions, but in our analysis we will consider a more general situation with different couplings to leptons, light quarks and the third quark family. The interaction in \eq{sml} should thus be generalized to
\beq
c_F V\cdot J_F\;\;\rightarrow\;\;c_l V\cdot J_l+c_q V\cdot J_q+c_3 V\cdot J_3\,.
\eeq
Given the strong constraints on additional sources of lepton and light quark flavor violation further generalizations seem unnecessary. The proliferation of fermionic parameters is a complication, but the effects of $c_l$, $c_q$, and $c_3$ can be easily disentangled by combining searches in different decay channels including third family quarks.\\
Finally, the third line of \eq{sml} contains $3$ new operators and free parameters, $c_{VVV}$, $c_{VVHH}$ and $c_{VVW}$. None of them, however, contains vertices of one $V$ with light SM fields, thus they do not contribute directly to $V$ decays\footnote{A priori, they could contribute to cascade decays. However, as we will see below, the mass splitting between the neutral and the charged state is very small and cascade decays are extremely suppressed.} and single production processes which are the only relevant for LHC phenomenology. As we will discuss in the following Section, they do affect the above processes only through the mixing of $V$ with the $W$, but since the mixing is typically small their effect is marginal. Therefore to a first approximation the operators in the third line can be disregarded and the phenomenology described entirely by the four parameters $c_H$, $c_l$, $c_q$ and $c_3$, plus the mass term $m_V$.

In \eq{sml} we adopted a rather peculiar parametrization of the interaction terms, with a coupling $\gst $ weighting extra insertions of $V$, of $H$ and of the fermionic fields. Similarly, the insertions of $W$ in the last line is weighted by the $SU(2)_L$ coupling $g$. We take $\gst $ to represent the typical strength of $V$ interactions while the dimensionless coefficients ``$c$'' parametrize the departures from the typical size. The parametrization of the fermion couplings is an exception to this rule. In this case one extra factor of $g^2/\gst^2$ has been introduced. This is convenient because in all the explicit models we will be interested in, both of weakly- and strongly-coupled origin, this factor is indeed present and the $c_F$'s, as defined in \eq{sml}, are of order one. The other $c$'s are typically of order one, except for $c_H$ which is of order one in the strongly-coupled scenario but can be reduced in the weakly coupled case as described in Section~4. In all cases, the $c$'s are never parametrically larger than one, with the notable exception of the third family coupling $c_3$, which could be enhanced in strongly-coupled scenarios where the top quark mass is realized by the mechanism of Partial Compositeness, see for instance \cite{Giudice:1024017}. The coupling $\gst $ can easily vary over one order of magnitude in different scenarios, ranging from $\gst \sim g\sim 1$ in the ``typical'' weakly-coupled case up to $\gst \simeq 4\pi$ in the extreme strong limit. Therefore it is useful to factor it out of the operator estimate. Notice that there is no sharp separation between the weak and strong coupling regimes as nothing forbids to consider theories with a ``weak'' UV origin but with large $\gst $, of the order of a few, and ``strong'' models where $\gst $ is reduced by the large number of colors of the strong sector, $\gst =4\pi/\sqrt{N_c}$. This provides one additional motivation for our approach which interpolates between the two cases.

Our parametrization of the operators is useful at the theoretical level but obviously redundant as $\gst $ could be reabsorbed in the $c$'s and is not a genuine new parameter of the model. For instance, one could resolve the redundancy by setting $c_{VVV}=1$ and thus define $\gst $ as the $V$ self-interaction strength. However for practical purposes, and in particular for presenting the experimental limits of the model, it could be easier to treat $\gst c_H$ and $g^2/\gst c_F$, the combinations that enter in the vertices, as fundamental parameters.

In the Bridge approach, as discussed in the Introduction, the Simplified Model does not need to fulfill any particular theoretical requirement and its only goal is to be simple enough while still capable to reproduce a large set of explicit models. Therefore a complete justification of our phenomenological Lagrangian has to be postponed to Section~\ref{4} where the matching with explicit constructions will be discussed. However we can already appreciate the general validity of the description by noticing that \eq{sml}  is the most general Lagrangian compatible with the SM gauge invariance and with the $CP$ symmetry restricted to operators of energy dimension below or equal to $4$. Assuming $CP$, which we take to act on $V$ as on the SM $W$
\beq
V^a(\vec{x},t) \rightarrow -(-)^{\delta_{a2}}V^a(-\vec{x},t)
\;\;\Leftrightarrow\;\;
\left\{
\bry{l}
V^\pm(\vec{x},t)\rightarrow -V^\mp(-\vec{x},t)\\
V^0(\vec{x},t)\rightarrow -V^0(-\vec{x},t)
\ery
\right.\,,
\eeq
is very convenient as it avoids the proliferation of operators constructed with the Levi-Civita tensor. Furthermore, it leads to a unique coupling of $V$ to the Higgs parametrized by only one real coefficient $c_H$. $CP$ is often also a good symmetry of explicit models  so that it is not a too restrictive assumption. It is important to note that the Lagrangian with the imposed $CP$ symmetry is also accidentally invariant under the custodial group $SO(4)=SU(2)_L\times SU(2)_R$, with $V$ in the $(\mathbf{3},\mathbf{1})$ representation. The custodial symmetry is of course broken, but only by the gauging of the hypercharge. This makes our setup very efficient in reproducing strongly-coupled scenarios where custodial symmetry is imposed by construction.

One invariant low-dimensional operator, the $W$-$V$ kinetic mixing
\beq\label{wv}
D_{[\mu}V^a_{\nu ]}W^{\mu\nu a}\,,
\eeq
is not reported in \eq{sml} because, following Ref.~\cite{Aguila:2010p1781}, it can be eliminated from the Lagrangian by a field redefinition of the form
\beq\label{fieldred}
\left\{
\bry{l}
W_\mu^a\rightarrow W_\mu^a + \alpha V_\mu^a\\
V_\mu^a\rightarrow \beta V_\mu^a
\ery
\right.\,.
\eeq
More details on this can be found in Appendix \ref{AppA}.
We also ignored dimension four quadrilinear $V$ interactions because they are irrelevant for the LHC phenomenology.

The choice of restricting to low-dimensional operators is clearly well-justified in the weakly-coupled case where the underlying model is a renormalizable theory, but it is questionable in the strongly-coupled one where higher-dimensional operators are potentially relevant. However in all strongly-coupled scenarios that obey the SILH paradigm \cite{Giudice:1024017} we do have a reason to stop at $d=4$. In the SILH power-counting the most relevant higher dimensional operators are those involving extra powers of the Higgs or the $V$ field which are weighted by the Goldstone-Boson-Higgs decay constant $f$. Their effects are generically suppressed by the parameter
$$
\displaystyle
\xi=\frac{v^{2}}{f^{2}}\,,
$$
where $v\simeq 246$~GeV is the EWSB scale. Since $\xi$ controls the departures from the Standard Higgs model, compatibility with the ElectroWeak Precision Tests (EWPT) and the LHC Higgs coupling measurements \cite{Ciuchini:2013vb, Contino:2013un} requires $\xi \lesssim 0.2$. If the higher dimensional operators do not induce any qualitatively new effect and only give relative corrections of order $\xi$ to the vertices, they can be safely ignored given the limited accuracy of the LHC direct searches. This will be confirmed by the analysis of Section~\ref{4}.\footnote{Notice that this does not need to be the case a priori. There are plenty of examples concerning for instance the LHC phenomenology of Composite Higgs Top Partners \cite{Panico:1359049,Matsedonskyi:2012ws, DeSimone:2012ul} where the Higgs non-linearities can not be ignored.}

There exist however other scenarios where higher dimensional operators are unsuppressed and the parametrization of \eq{sml} is insufficient. These are technicolor-like models where the strong sector condensate breaks the EW symmetry directly and the observed Higgs boson is a light composite particle with couplings compatible with the SM expectations. This might occur by accident or in hypothetical scenarios with a light Higgs-like dilaton \cite{Chacko:2012wh,Bellazzini:2012td}. In spite of the tension with EWPT and with the Higgs data it would be interesting to generalize our framework in order to test also these ideas.

\subsection{Basic phenomenology}\label{2.1}

\subsubsection*{Masses and Mixings}

Having introduced our Simplified Model in \eq{sml}, let us discuss its phenomenology starting from the mass spectrum. After EWSB, the only massless state is the photon which can be identified as the gauge field associated with the unbroken \mbox{U$(1)_{\textrm{em}}$}. It is given by the SM-like expression\footnote{This only holds in the field basis where the $W$-$V$ mixing of \eq{wv} is set to zero, otherwise the photon acquires a component along $V^0$.}
\beq\label{thetaW}
\displaystyle
A_\mu=\cos{\theta_W}B_\mu+\sin{\theta_W} W_\mu^3\,,\;\;\;\;\;\textrm{where}\;\;\tan{\theta_W}=\frac{g'}g\,.
\eeq
The orthogonal combination, the $Z$ field, instead acquires a mass and a mixing with $V^0$. Notice that since the photon is given by the canonical SM expression, its couplings are also canonical. The electric charge in our model is therefore simply given by
\beq
\displaystyle
e=\frac{gg'}{\sqrt{g^2+{g'}^2}}\,,\;\;\;\Rightarrow\;\;\;\left\{
\bry{l}
g^{\,}={e}/{\sin\theta_W}\\
g'={e}/{\cos\theta_W}
\ery
\right.\,.
\label{pars}
\eeq
In what follows we will trade $g$ and $g'$ for $e$ and $\sin\theta_W$, taking $e$ as an input parameter and setting it to the experimental value $e\approx\sqrt{4\pi/137}$. 

The two other neutral mass eigenstates are the SM $Z$ boson and one heavy vector of mass $M_0$ which are obtained by diagonalizing the mass matrix of the $(Z,\,V^0)$ system by a rotation
\beq
\left(
\bry{l}
Z\\ V^0
\ery
\right)\;\rightarrow\;
\left(
\bry{cc}
\cos\theta_N & \sin\theta_N\\ 
-\sin\theta_N &\cos\theta_N
\ery
\right)
\left(
\bry{l}
Z\\ V^0
\ery
\right)\,.
\eeq
The mass matrix is
\beq
\displaystyle
{\mathcal{M}}_N^2=\left(
\bry{cc}
{\hat{m}}^2_Z & c_H \zeta {\hat{m}}_Z {\hat{m}}_V\\
c_H \zeta {\hat{m}}_Z {\hat{m}}_V &  {\hat{m}}_V^2
\ery
\right)\,,\;\;\;\;\;
{\textrm{where}}\;\;
\left\{
\bry{l}
\displaystyle
{\hat{m}}_Z = \frac{e}{2\sin\theta_W\cos\theta_W}\hat{v}\vspace{1mm}\\
\displaystyle
{\hat{m}}_V^2=m_V^2+\gst ^2c_{VVHH}\hat{v}^2\vspace{1mm}\\
\displaystyle
\zeta=\frac{\gst \hat{v}}{2\,{\hat{m}}_V}
\ery
\right..
\eeq
In the above equations $\hat{v}$ denotes the Higgs field Vacuum Expectation Value (VEV) defined by $\langle H^\dagger H\rangle=\hat{v}^2/2$, which in our model can differ significantly from the physical EWSB scale $v=246$~GeV. The mass eigenvalues and the rotation angles are easily obtained by inverting the relations
\bea
&&\textrm{Tr}\left[{\mathcal{M}}^2_N\right]={\hat{m}}_Z^2+{\hat{m}}_V^2=m_Z^2+M_0^2\,,\nn\\
&&\textrm{Det}\left[{\mathcal{M}}^2_N\right]={\hat{m}}_Z^2{\hat{m}}_V^2\left(1-c_H^2\zeta^2\right)=m_Z^2M_0^2\,,\nn\\
&&\tan2\theta_N=\frac{2\,c_H\zeta {\hat{m}}_Z{\hat{m}}_V}{{\hat{m}}_V^2-{\hat{m}}_Z^2}\,. 
\label{ndiag}
\eea
Notice that the tangent can be uniquely inverted because the angle $\theta_N$ is in the range $[-\pi/4,\pi/4]$ in the parameter region we will be interested in, where $ {\hat{m}}_Z<{\hat{m}}_V$.

The situation is similar in the charged sector where the mass matrix of the $(W^\pm,\,V^\pm)$ system reads
\beq
\displaystyle
{\mathcal{M}}_C^2=\left(
\bry{cc}
{\hat{m}}^2_W & c_H \zeta {\hat{m}}_W {\hat{m}}_V\\
c_H \zeta {\hat{m}}_W {\hat{m}}_V &  {\hat{m}}_V^2
\ery
\right)\,,\;\;\;
{\textrm{where}}\;\;
{\hat{m}}_W = \frac{e}{2\sin\theta_W}\hat{v}=\cos\theta_W{\hat{m}}_Z\,,
\eeq
and it is diagonalized by
\bea
&&\textrm{Tr}\left[{\mathcal{M}}^2_C\right]={\hat{m}}_W^2+{\hat{m}}_V^2=m_W^2+M_+^2\,,\nn\\
&&\textrm{Det}\left[{\mathcal{M}}^2_C\right]={\hat{m}}_W^2{\hat{m}}_V^2\left(1-c_H^2\zeta^2\right)=m_W^2M_+^2\,,\nn\\
&&\tan2\theta_C=\frac{2\,c_H\zeta {\hat{m}}_W{\hat{m}}_V}{{\hat{m}}_V^2-{\hat{m}}_W^2}\,. 
\label{cdiag}
\eea
The charged and neutral mass matrices are connected by custodial symmetry, which can be shown in full generality to imply
\beq
\label{cusmass}
{\mathcal{M}}^2_C=
\left(
\begin{array}{cc}
\cos\theta_W & 0\\
0 & 1
\end{array}
\right)
{\mathcal{M}}^2_N
\left(
\begin{array}{cc}
\cos\theta_W & 0\\
0 & 1
\end{array}
\right)\,.
\eeq
By taking the determinant of the above equation, or equivalently by comparing the charged and neutral determinants in \eq{ndiag} and \eq{cdiag}, we obtain a generalized custodial relation among the physical masses
\beq
m_W^2M_+^2=\cos^2\theta_Wm_Z^2M_0^2\,.
\label{cust}
\eeq

From the simple formulas above we can already derive interesting features of our model. First of all, we can identify the physically ``reasonable'' region of its parameter space. We aim at describing new vectors with masses at or above the TeV scale, but of course we also want the SM masses $m_{W,Z}\sim100$~GeV to be reproduced. Therefore we require a hierarchy in the spectrum, which can only occur, barring unnatural cancellations in the determinant of the mass matrices, if ${\hat{m}}_{W,Z}$ and ${\hat{m}}_V$ are hierarchical, {\it{i.e.}}
\beq
\frac{{\hat{m}}_{W,Z}}{{\hat{m}}_V}\sim\frac{m_{W,Z}}{M_{+,0}}\lesssim 10^{-1}\ll1\,.
\label{hi}
\eeq
The parameter $\zeta$, instead, can be either very small or of order one. Both cases are realized in explicit models. While $\zeta\ll1$ is the most common situation, $\zeta\sim1$ only occurs in strongly coupled scenarios at very large $\gst $.

In the limit of \eq{hi} we obtain simple approximate expressions for $m_W$ and $m_Z$
\beq
\bry{lll}
m_Z^2 &=&\dst {\hat{m}}_{Z}^2\left(1-c_H^2\zeta^2\right)\left(1+\mc{O}({\hat{m}}_{Z}^2/{\hat{m}}_{V}^2)\right)\,,\nn\vspace{2mm}\\
m_W^2 &=&\dst {\hat{m}}_{W}^2\left(1-c_H^2\zeta^2\right)\left(1+\mc{O}({\hat{m}}_{W}^2/{\hat{m}}_{V}^2)\right)\,.
\ery
\eeq
Since ${\hat{m}}_W=\cos\theta_W{\hat{m}}_{Z}$, the $W$-$Z$ mass ratio is thus given, to percent accuracy, by
\beq
\displaystyle
\frac{m_W^2}{m_Z^2}\simeq\cos^2\theta_W\,.
\label{rho}
\eeq
In order to reproduce the observed ratio, which satisfies the $\rho=1$ SM tree--level relation to $\sim 1\%$, we need  \footnote{The reader might be confused by the fact that $m_Z^2/(\cos^2\theta_W m_W^2)$ is not strictly equal to one at tree--level in our model, as \eq{rho} shows, in spite of custodial symmetry. The reason is that custodial symmetry provides a relation, reported in \eq{cusmass}, among the charged and neutral mass matrices and it does not directly imply a relation among the $W$ and $Z$ mass eigenvalues appearing in \eq{rho}. Moreover, $\theta_{W}$ defined by Eq.~\eqref{thetaW} does not correspond to the physical one. Custodial Symmetry also implies that the $\widehat{T}$ parameter of EWPT, defined in terms of zero--momentum correlators and not of the pole masses, vanishes. This fact is explicitly verified to hold in our model in Appendix~\ref{AppB}.}
\beq
\label{wan}
\cos^2\theta_W\approx\left(\cos^2\theta_W\right)_{\textrm{exp}}=1-0.23\,.
\eeq
Similarly to the electric charge, also the weak mixing angle $\theta_W$ defined by \eq{thetaW}, and therefore in turn the couplings $g$ and $g'$, has to be close to the SM tree-level value. \eq{rho} also has one important implication on the masses of the new vectors. When combined with the custodial relation \eqref{cust}, it tells us that the charged and neutral $V$s are practically degenerate
\beq
\displaystyle
M_+^2=M_0^2 \left(1+\mc{O}(\%)\right)\,,
\eeq
and therefore they are expected to have comparable production rates at the LHC. Combining experimental searches of charged and neutral states could thus considerably improve the reach, as discussed in Ref.~\cite{deBlas:2012tc}. Furthermore, the small mass splitting implies a phase-space suppression of cascade decays, which can be safely ignored. In the following, when working at the leading order in the limit \eqref{hi}, we will ignore the mass splitting and denote the mass of the charged and the neutral states collectively as $M_V$. It is easy to check that in that limit $M_V=\hat{m}_V$.

Because of the hierarchy in the mass matrices, the mixing angles are naturally small. By looking at \eqs{ndiag}{cdiag} we estimate
\beq
\displaystyle
\theta_{N,C}\simeq c_H\zeta \frac{{\hat{m}}_{W,Z}}{{\hat{m}}_V}\lesssim10^{-1}\,.
\eeq
The couplings of the physical states are thus approximately those of the original Lagrangian  before the rotation. In particular, the $W$ and $Z$ couplings to fermions and among themselves mainly come from the SM Lagrangian and thus are automatically close to the SM prediction thanks to the hierarchy \eqref{hi} and to the parameter choice \eqref{wan}. Obviously this is not enough to ensure the compatibility of the model with observations. The $W$ and $Z$ couplings are very precisely measured, and the deviations due to new physics are constrained at the per mil level. These measurements translate into limits on the so-called EWPT observables \cite{Peskin:1992p1662,Barbieri:2004p1607,Cacciapaglia:2006pk}, which we will compute in Appendix~\ref{AppB}. This will allow us to quantify the additional restrictions on the parameter space, besides \eq{hi}.\footnote{In the following we will not ask EWPT to be strictly satisfied since this would be in contrast with the spirit of the Simplified Model approach adopted in this paper. We will take care of additional contributions to EWPT, not calculable within the Simplified Model, by considering bounds looser than the strict $95\%$ CL limits.}

\subsubsection*{Decay Widths}

Let us now turn to the resonance decays. The relevant channels are di-lepton, di-quark and di-boson. The latter category includes final states with $W$s, $Z$s and the Higgs boson. Decay into $W\gamma$ is also possible, but always with a tiny BR, as we will show below.

After rotating to the mass basis, the couplings of the neutral and charged resonances to left- and right-handed fermion chiralities can be written in a compact form\footnote{Because of quark mixings, the charged vector couplings should actually be multiplied by the appropriate CKM matrix elements.}
\bea\label{fermioncouplings}
\displaystyle
&&\left\{\bry{l}
\displaystyle
g_L^N= \dst \frac{g^2}{\gst }\frac{c_F}{2}\cos\theta_C+\left(g^Z_L\right)_{SM}\sin\theta_N\simeq\frac{g^2}{\gst }\frac{c_F}{2}\\
g_R^N= \dst \left(g^Z_R\right)_{SM}\sin\theta_N\simeq0
\ery
\right.\,,\nn\\
&&\left\{\bry{l}
\displaystyle
g_L^C=\frac{g^2}{\gst }\frac{c_F}{\sqrt{2}}\cos\theta_C+\left(g^W_L\right)_{SM}\sin\theta_C\simeq\frac{g^2}{\gst }\frac{c_F}{\sqrt{2}}\\
g_R^C=0
\ery
\right.\,,
\eea
for each fermion species $F=\{l,q,3\}$. In the above equation, $\left(g^{W,Z}_{L,R}\right)_{SM}$ denote the ordinary SM $W$ and $Z$ couplings (with the normalisation given by $g^{W}_{L}=g/\sqrt{2}$) that originate from the fermion covariant derivatives and contribute to the $V$ interactions because of the rotation. Given that the rotation angles are small, the couplings further simplify, as also shown in the equation. We see that the $V$s interact mainly with left-handed chiralities and that all the couplings for each fermion species are controlled by the parameter combination $g^2/\gst c_F$. This gives tight correlations among different channels 
\beq
\displaystyle
\Gamma_{V_\pm\to f\overline{f}'}\simeq 2\, \Gamma_{V_0\to f\overline{f}}
\simeq N_c[f]\,\left(\frac{g^2 c_F}{\gst }\right)^2\frac{M_V}{48\pi}\,,
\label{gaF}
\eeq
where $N_c[f]$ is the number of colors and is equal to $3$ for the di-quark and to $1$ for the di-lepton decays. The parameters $c_F=\{c_l,c_q,c_3\}$ control the relative BRs to leptons, light quarks and the third family. Furthermore through the partial width to $q\overline{q}$, $c_q$ controls the DY production rate, as we will discuss in the following Section.

The analysis is more subtle in the case of di-bosons. Obviously it is straightforward to compute the $V$ couplings to $W$, $Z$ and Higgs in the Unitary Gauge, after rotating to the mass eigenstates, and to obtain exact analytical formulas for the widths. We will not report the resulting expressions because they are rather involved and not particularly informative. It is instead useful to derive approximate decay widths in the limit of \eq{hi}, but the Unitary Gauge is not suited for this purpose. In the Unitary Gauge there are no direct couplings of $V$ to the SM vectors, these interactions only emerge from the mixing and are thus suppressed by the small mixing angles $\theta_{N,C}\lesssim10^{-1}$. On this basis, one would naively expect small di-boson widths and negligible BR. While this conclusion is correct for the processes involving transversely polarized SM vectors, the decay to zero-helicity longitudinal states is actually unsuppressed and potentially dominant. This is because the longitudinal polarization vectors grow with the energy of the process and even a tiny Unitary Gauge coupling can have a large effect in a high-energy reaction such as the decay of $V$. Rather than in the Unitary Gauge, one could work in an ``Equivalent Gauge'' \cite{Wulzer:2013tp} where the growth of the polarization vectors is avoided and the decay to longitudinals is straightforwardly estimated. However for the present analysis, it is sufficient to rely on a well-known result, the "Equivalence Theorem" \cite{Chanowitz:1985hj}, according to which the longitudinal $W$ and $Z$ are equivalent to the corresponding Goldstone Bosons in the high energy limit. Namely, the theorem states that if we parametrize the Higgs doublet as
\beq
\displaystyle
H=\left(\bry{c}
\displaystyle
\frac{\pi_2+i\,\pi_1}{\sqrt{2}}\\
\displaystyle
\frac{\hat{v}+h-i\,\pi_3}{\sqrt{2}}
\ery
\right)\,\equiv\,\left(
\bry{c}
\displaystyle
i\,\pi_+\\
\displaystyle
\frac{\hat{v}+h-i\,\pi_0}{\sqrt{2}}
\ery
\right)\,,
\eeq
the longitudinal $W$s and $Z$s will be described by $\pi_+$ and $\pi_0$, respectively, with $h$ being the physical Higgs boson. The vector fields $W_\mu$ and $Z_\mu$ can be safely ignored, and the terms in the Lagrangian \eqref{sml} which are relevant for the decay process are only
\bea\label{egV}
{\mathcal{L}}_\pi=&&-\frac14 \partial_{[\mu}V_{\nu ]}^a \partial^{[\mu}V^{\nu ]\;a}+\frac{M_V^{2}}2V_\mu^a V^{\mu\;a} -c_H\zeta M_V V_\mu^a \partial^\mu\pi^a\nn\\
&&+\frac{\gst  c_H}{2}V_\mu^a\left(\partial^\mu h\,\pi^a - h\,\partial^\mu \pi^a+\epsilon^{abc}\pi^b\partial^\mu\pi^c\right)
\nn\\
&&+\,2\,\gst  c_{VVHH}\zeta M_V h\,V_\mu^a V^{\mu\,a}\,+\,\frac{\gst }2 c_{VVV} \epsilon_{abc}V_\mu^a V_\nu^b \partial^{[\mu}V^{\nu]\; c}\,.
\label{lpi}
\eea
We omitted the kinetic term of the massless Goldstones and of the physical Higgs for shortness and we used $\hat{m}_V\equiv M_V\approx M_{\pm, 0}$.

We now see clearly why the longitudinal decays are unsuppressed. The second line of the Lagrangian \eqref{lpi} contains a direct interaction of the resonance with the Goldstones. This term gives a universal contribution to di-boson decays of the charged and neutral resonances, which are all controlled by the same parameter combination $\gst c_H$. If it dominates, all the relative BRs in these channels are fixed and uniquely predicted. This is indeed what happens in most of the parameter space of our model where, as discussed in the previous Section, $\zeta$ is small. When instead $\zeta$ is of order one, the widths receive additional contributions because the $V$-$\pi$ mixing in the first line of \eq{lpi} can not be ignored and must be eliminated by a field redefinition before reading the physical couplings. The redefinition is performed in two steps, first we shift
\beq\label{fr1}
\displaystyle
V_\mu^a\;\rightarrow\; V_\mu^a + \frac{c_H\zeta}{M_V}\partial_\mu\pi^a\,,
\eeq
and we cancel the mixing from the variation of the mass term. Second, in order to restore the canonical normalization of the $\pi$ kinetic term we rescale
\beq\label{fr2}
\displaystyle
\pi^a\;\rightarrow\; \frac{1}{\sqrt{1-c_H^2\zeta^2}}\pi^a\,.
\eeq
Notice that $1-c_H^2\zeta^2$ is necessarily positive to avoid negative-defined mass matrices, see \eqs{ndiag}{cdiag}.

After these redefinitions, the relevant interactions become\footnote{To obtain the equations that follow we also made use of the $V$ equations of motion or, equivalently, of further field redefinitions.}
\bea\label{egV1}
\displaystyle
&&\frac{\gst c_H}{2(1-c_H^2\zeta^2)}\left[1+c_Hc_{VVV}\zeta^2\right]\epsilon^{abc}V_\mu^a\pi^b\partial^\mu\pi^c\nn\\
&&-\,\frac{\gst c_H}{\sqrt{1-c_H^2\zeta^2}}\left[1-4c_{VVHH}\zeta^2 \right]\,h\,V_\mu^a\partial^{\mu}\pi^a\,,
\eea
and the partial widths are immediately computed 
\beq\label{bbw}
\begin{array}{lllll}
\dst \Gamma_{V_0\to W^+_LW^-_L} & \simeq & \dst \Gamma_{V_\pm\to W^\pm_LZ_L} & \simeq &\dst \frac{ \gst ^2 c_H^2 M_V}{192\pi}\frac{(1+ c_Hc_{VVV} \zeta^2)^2}{(1- c_H^2 \zeta^2)^2}= \frac{ \gst ^2 c_H^2 M_V}{192\pi}\left[1+{\mathcal{O}}(\zeta^2)\right]\,,\vspace{2mm}\\
\dst \Gamma_{V_0\to Z_Lh} & \simeq &\dst \Gamma_{V_\pm\to W^\pm_Lh} & \simeq & \dst\frac{\gst ^2 c_H^2 M_V}{192\pi }\frac{(1-4 c_{VVHH} \zeta^2)^2}{1- c_H^2\zeta^2}=\frac{\gst ^2 c_H^2 M_V}{192\pi }\left[1+{\mathcal{O}}(\zeta^2)\right]\,.
\end{array}
\eeq
We checked that these expressions reproduce the exact widths up to ${\mathcal{O}}(\hat{m}_{W,Z}^2/\hat{m}_V^2)$ corrections, as expected. The channels which are not reported in the above equations are either forbidden, like $hh$ and $\gamma\gamma$ decays, or suppressed like the decays to transverse polarizations which follow the estimate based on the Unitary Gauge and experience mixing angle suppressions. In particular, the $W\gamma$ final state is generically suppressed, and exactly vanishes in the explicit models described in Section~4 that obey Minimal Coupling \cite{Grojean:2011vu}. Notice that the dominance of the longitudinal polarizations in the di-boson decays is an important simplification for the interpretation of experimental searches. Indeed, the boosted vector boson reconstruction could slightly depend on the helicity because different helicities would lead to different kinematical distributions of the final decay products. In our case one can safely restrict to the longitudinal case when computing the efficiencies and ignore the transverse.

From the analysis of the present Section a very simple picture emerges. At small $\zeta$, all the decay widths are fixed, for a given resonance mass, by the couplings $g^2c_F/\gst $ and $\gst c_H$, which therefore control the BRs in all the relevant channels. Furthermore, since the dominant processes are $2\to1$ reactions and can be parametrized, as we will do in the following Section, in terms of the corresponding decay widths, the two parameters $g^2c_F/\gst $ and $\gst c_H$ also control the production rate. Therefore the phenomenology of the model is entirely described, to a good approximation, in terms of the two couplings  $g^2c_F/\gst $ and $\gst c_H$ and the mass $M_{V}$, making, as anticipated in Section~2.1, $c_{VVV}$, $c_{VVHH}$ and $c_{VVW}$ basically irrelevant. When studying our model at the LHC the latter parameters can be safely ignored, or set to benchmark values inspired by explicit models, and the limits can be presented in terms of the relevant couplings. Additional plausible assumptions, like the universality of lepton and quark couplings, could further simplify the analysis. 

Now that the general picture is clear, we can get an idea of the expected widths and BRs by studying explicit models. We consider two benchmark models, A and B, which correspond to two explicit models describing the heavy vectors, namely those in Refs.~\cite{1980PhRvD..22..727B} and \cite{Contino:2011np}, respectively. As discussed in detail in Section~\ref{4}, all the $c$'s are fixed to specific values in these models and the only free parameters are the resonance coupling $\gst$ and its mass $M_V$.\footnote{Actually the model of Ref.~\cite{Contino:2011np} has an additional freedom in the choice of $c_H$, which depends on the additional parameter $a_\rho=m_\rho/(g_\rho f)$ as in \eq{mathchingNSM}, we define the benchmark model B by setting $a_{\rho}=1$.} We refer to the benchmark points at fixed $\gst=\bar{g}_{V}$ with the notation A$_{\gst=\bar{g}_{V}}$ and B$_{\gst=\bar{g}_{V}}$. Moreover, since models A and B are inspired, respectively, by weakly coupled extensions of the SM gauge group and strongly coupled scenarios of EWSB, {\it{i.e.}}~Composite Higgs models, we will consider them in different regions of $\gst$, relatively small, $\gst\lesssim 3$, and relatively large, $\gst\gtrsim 3$, respectively.

\begin{figure}[t!]
\label{fig:widthsBRs}
\begin{center}
\includegraphics[scale=0.246]{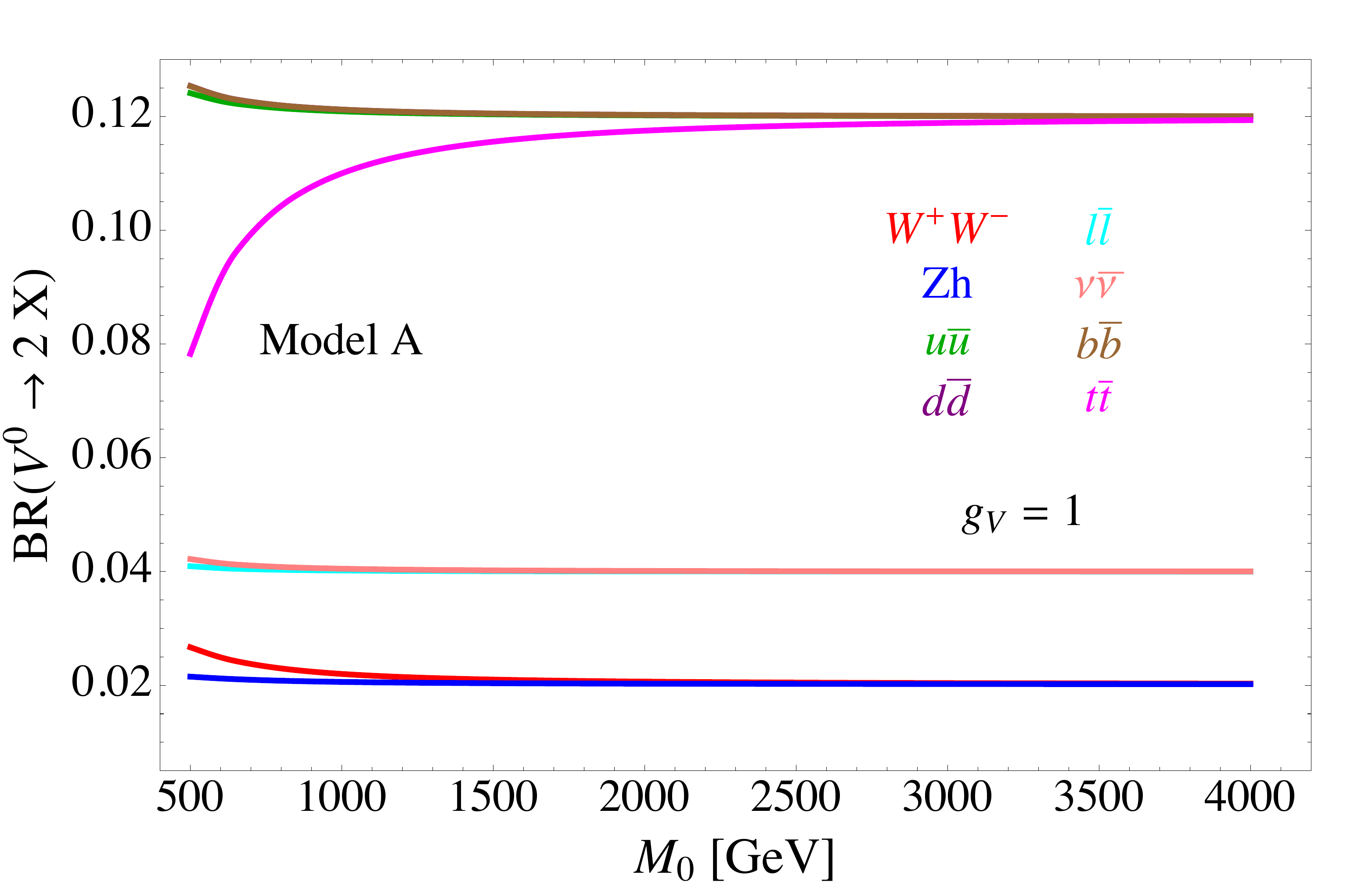}
\includegraphics[scale=0.246]{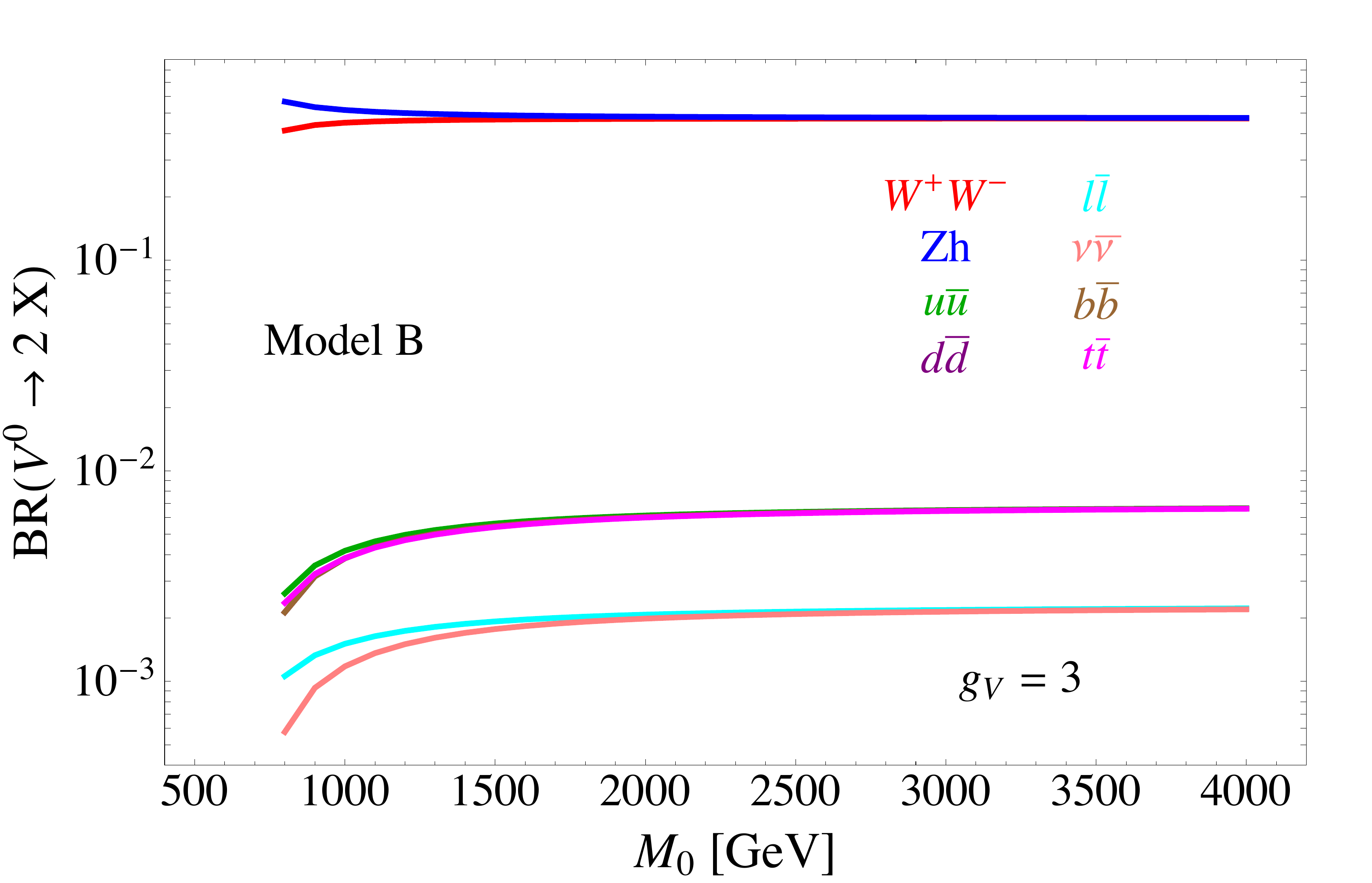}\\
\hspace{1.2mm}\includegraphics[scale=0.242]{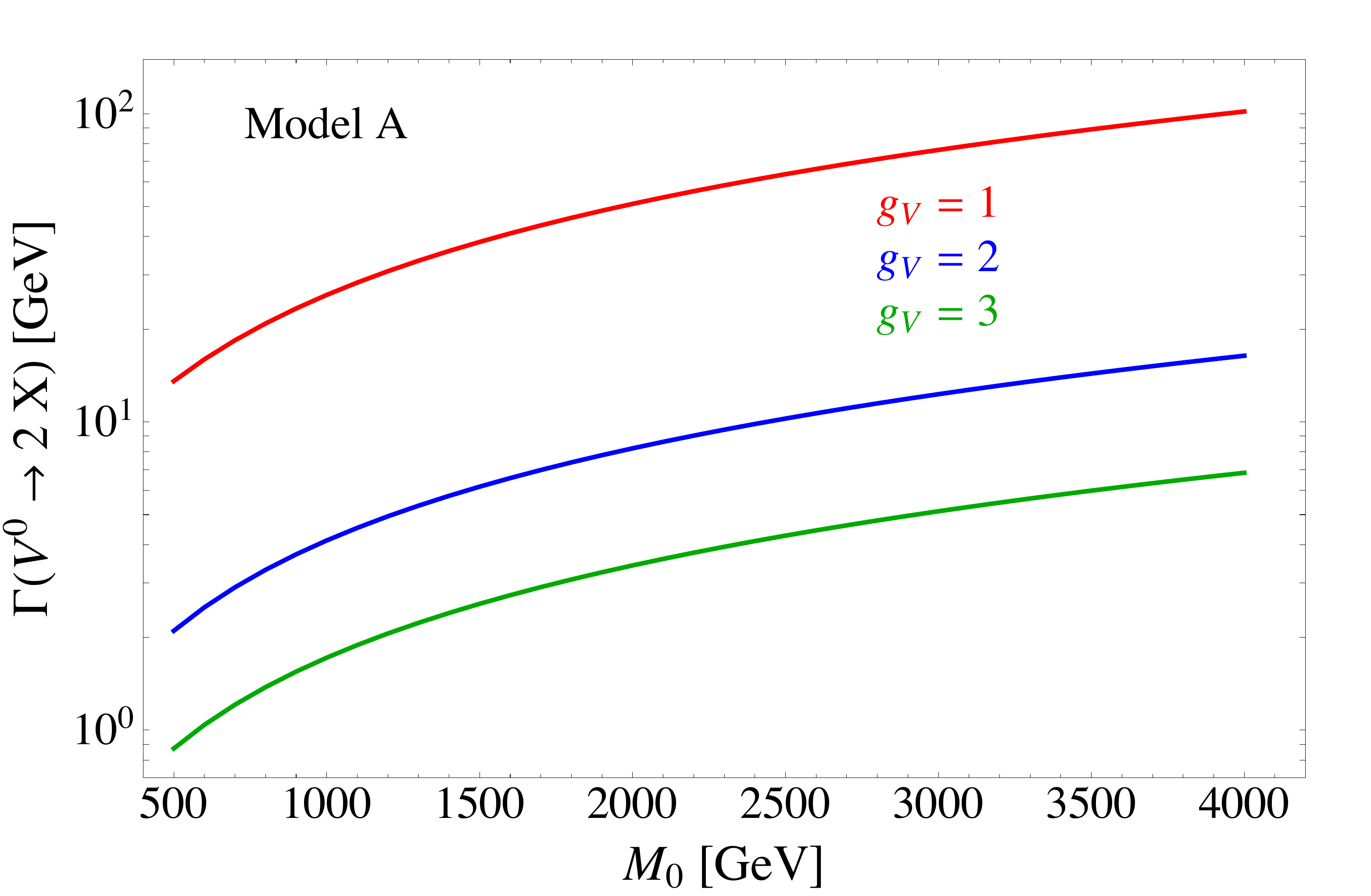}\hspace{1.2mm}
\includegraphics[scale=0.242]{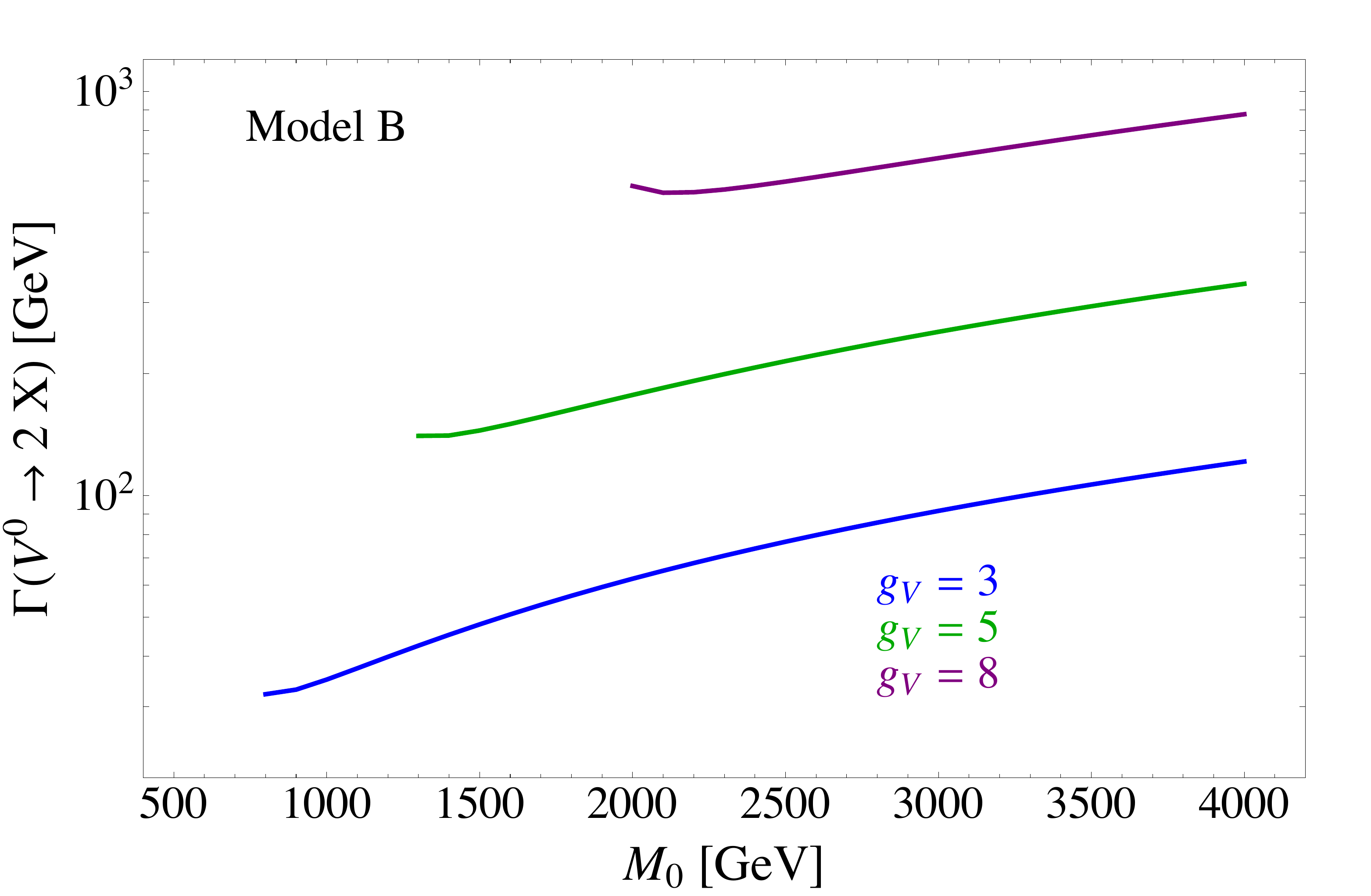}
\caption{\small Upper panel: Branching Ratios for the two body decays of the neutral vector $V^{0}$ for the benchmarks A$_{\gst=1}$ (left) and B$_{\gst=3}$ (right). Lower panel: Total widths corresponding to different values of the coupling $\gst$ in the models A (left) and B (right).}
\end{center}
\end{figure}

In Figure \ref{fig:widthsBRs} we show the BRs (upper panels) and the total widths (bottom panels) as functions of the mass in models A (left panels) and B (right panels) for different values of $\gst$. As expected from the discussion above, model A$_{g_{V}=1}$, which predicts
\beq\label{BRwidthLSM}
\gst c_{H}\simeq g^2c_F/\gst \simeq g^{2}/\gst\,,
\eeq 
has comparable BRs into fermions and bosons, with a factor of two difference coming from a numerical factor in the amplitude squared (cfr.~\eq{bbw} with \eq{gaF}). The difference between the BRs into leptons and quarks is due to the color factor, since $c_{F}$ is universal both in A and B. The total width in model $A$ decreases with increasing $\gst$ because of the overall suppression $g^{2}/\gst$ in \eq{BRwidthLSM}. In model B, on the contrary, $c_H$ is unsuppressed
\beq\label{BRwidthNSM}
\gst c_{H}\simeq -\gst\,,\qquad g^2c_F/\gst \simeq g^{2}/\gst\,.
\eeq 
Therefore, for model B$_{\gst= 3}$ the dominant BRs are into di-bosons and the fermionic decays are extremely suppressed, of the order of one percent to one per mil. Moreover, the total width increases with increasing $\gst$ since it is dominated by the di-boson width which grows with $\gst$ as expected from \eq{BRwidthNSM}. Finally, in model B we see that a very large coupling $\gst$ (the case of $\gst=8$ is shown in the Figure) leads to an extremely broad resonance, with $\Gamma/M\gg0.1$, for which the experimental searches for a narrow resonance are no longer motivated. For this reason we expect, if no further suppression is present in the parameter $c_{H}$, to be able to constrain heavy vector models from direct searches only up to $\gst$ of the order $6-7$. For larger couplings different searches become important, like for instance those for four fermion contact interactions (see for instance Refs.~\cite{Domenech:2012ai,deBlas:2013qqa}).

\subsection{Production rates parameterized}
\label{2.2}

The two main production mechanisms of the new vectors are DY and VBF.\footnote{We ignore the production in association with a gauge boson because it is always negligible for $V$ masses and couplings in the interesting region.} They are both $2\to1$ processes, therefore their cross-section can be expressed as
\beq
\sigma(p p \to V+X) = \sum_{i,j \,\in\, p}\f{\Gamma_{V\,\to \,i j}}{M_{V}} \f{16\pi^{2}(2J+1)}{(2S_{i}+1)(2S_{j}+1)}\f{C}{C_{i}C_{j}}\f{dL_{i j}}{d\hat{s}}\Bigg|_{\hat{s}=M_{V}^{2}}\,,
\label{CS}
\eeq
in terms of the partial widths $\Gamma_{V\,\to\, i j}$ of the inverse decay process $V\to ij$. In the equation, $i,j=\{q,\overline{q},W,Z\}$ denote the colliding partons in the two protons, and $dL_{i j}/d\hat{s}|_{\hat{s}=M_{V}^{2}}$ is the corresponding parton luminosity evaluated at the resonance mass. The factor $J$ is the spin of the resonance and $C$ its color factor, $S_{i,j}$ and $C_{i,j}$ are the same quantities for the initial states. If needed, the cross-section above could be corrected by a $k$-factor taking into account higher order QCD corrections. It is important to remark that the only terms in the above equation that carry some dependence on the model parameters are the partial widths $\Gamma_{V\,\to\, i j}$, while the parton luminosities are completely model-independent factors that can be encoded in universal fitted functions. Since the widths are analytical functions of the parameters, this allows us to compute the production rates analytically, making the exploration of different regions of the parameter space extremely fast as we will do in the following Section. Simple approximated expressions of the partial widths are reported in \eqs{gaF}{bbw}, however in the following we will make use of the exact expressions embedded in a {\sc Mathematica} code and available through a web interface in \cite{projectwebpage}.

While the meaning of \eq{CS} is completely obvious for DY, and the corresponding luminosities are immediately computed by convoluting the quark and anti-quark Parton Distribution Functions (PDF), the case of VBF requires additional comments. In \eq{CS} we are regarding the $W$ and $Z$ bosons as ``partons'', or constituents of the proton, relying on the validity of the Effective $W$ Approximation (EWA) \cite{1985NuPhB.249...42D}. By the EWA, the vector bosons' PDFs and in turn the corresponding luminosities are obtained by convoluting those of the initial quarks with appropriate splitting functions. More details can be found, for instance, in Ref.~\cite{2011arXiv1110.3906T}. A priori, all the $W$ and $Z$ polarizations should be taken into account, as well as photons. However we saw above that the only sizable partial widths are those with longitudinal vectors, thus we can safely restrict to the $W^+_L W^-_L$ and $W^\pm_L Z_L$ initial states for the production of the neutral and the charged $V$, respectively. It is also important to remark that the EWA has a limited range of validity and it is not expected to reproduce the full partonic process $pp\to Vjj$ in all possible regimes \cite{Borel:2012wg}. It might fail if the $W$ collision is not sufficiently hard, which however is never the case for $M_V\gtrsim1$~TeV, or if other kinematically enhanced configurations exist, besides the standard VBF ones with forward energetic jets, and contribute to the partonic process. Also this second issue does not arise in our case. Finally, it might happen that other processes give a sizable contribution to the $Vjj$ final state. This occurs in our case when the $V$ coupling to fermions is much larger than the one to vector bosons. In this case the $Vjj$ final state could arise, for instance, by dressing the DY process with QCD initial state emissions. However when this happens the ordinary DY, without extra parton-level jets, is necessarily the dominant production mechanism and the failure of the EWA is irrelevant at the practical level. We checked that the partonic cross-section is extremely well reproduced by \eq{CS} in all regions of parameter space, where the VBF rate is not completely negligible, up to order $1\%$ of the DY one.

\begin{figure}[t!]
\label{fig:partonlumi}
\begin{center}
\includegraphics[scale=0.368]{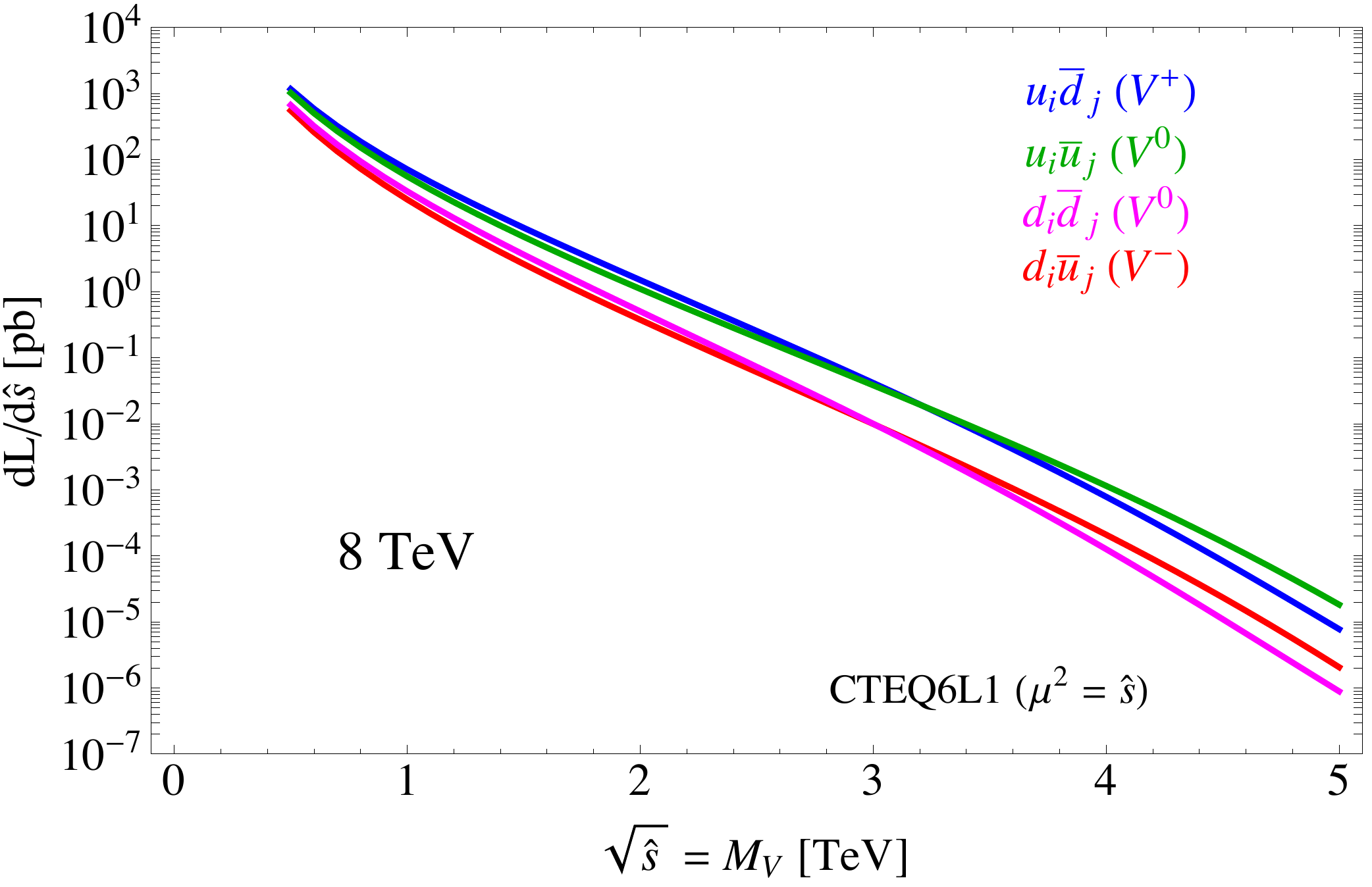}
\includegraphics[scale=0.368]{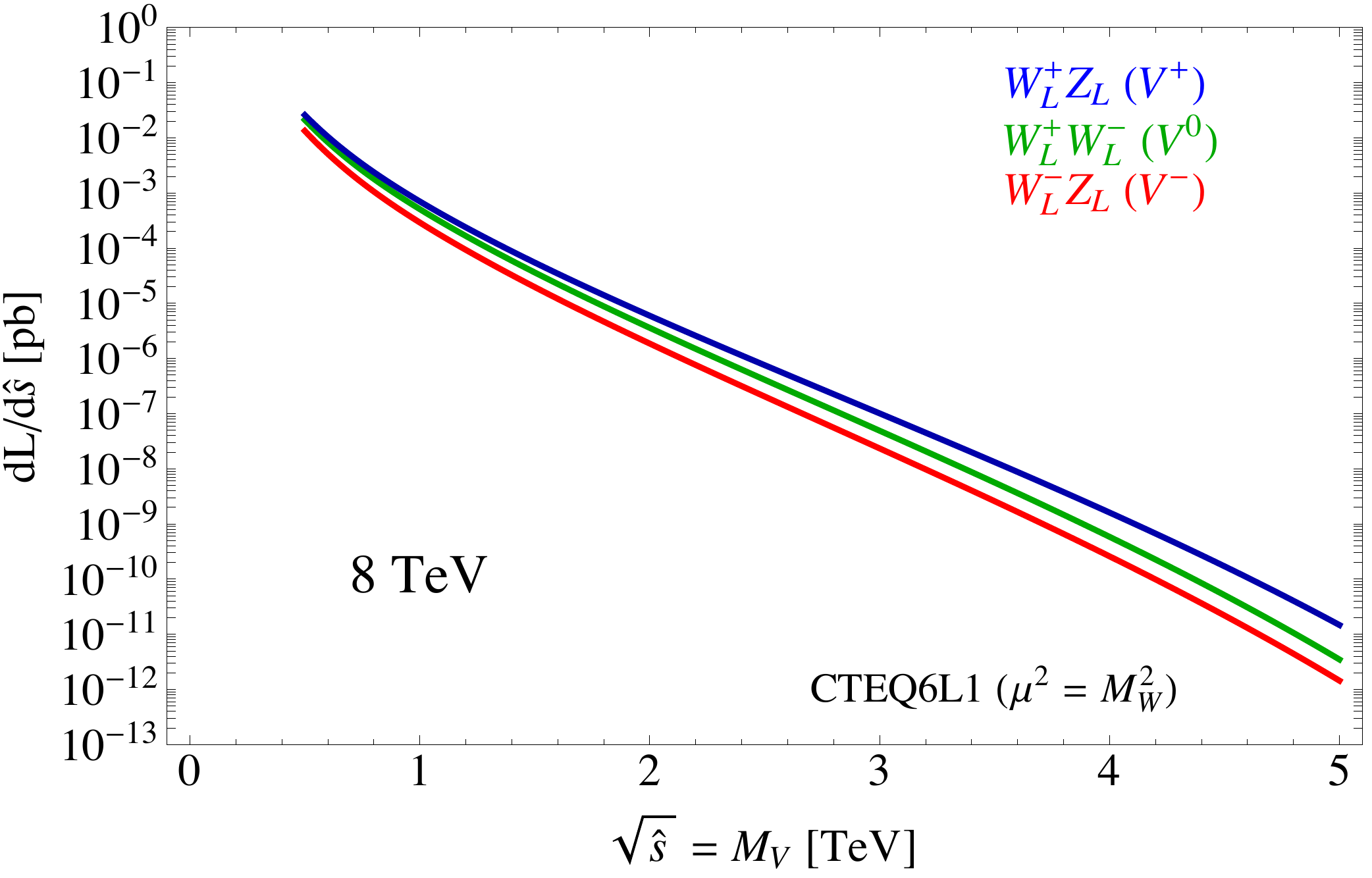}
\includegraphics[scale=0.368]{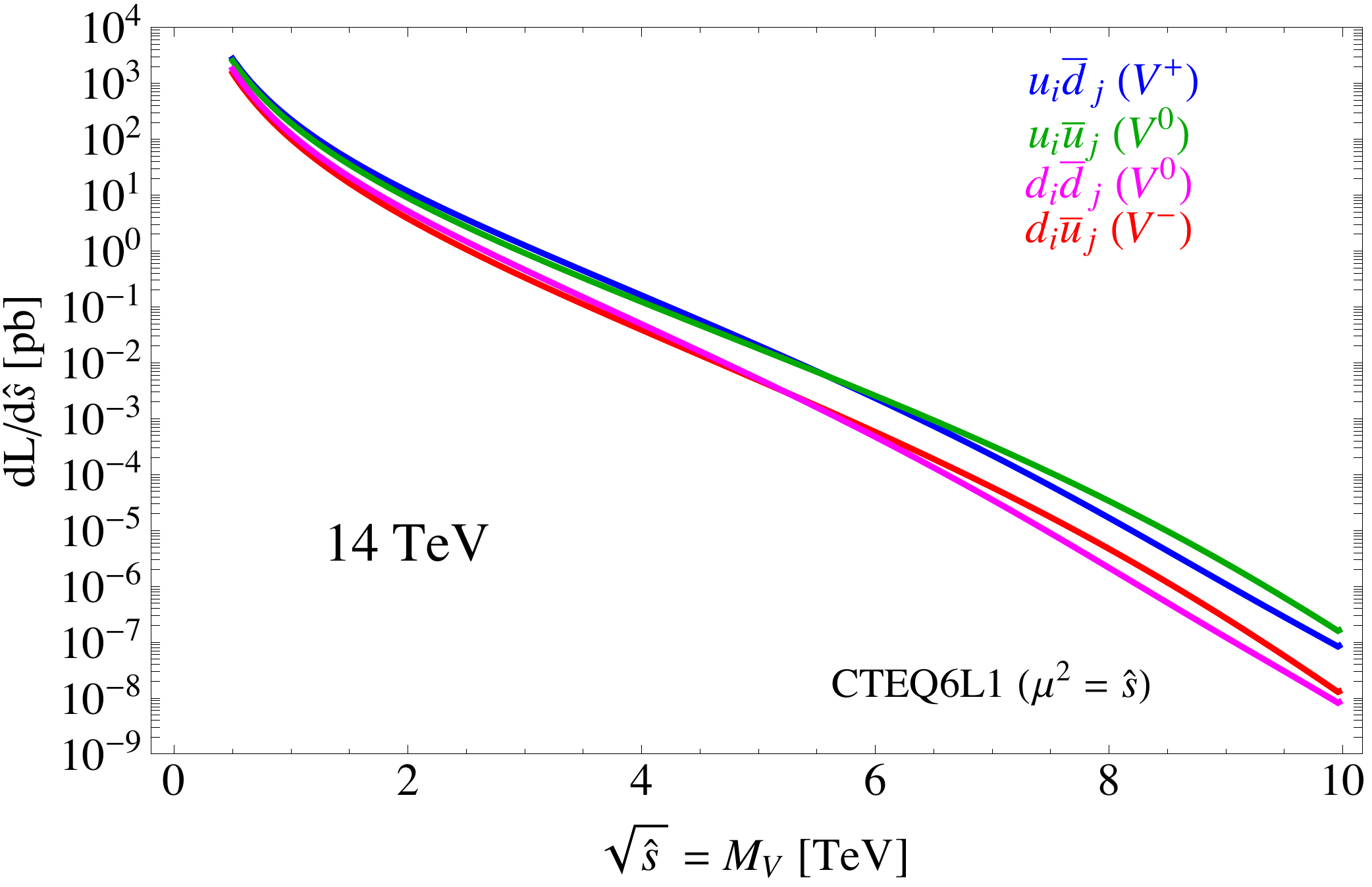}
\includegraphics[scale=0.368]{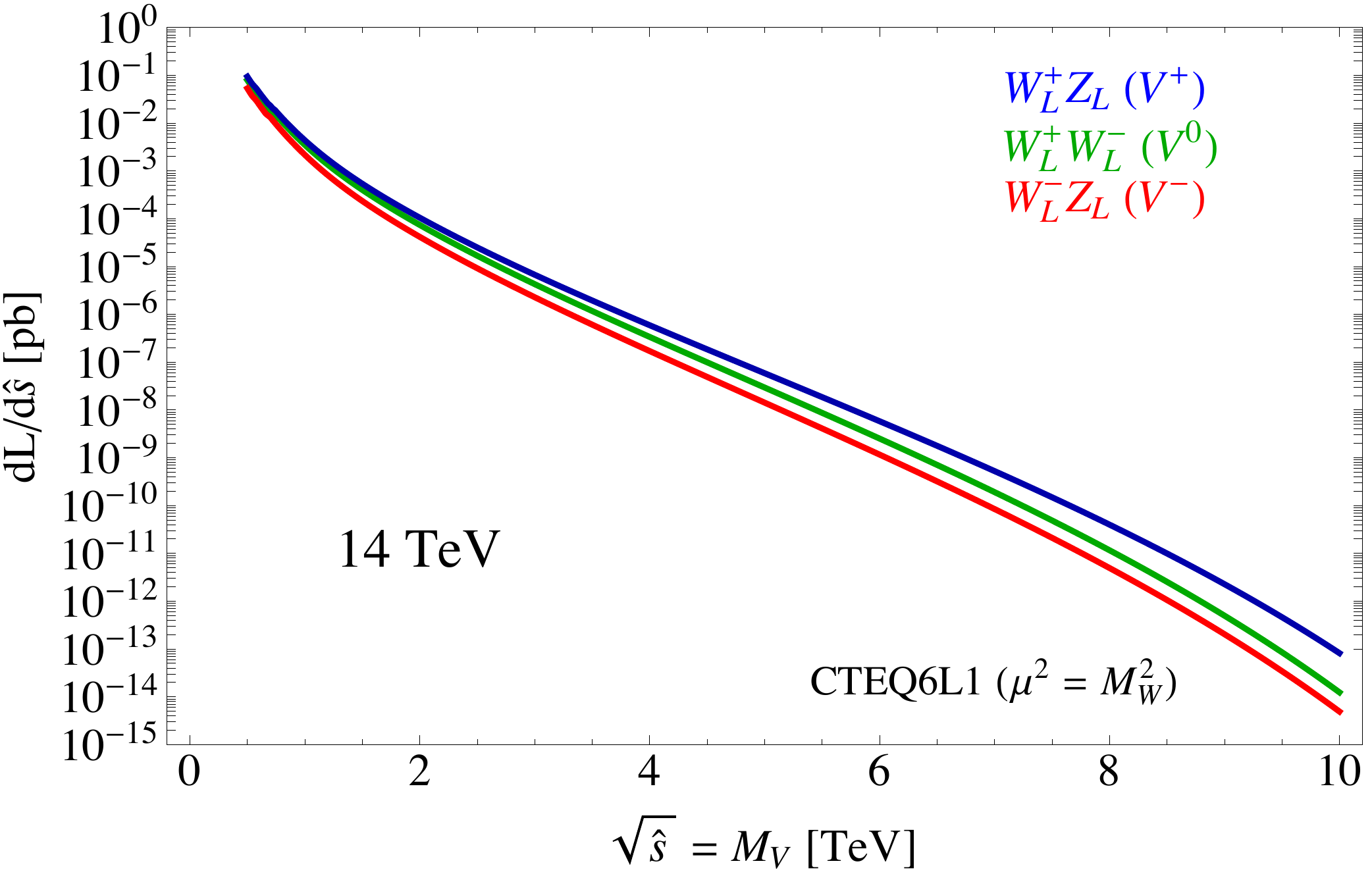}
\includegraphics[scale=0.368]{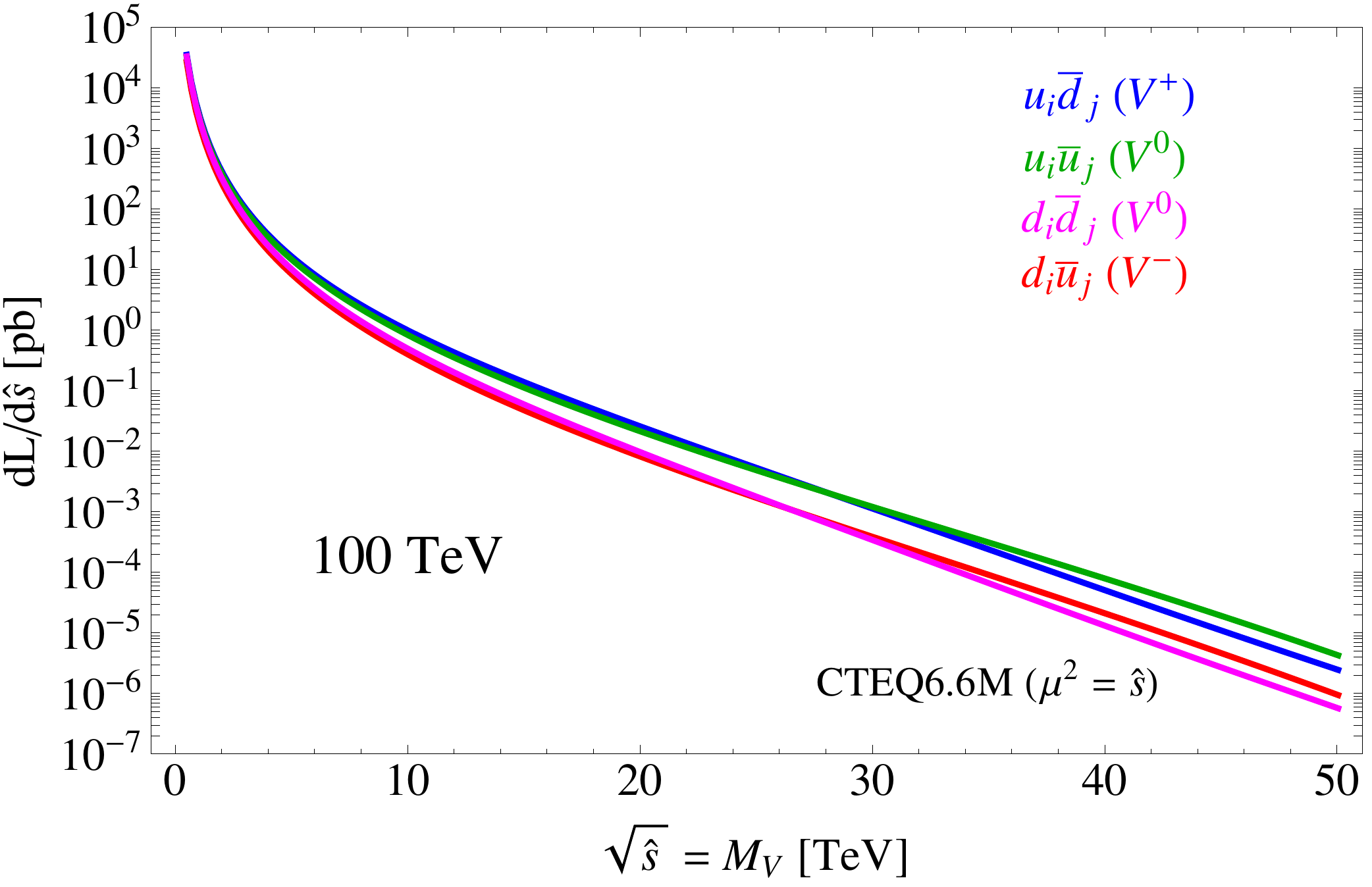}
\includegraphics[scale=0.368]{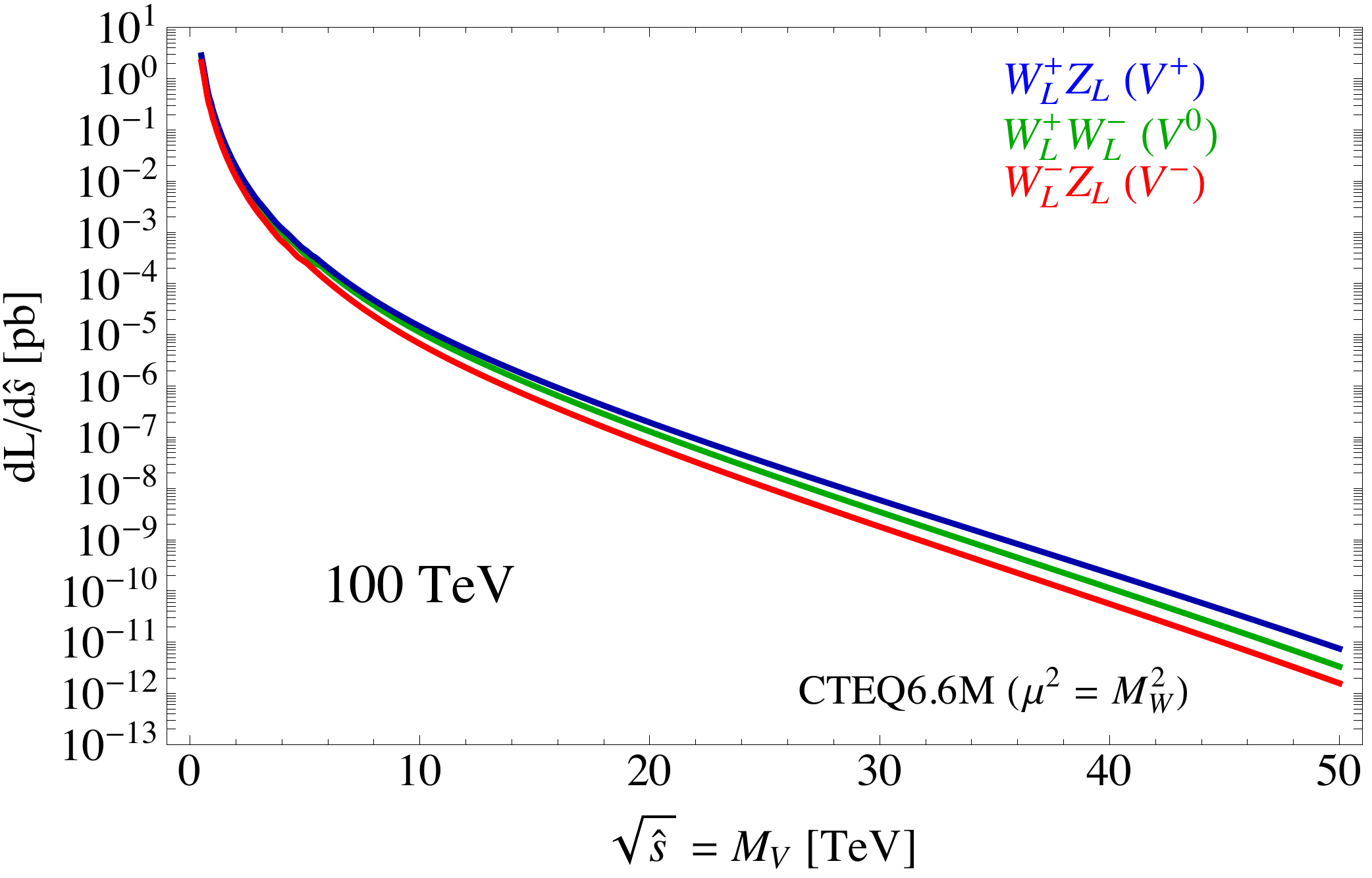}
\caption{\small Value of the $dL/d\hat{s}$ for quark initial states (left plots) and longitudinal vector boson initial states (right plots) for $8$ TeV (first row) and $14$ TeV (second row) LHC and for a hypothetical $100$ TeV (last row) $pp$ collider.}
\end{center}
\end{figure}

\begin{figure}[t!]
\begin{center}
\includegraphics[scale=0.368]{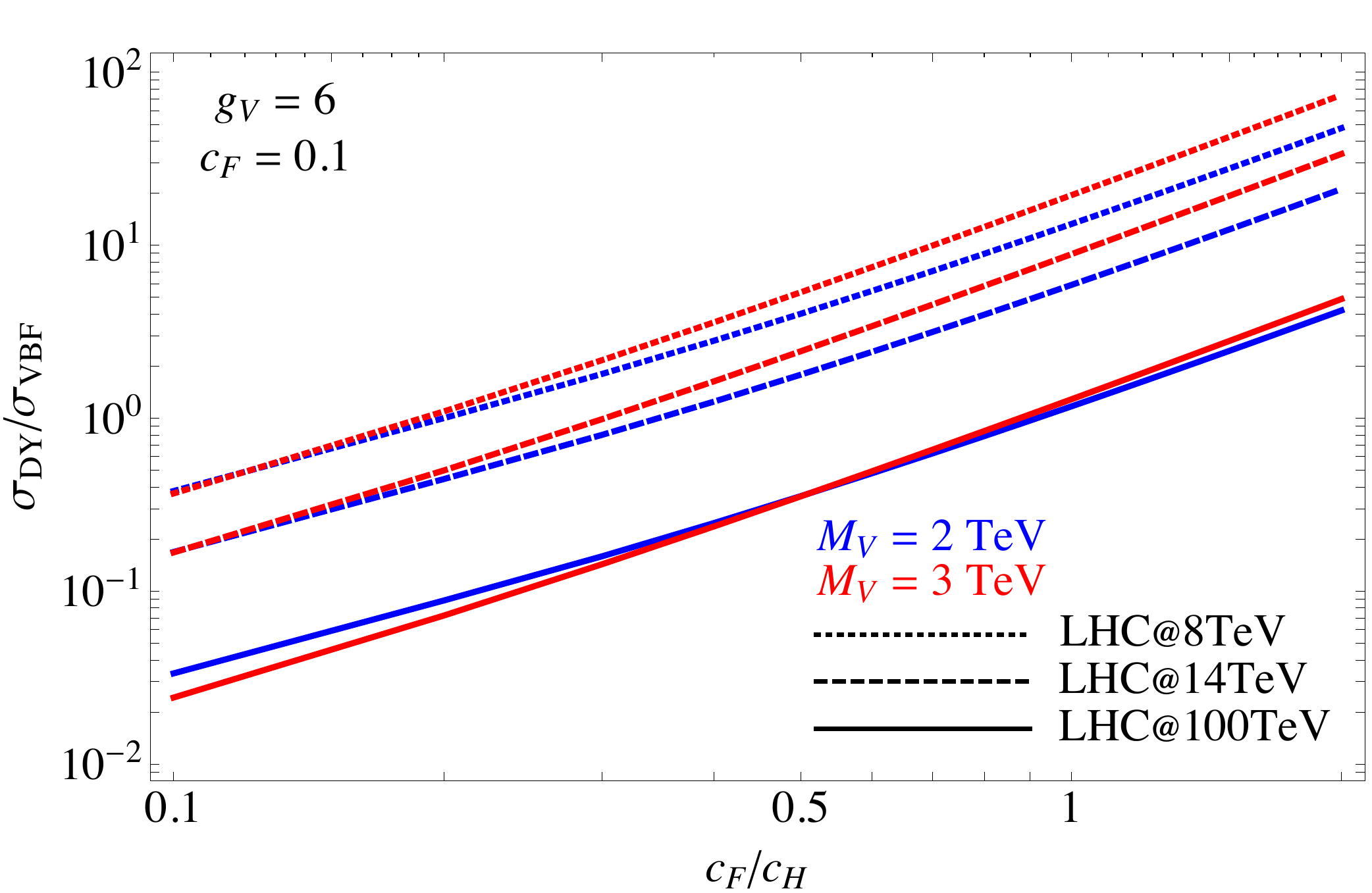}
\includegraphics[scale=0.368]{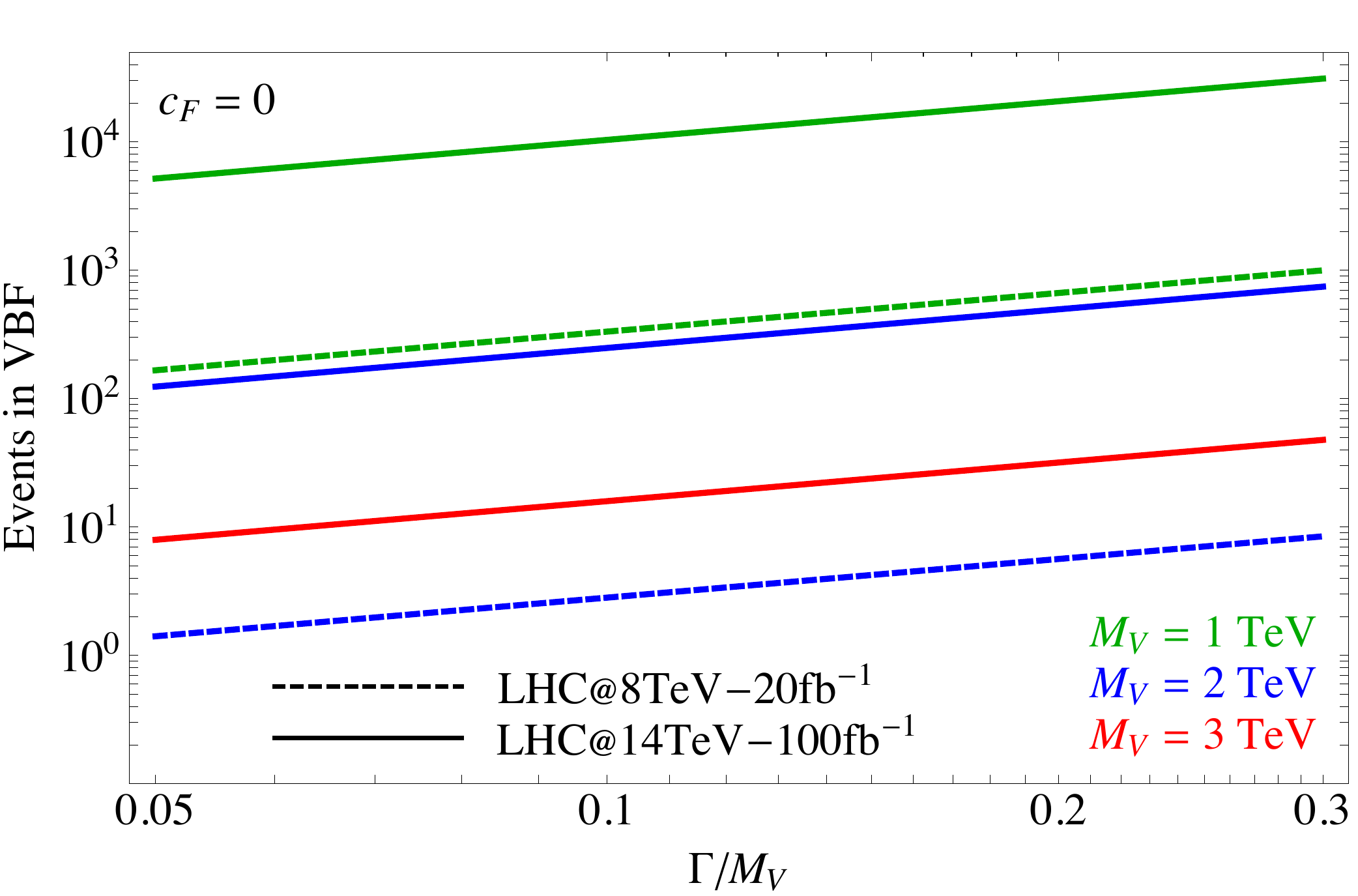}
\caption{\small Left panel: ratio of DY over VBF production cross-section for different masses and collider energies. Right panel: expected number of events from VBF production for different masses, collider energies and integrated luminosities, assuming $c_{F}=0$. For illustration we consider only $V^{+}$.}\label{fig:VBFvsDY}
\end{center}
\end{figure}

The parton luminosities for the various production processes are shown in Figure~\ref{fig:partonlumi}. The VBF luminosity is obviously much smaller than the DY one because of the $\alpha_{EW}$ suppression in the vector boson PDFs. Therefore VBF only has a chance of being comparable to DY if the widths in $q\overline{q}$ are much smaller than those in di-bosons. We see from \eqs{gaF}{bbw} that this can happen only at large $\gst$, and in the strongly coupled scenario, {\it{i.e.}} model B, where $c_{H}$ is not suppressed. In the left panel of Figure \ref{fig:VBFvsDY} we show the ratio of the production cross-section by DY and VBF (for the $V^{+}$ for illustration) as a function of the $c_{F}/c_{H}$ ratio, for different masses at the LHC at $8$ TeV and $14$ TeV and at a hypothetical $100$ TeV $pp$ collider\footnote{Studies of new vector resonances at future hadron collider were done in the context of the Snowmass 2013, see Refs.~\cite{Han:2013wa,Hayden:2013ve,Godfrey:2013uu}.}. Since, as shown by \eq{CS} the production cross-section only depends on the corresponding partial widths, we expect the ratio of the cross-sections to depend only on $c_{F}/c_{H}$, up to small corrections of order $\zeta^{2}$. The overall normalization of the $c_{H},c_{F}$ parameters has been set to $c_{F}=0.1$. This is necessary because for large $\gst$, which is the case of interest, when the ratio $c_{F}/c_{H}$ becomes small, a $c_{F}$ of order one would imply a large $c_{H}$, which would lead to an unacceptably large total width. The left panel of Figure \ref{fig:VBFvsDY} shows that even for a large coupling $\gst=6$, a ratio of the order of $3(5)$ at $14(8)$ TeV is needed for VBF to become comparable with DY. This ratio can be regarded as the needed suppression in $c_{F}$ with $c_{H}$ still being of order one. This further suppression is not expected in general in explicit models, making VBF typically less relevant than DY. For this reason we will ignore VBF in the analysis of the following section and consider only DY production. For the $100$ TeV option the situation is different, since for $c_{F}\approx c_{H}$ the DY and VBF production cross-sections are comparable for resonances with masses in the few TeV region. Obviously, for higher masses in the range of a few tens of TeV, close to the reach of the $100$ TeV collider, we expect VBF to be again subleading with respect to DY.

If the coupling to fermions is suppressed for some reason, $c_{F}\approx 0$, VBF becomes the dominant production mechanism and it is worth asking ourselves what the mass reach of the LHC would be in this case. In order to answer, we notice that for $c_{F}\approx 0$ the fermionic decays are suppressed and thus the total resonance width $\Gamma$ is simply twice the di-boson one which controls, by \eq{CS}, the production rate. Therefore, for a given mass, the expected number of produced vectors (again $V^{+}$ for illustration) can be expressed as a function of the $\Gamma/M_V$ ratio as shown in the right plot of Figure \ref{fig:VBFvsDY} for different masses, collider energies and integrated luminosities. The $\Gamma/M_V$ ratio is an important parameter, as it quantifies to what extent the resonance can be reasonably regarded as a particle. By requiring, for instance, $\Gamma/M_V\lesssim0.3$ we can obtain an upper bound on the expected signal. We see that at the $8$~TeV LHC with $20$ fb$^{-1}$ a reasonable number of events can be obtained only for very low masses, around $1$ TeV, where however we expect the resonance to be excluded already by EWPT. At the LHC at $14$ TeV with $100$ fb$^{-1}$, a sizable number of events for a narrow resonance ($\Gamma/M\lesssim 0.1$) seems possible even for relatively high masses, up to around $2.5$~TeV. This makes VBF more interesting at the LHC at $14$ TeV, at least to explore specific scenarios where the coupling to fermions is suppressed.

\section{Data and Bounds}
\label{3}

The ATLAS and CMS collaborations have performed a number of searches for heavy resonances, not only spin-1, decaying into different final states both at the $7$ and $8$ TeV LHC. A summary of the relevant searches for the study of a heavy vector boson, either charged or neutral, is given in Table \ref{Table:Searches}. Most of those analyses present limits on the production cross-section times BR, \mbox{$\sigma\times$BR}, as a function of the resonance mass.\footnote{We will not consider the ATLAS and CMS di-jet searches \cite{Aad:2014aqa, CMS-PAS-EXO-12-059} because the limits are given in terms of an acceptance factor. Furthermore the sensitivity of the latter channels is rather reduced.} If taken at face value, these results are thus very easy to interpret because, as explained in the previous Section, both $\sigma$ and BR can be expressed in analytical form. This allows us to draw exclusion contours in the parameter space of our model in a very efficient way, as we will show in Section~3.2. However, an important message of our paper, on which we will elaborate in Section~3.3, is that in some cases the experimental limits, depending on the details of the analysis, are not properly set on \mbox{$\sigma\times$BR} because of the effects of the finite resonance width. This problem is particularly acute in strongly coupled scenarios, where the resonance is broader and should be more carefully investigated by the experimental collaborations.

\subsection{A first look at the LHC bounds}

\begin{table}[t]
\begin{center}
{
\begin{tabular}{c|c|c}
\hline
\hline
Experiment & Channel & Reference \\ 
\hline
ATLAS & \multirow{2}{*}{$l^{+}l^{-}$} 
& \cite{Aad:2014cka} \\ 
CMS & & \cite{CMS-PAS-EXO-12-061} \\
\hline
ATLAS & \multirow{2}{*}{$l \nu$} 
& \cite{ATLAS-CONF-2014-017} \\ 
CMS & & \cite{CMS-PAS-EXO-12-060} \\
\hline
ATLAS & $\tau \tau$ & \cite{ATLAS-CONF-2013-066} \\
\hline
ATLAS & \multirow{2}{*}{$WZ\to 3l\nu$} & \cite{Aad:2014pha} \\ 
CMS  && \cite{Khachatryan:2014xja} \\ 
\hline
CMS & $qW,qZ,WW/WZ/ZZ\to jj$ & \cite{CMS-PAS-EXO-12-024} \\ 
\hline
CMS & $WW\to l\nu jj$ & \cite{CMS-PAS-EXO-12-021} \\
\hline
\grey{ATLAS} & \grey{\multirow{2}{*}{$jj$}} 
& \cite{Aad:2014aqa} \\
\grey{CMS} && \cite{CMS-PAS-EXO-12-059} \\
\hline
\grey{CMS} & \grey{$b\bar{b}$, $bg$} & \cite{CMS-PAS-EXO-12-023} \\
\hline
ATLAS & \multirow{2}{*}{$t \bar{t}$} & \cite{ATLAS-CONF-2013-052} \\
CMS & & \cite{CMS-PAS-B2G-12-005} \\
\hline
ATLAS & \multirow{2}{*}{$t \bar{b}$} 
& \cite{ATLAS-CONF-2013-050} \\
CMS && \cite{CMS-PAS-B2G-12-010} \\
\hline
\hline
\end{tabular}
}
\\[0.1cm]
\caption{\small Summary of experimental searches relevant for heavy vector resonances. We have not mentioned the searches of Refs.~\cite{CMS-PAS-EXO-12-022,ATLAS-CONF-2012-150} in the $ZZ$ final state, since this channel is not present in our Simplified Model. The gray entries will not be used when showing bounds in Figure \ref{Fig:BoundsonCS} since the acceptances for a heavy vector are not reported in the experimental analyses.}\label{Table:Searches}
\end{center}
\end{table}

\begin{figure}[t!]
\label{Fig:BoundsonCS}
\begin{tikzpicture}
    \begin{scope}[xshift=7.9cm,yshift=-1cm]
    \node {\includegraphics[scale=0.24]{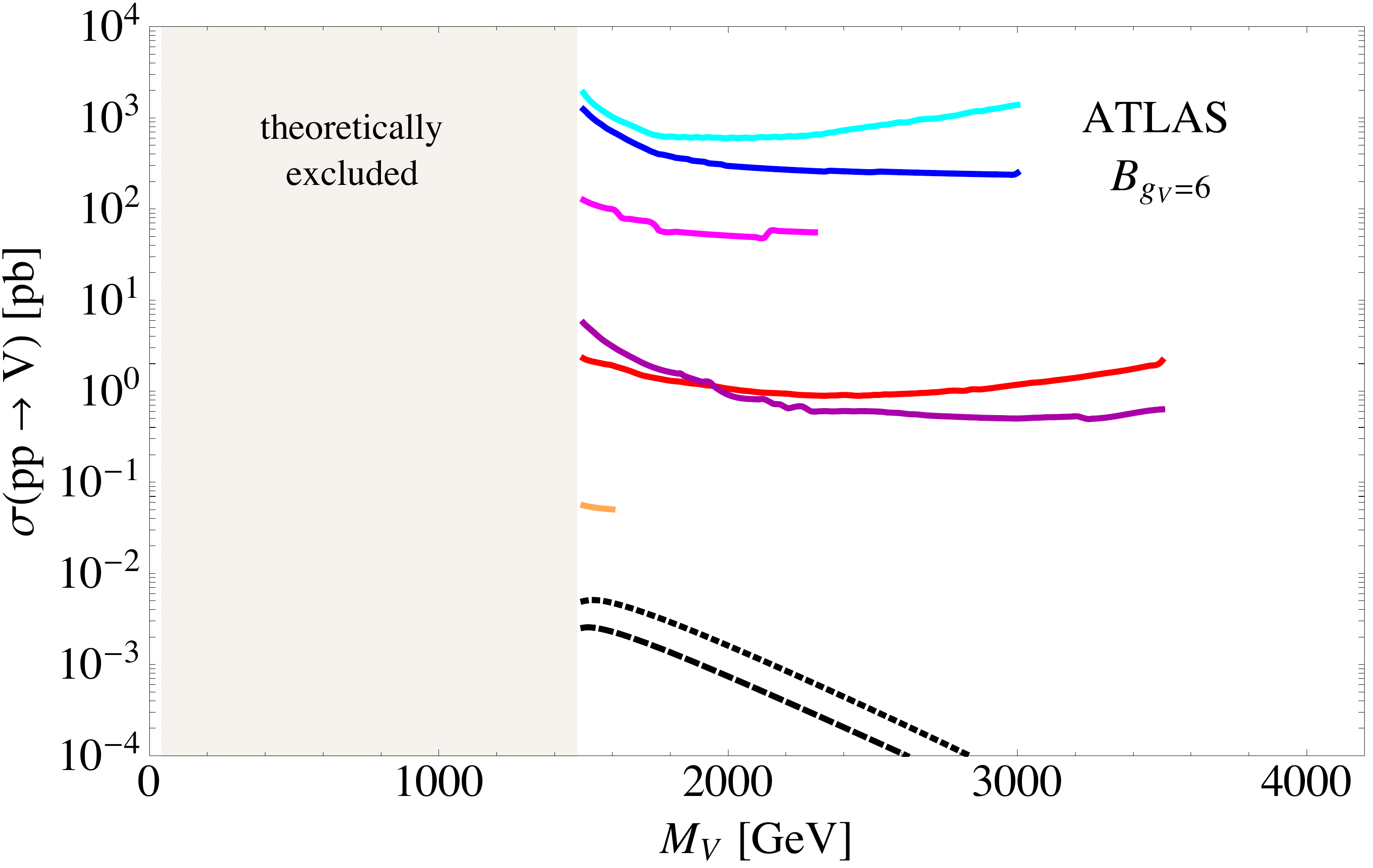}};
    \end{scope}
    \begin{scope}[yshift=-1cm]
    \node {\includegraphics[scale=0.24]{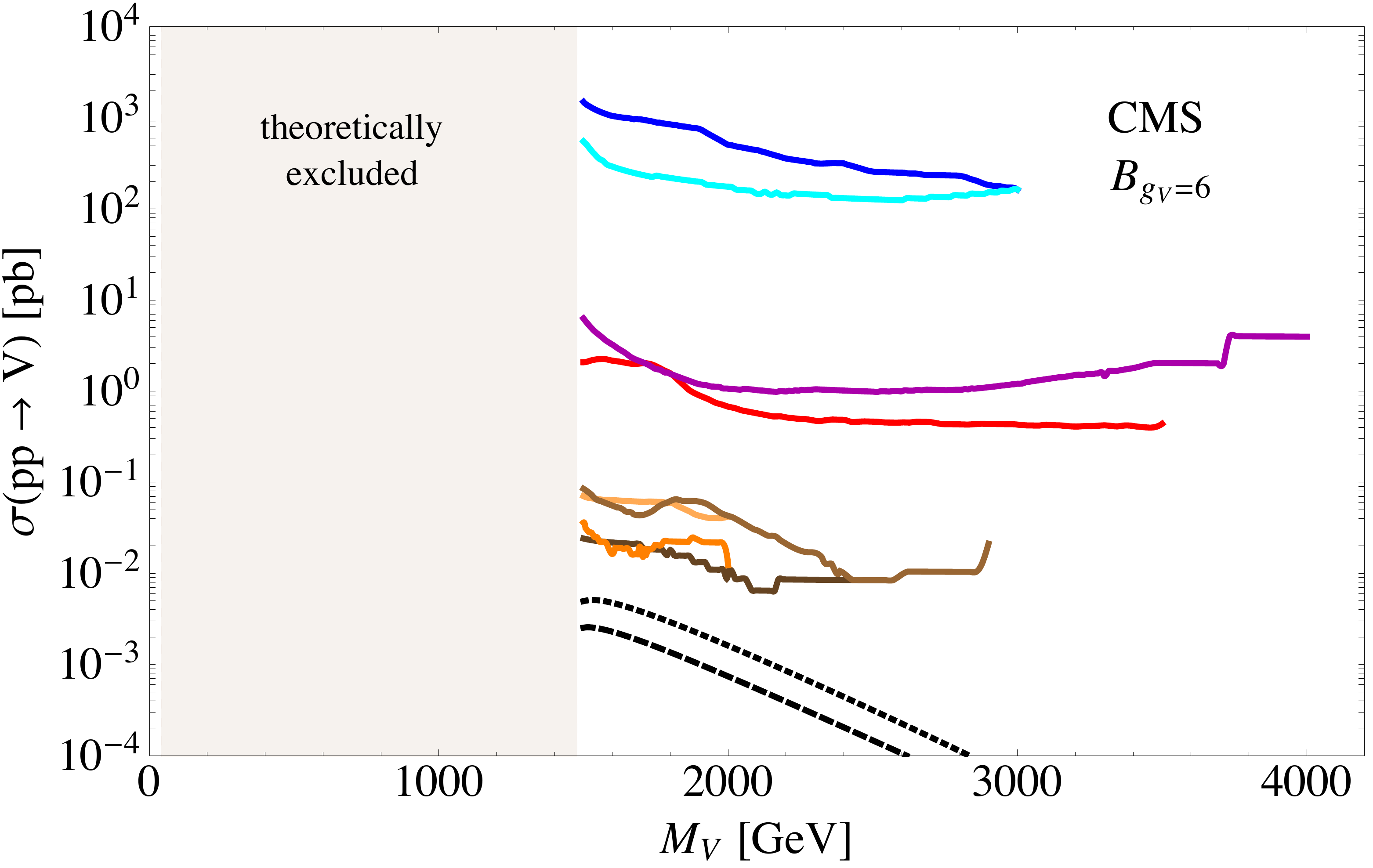}};
    \end{scope}
     \begin{scope}[xshift=7.9cm,yshift=3.9cm]
    \node {\includegraphics[scale=0.24]{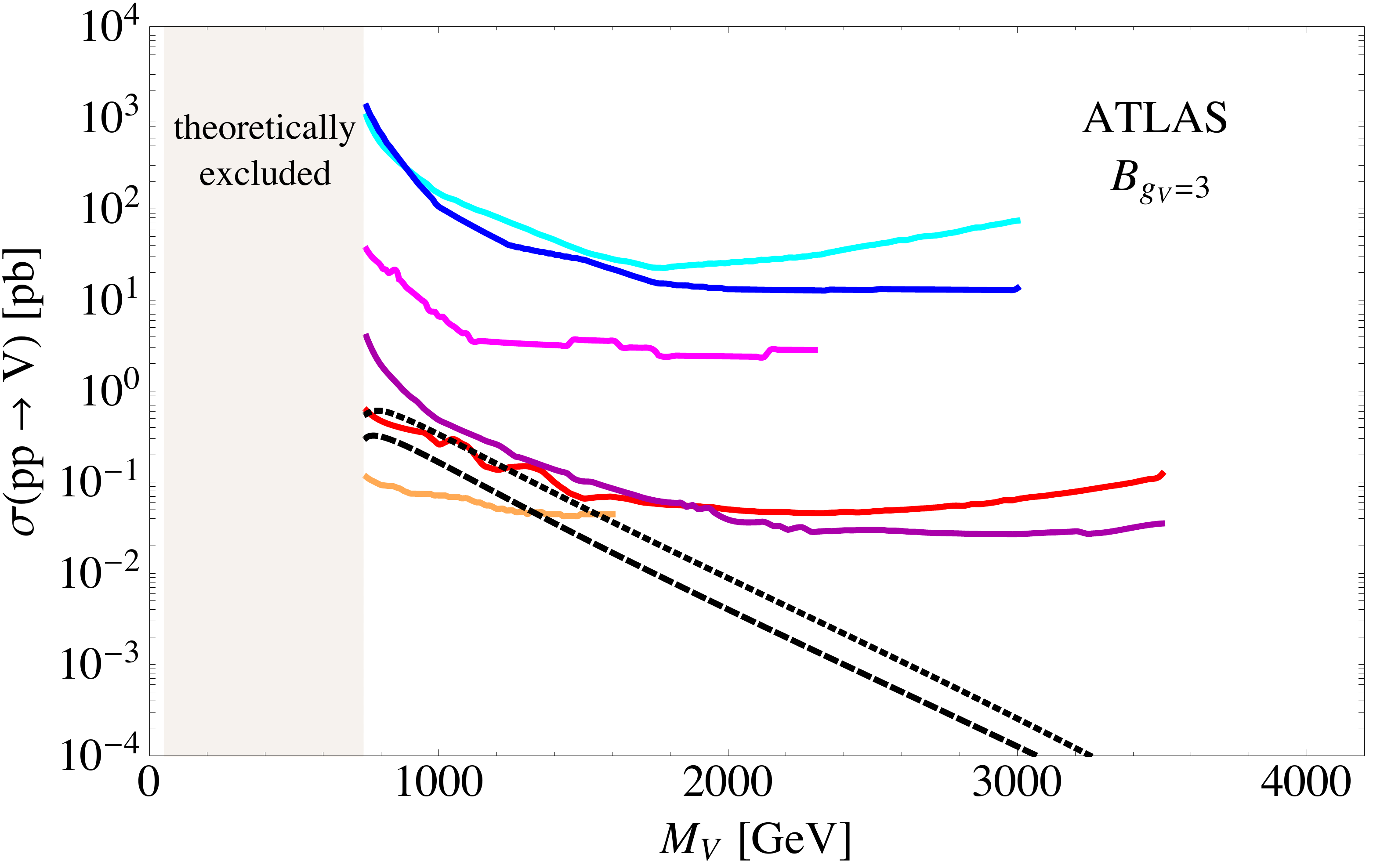}};
    \end{scope}
    \begin{scope}[yshift=3.9cm]
    \node {\includegraphics[scale=0.24]{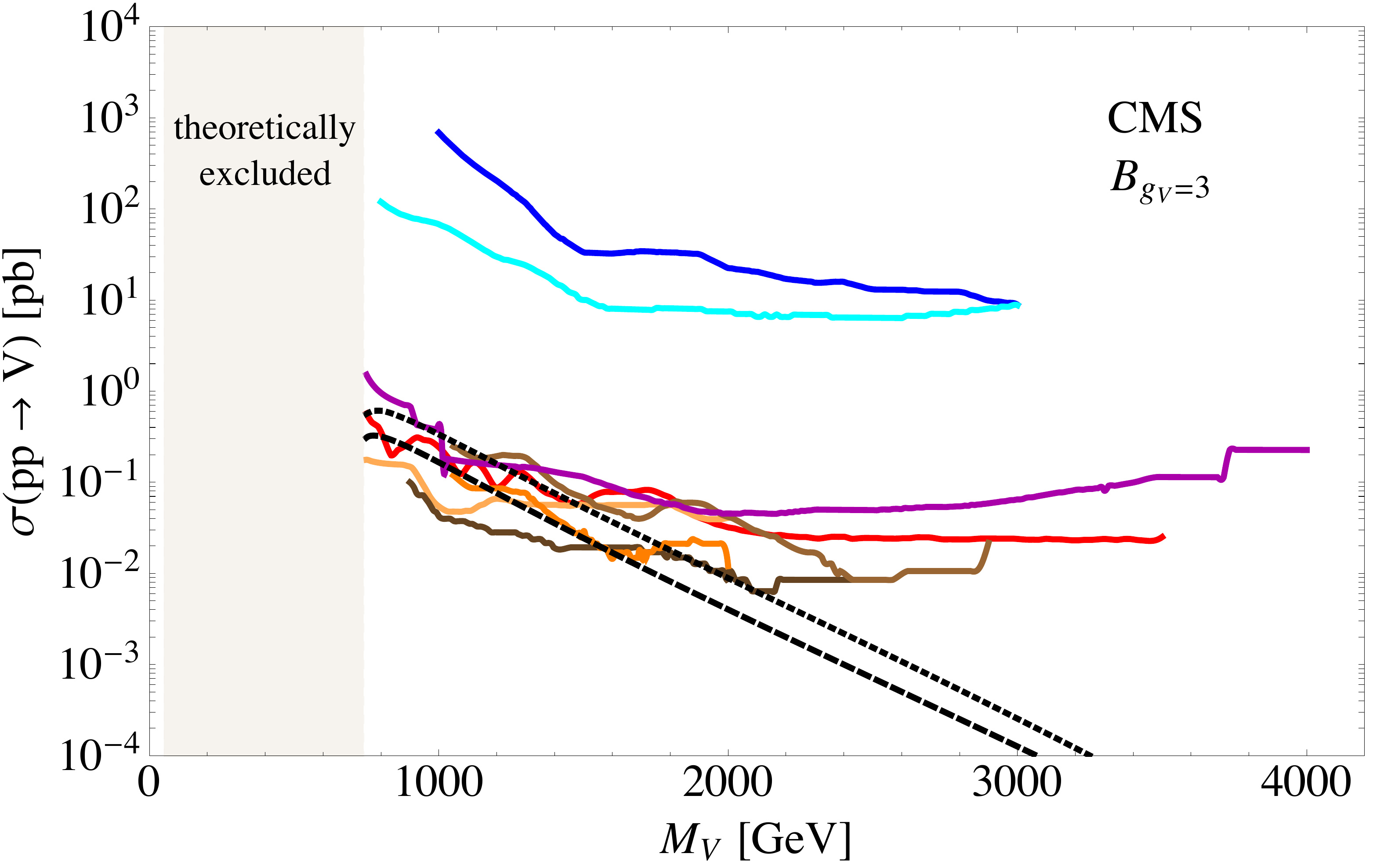}};
    \end{scope}
    \begin{scope}[yshift=10.6cm]
    \node {\includegraphics[scale=0.24]{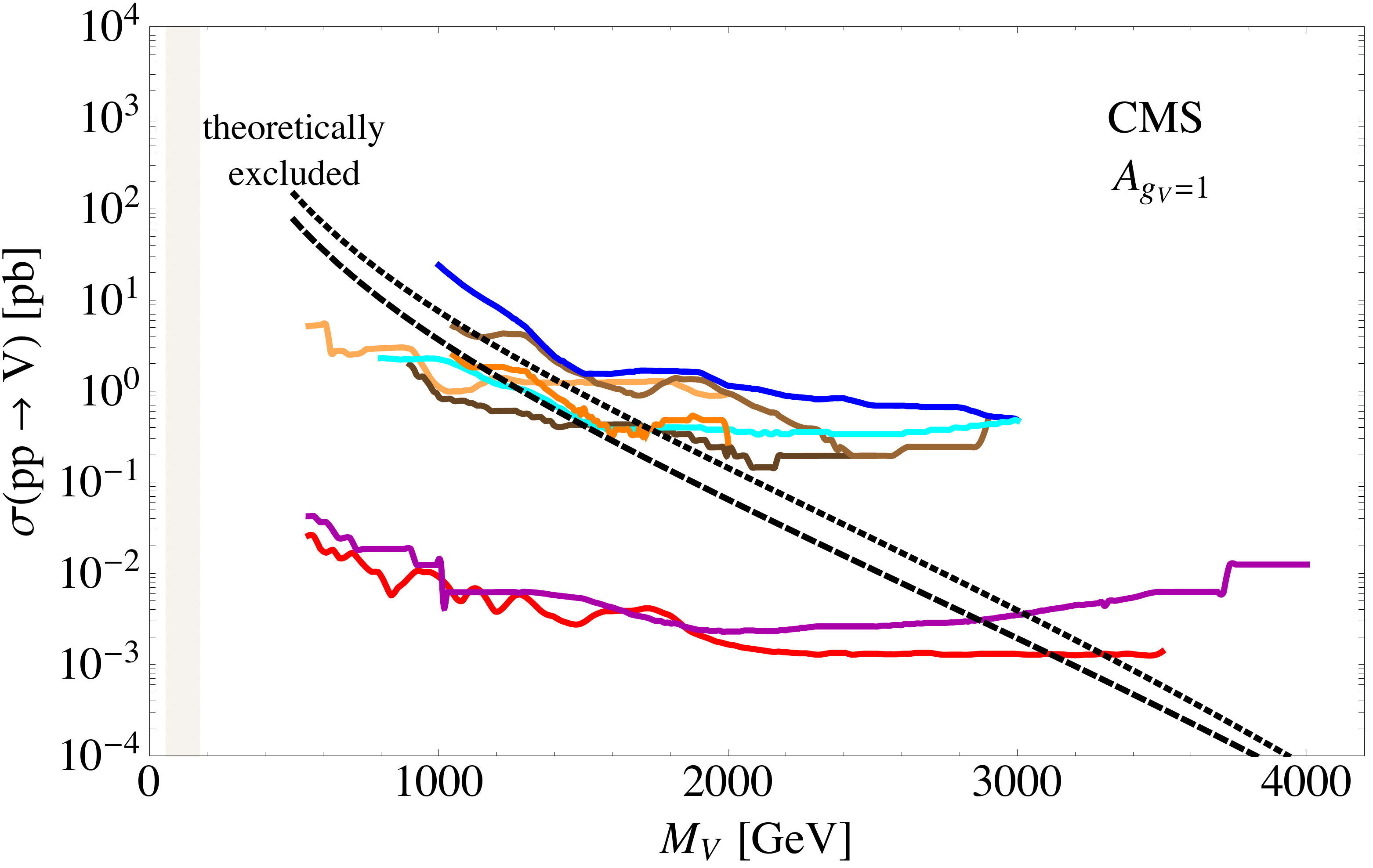}};
    \end{scope}
    \begin{scope}[yshift=10.6cm,xshift=7.9cm]
    \node {\includegraphics[scale=0.24]{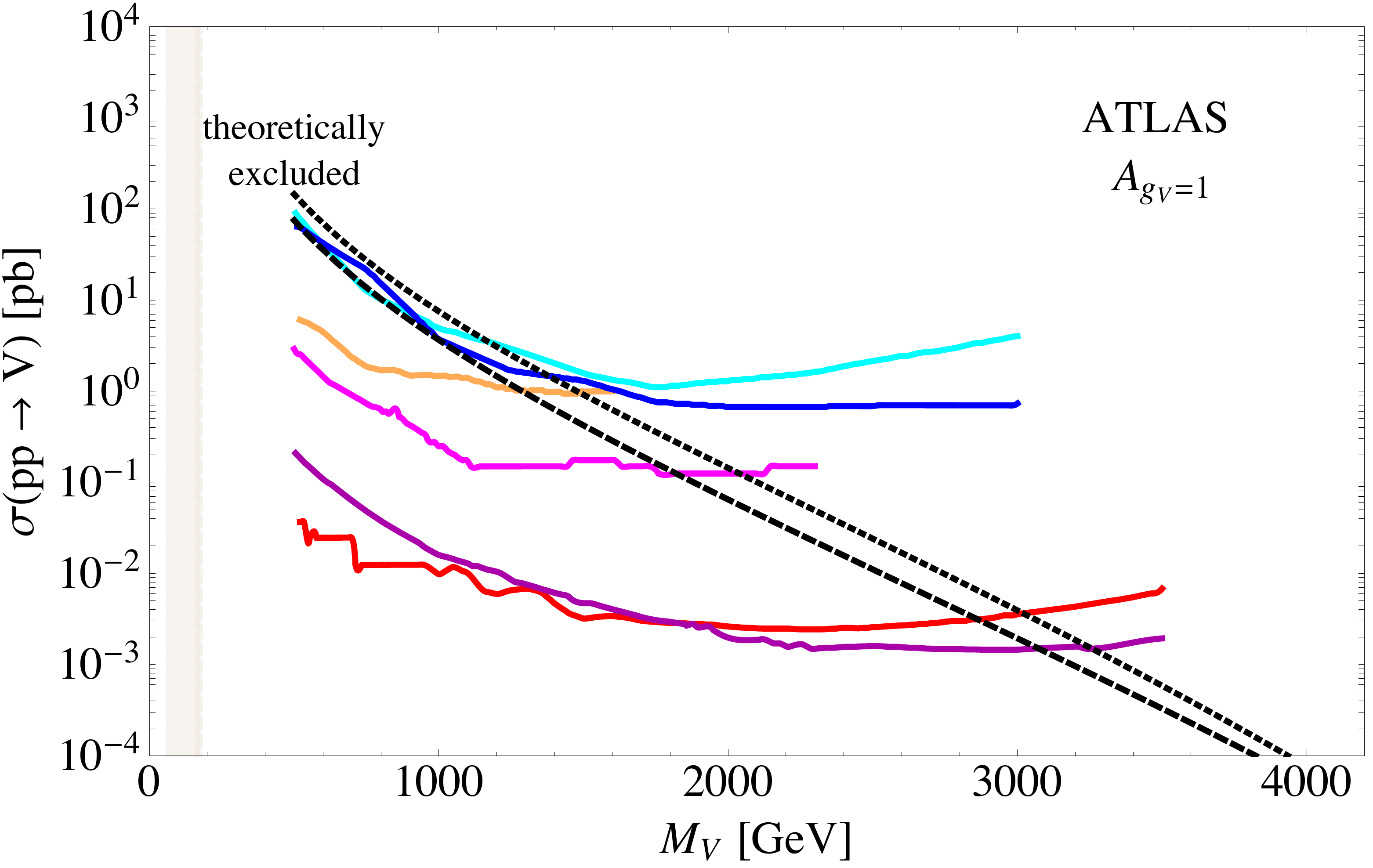}};
    \end{scope}
     \begin{scope}[xshift=5.1cm,yshift=7cm]
    \node {\includegraphics[scale=0.225]{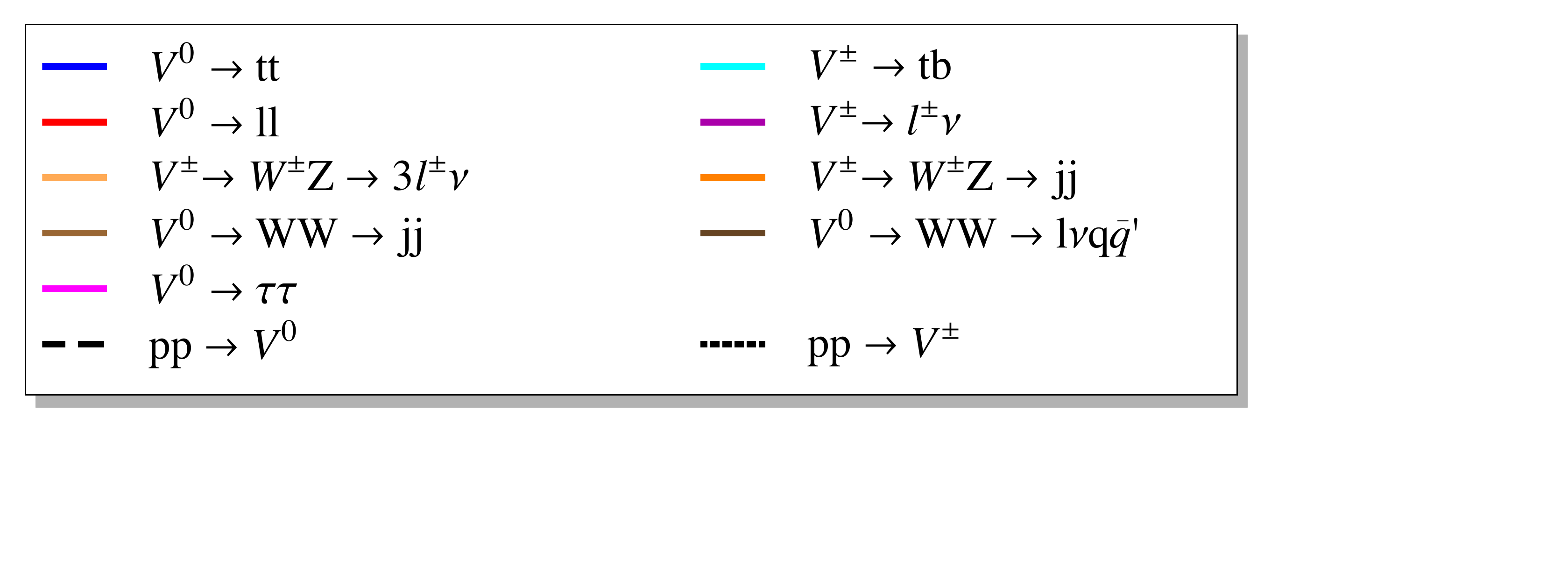}};
    \end{scope}
\end{tikzpicture}
\caption{\small Bounds on the production cross-section for some of the searches listed in Table \ref{Table:Searches} (except for the ones in grey) for the models A$_{\gst=1}$ (upper plots), B$_{\gst=3}$ (middle plots) and B$_{\gst=6}$ (lower plots) for the CMS (left) and ATLAS (right) collaborations. The black dashed curves represent the theoretical prediction for the corresponding benchmark points ($pp\to V^{\pm}$ in the legend stands for the sum of the production cross section of the $V^{+}$ and $V^{-}$ vectors).}
\end{figure}

Before starting a detailed analysis, let us try to get an idea of the present experimental bounds by discussing two illustrative examples. We consider the benchmark models A$_{\gst=1}$, B$_{\gst=3}$ and B$_{\gst=6}$ described in the previous Section (see e.g.~eqs.~\eqref{BRwidthLSM} and \eqref{BRwidthNSM} and the related discussion) as representatives of the ``typical'' weakly coupled (A) and strongly coupled (B) models with intermediate ($\gst=3$) and rather strong coupling ($\gst=6$). In the benchmarks, all the parameters of the model are fixed except for the resonance mass, so they can be very easily compared with the data by looking at Figure \ref{Fig:BoundsonCS} where we report the bounds on the production cross-sections obtained by rescaling the experimental bounds on $\sigma \times$BR by the corresponding BRs and superimpose the theoretical predictions for the production of the positively charged and neutral states. Let us discuss the results separately for the two cases.

\subsubsection*{Weakly coupled heavy vector}
This case is depicted in the upper plots of Figure \ref{Fig:BoundsonCS}. A weakly coupled vector resonance, arising for example as a new gauge boson from an extension of the SM gauge group, is excluded for masses below around $3$~TeV for $\gst=1$. The limit deteriorates for larger coupling because the DY rate is reduced according to \eq{gaF}. For this reason, much weaker bounds will be obtained in the strongly-coupled case described below. The bound is dominated by searches into di-lepton and lepton neutrino final states. The searches in di-bosons, {\it i.e.}~namely in hadronic and semileptonic $WW$ and hadronic and fully leptonic $WZ$ are less constraining, but still able to set a bound around $1- 2$~TeV.  Also relevant is the search of ATLAS into the $\tau\tau$ final state which sets a bound around $2$ TeV. Definitely less constraining are the searches involving the top quark in the final state, like $t\bar{t}$ and $tb$.

Using the ratio of the parton luminosities shown in Figure \ref{fig:partonlumi} we can obtain a naive estimate of the mass reach of the $14$ TeV LHC and of a hypothetical $100$ TeV $pp$ collider. The exclusion in the weakly coupled region $M_{V}\sim 3$ TeV for $20$ fb$^{-1}$ corresponds to a parton luminosity $\sim 4\cdot 10^{-2}$ pb (see Figure \ref{fig:partonlumi}). Using this number and rescaling to a luminosity of $300$ fb$^{-1}$ at the LHC at $14$ TeV and to $1$ ab$^{-1}$ at the $100$ TeV collider we naively find a sensitivity up to $M_{V}\sim 6$ TeV and $M_{V}\sim 30$ TeV respectively.

\subsubsection*{Strongly coupled heavy vector}
This case is depicted in the middle and lower plots of Figure \ref{Fig:BoundsonCS} for an intermediate, $\gst=3$, and rather stronger, $\gst=6$, coupling. A strongly coupled vector resonance like a new composite vector boson, analogous to the $\rho$ in QCD, arising for example in Composite Higgs models is excluded up to $\sim 1.5-2$ TeV for intermediate coupling of the strong sector and almost unconstrained for large enough coupling ($\gst\gtrsim 5$). The most constraining searches are those into di-boson final states because, as described above, the BRs into vector bosons are much larger than those into fermions. $t\bar{t}$ and $tb$ searches are not particularly sensitive. Notice, however, that we are working under the assumption of a universal coupling to fermions. In potentially realistic strongly coupled scenarios the parameter $c_3$ is actually expected to be enhanced, improving the sensitivity of third family searches. A careful assessment of this interesting effect is left to future work. Notice that a large portion of the mass range is theoretically excluded, as shown in the plots. This corresponds to regions where it is not possible to reproduce the SM input parameters $\alpha_{\text{EW}}$, $G_{F}$ and $M_{Z}$ for such a small physical mass and large $\gst$ coupling.

Assuming a rather weak strong coupling $g_{V}=3$, the same naive rescaling made for the weakly coupled vector gives a naive reach of $M_{V}\sim 3-4$ TeV and $M_{V}\sim 15-20$ TeV at the LHC at $14$ TeV with $300$ fb$^{-1}$ and the $100$ TeV collider with $1$ ab$^{-1}$ respectively.

\subsection{Limits on the Simplified Model parameters}
The experimental limits on $\sigma\times \text{BR}$ can be simply converted into limits on the relevant parameters of the Simplified Model. In Section \ref{2} we showed that the most relevant parameters are the mass of the resonance, the overall scale of its couplings $\gst$ and the parameters $c_{F}$ and $c_{H}$ describing the interactions with SM fermions and bosons respectively. In order to give an idea of the bounds coming from present analyses we make the simple choice $c_{F}=c_{q}=c_{l}=c_{3}$ and show the bounds, for given mass and coupling, in the two-dimensional $(c_{H},c_{F})$ plane. The results, as expected from the discussion of Section \ref{2}, are very weakly sensitive to the other parameters $c_{VVW}$, $c_{VVV}$ and $c_{VVHH}$. In the plots we fixed the latter to their values in model A (see Section~\ref{4.1}) and checked explicitly that the results do not change significantly by setting them to model B.

\begin{figure}[t!]
\begin{center}
\includegraphics[scale=0.1692]{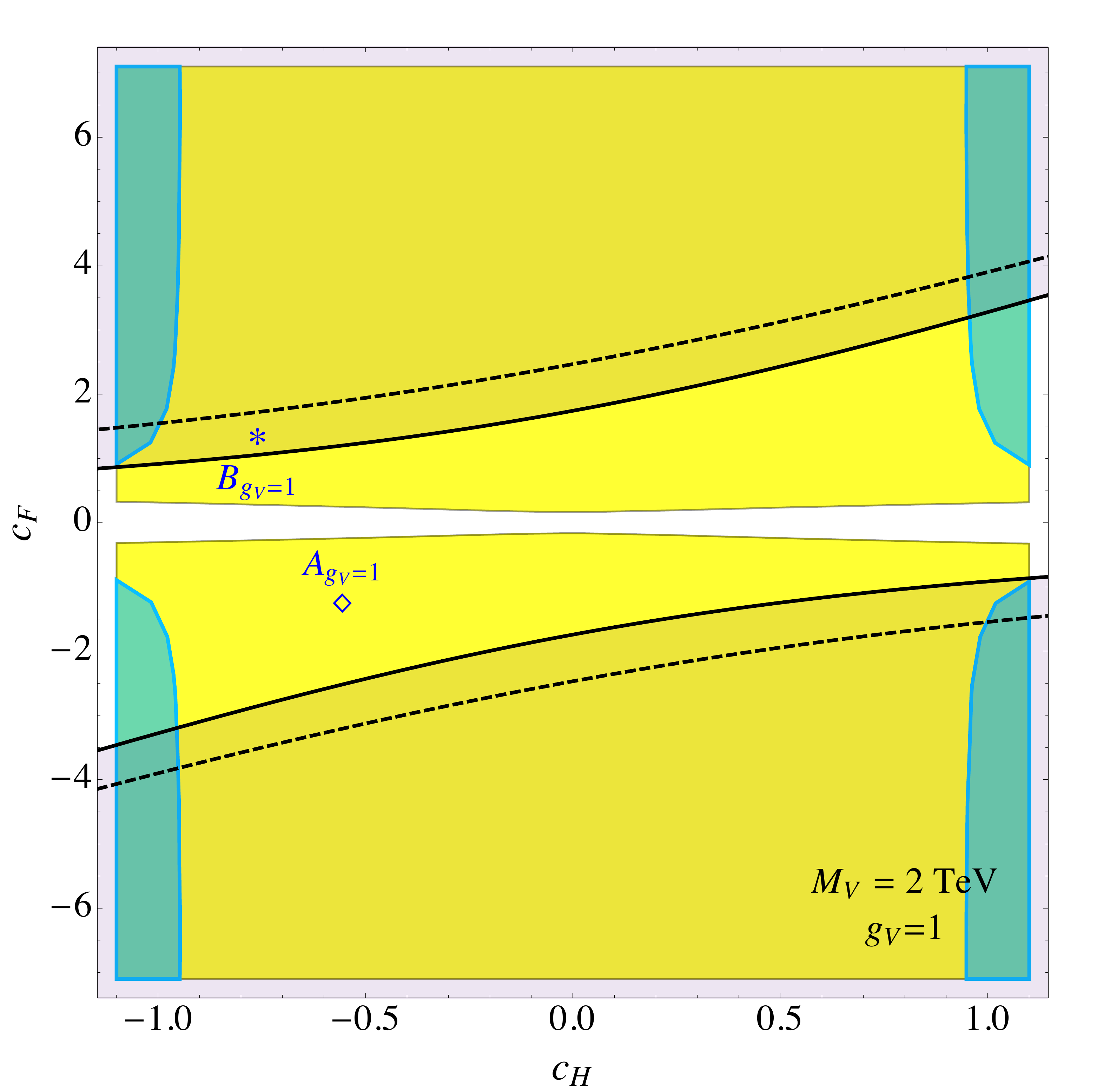}\hspace{-3mm}
\includegraphics[scale=0.1692]{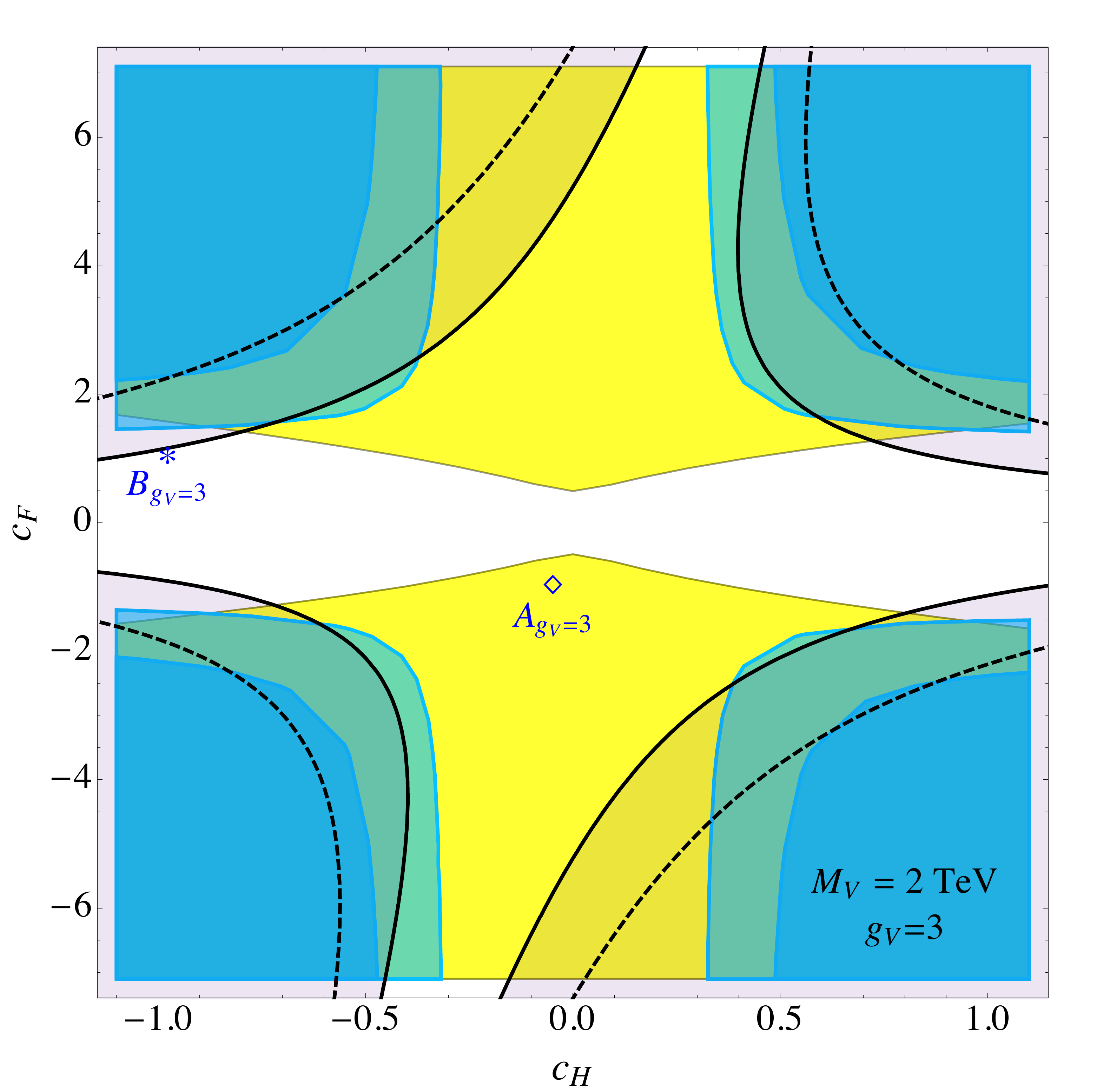}\hspace{-3mm}
\includegraphics[scale=0.1692]{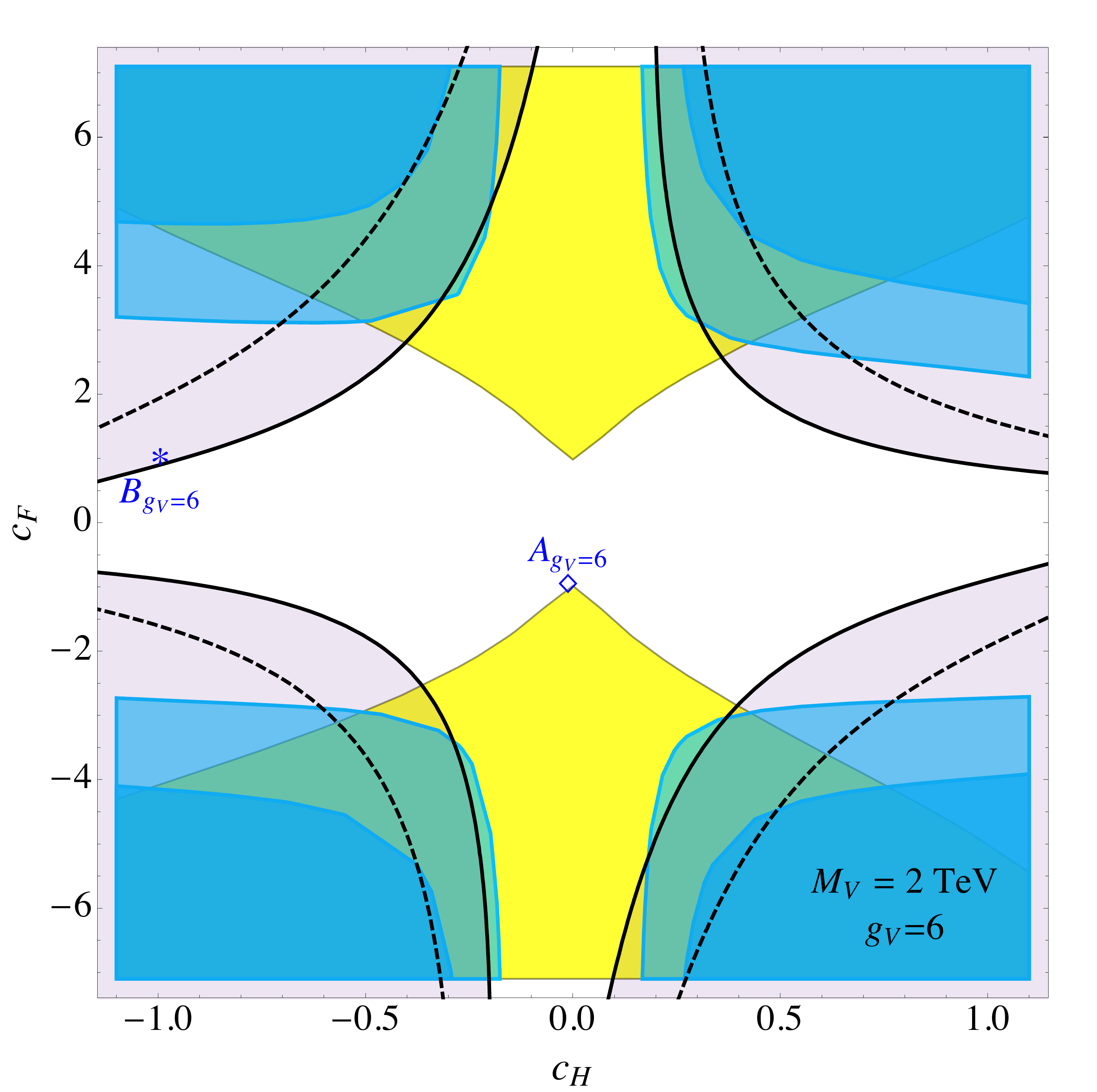}
\caption{\small Current experimental constrains in the $(c_{H},c_{F})$ plane for the four benchmark points at $2$ TeV. The yellow region shows the exclusion from $V \to l \nu$ searches \cite{CMS-PAS-EXO-12-060} while in blue are regions excluded by $V \to WZ$ searches with $WZ\to jj$ \cite{CMS-PAS-EXO-12-024} in light blue and $WZ\to 3l\nu$ \cite{Khachatryan:2014xja} in dark blue. The solid black lines depict constrains from EWPT at $95\%$ CL and the dashed black line twice this limit. The points corresponding to models A and B for the different values of $\gst$ are also shown.}\label{Fig:Boundscfch}
\end{center}
\end{figure}

In Figure \ref{Fig:Boundscfch} we show the allowed and excluded regions in the $(c_{H},c_{F})$ plane for fixed $M_{V}$ and $\gst=1,3,6$ corresponding, respectively, to weak, intermediate and strong coupling. 
As an illustrative example we chose $M_{V}=2$ TeV as an intermediate mass scale where the experimental constraints are neither too strong nor too weak and thus more interesting. For simplicity, we did not report all the relevant limits in the plots, but only the ones from charged vector searches. The neutral ones could be easily added but would just give comparable constraints and not change the result significantly. Obviously, the situation could have changed if we had performed a statistical combination of the limits in the different channels rather than a superposition of the corresponding excluded regions. However, we think that correlations among the different channels should be taken into account by the experimental collaborations. In the plots, the yellow region represents the exclusion from the CMS $l^{+}\nu$ analysis of Ref.~\cite{CMS-PAS-EXO-12-061}, while the dark and light blue ones show the limits from CMS $WZ\to 3l\nu$ \cite{Khachatryan:2014xja} and \mbox{$WZ\to jj$} with \mbox{$W/Z$} tagged jets \cite{CMS-PAS-EXO-12-024} respectively.\footnote{For recent theoretical developments in the search for vector resonances using boosted techniques see, for instance, in Refs.~\cite{Katz:2010jx,Katz:2010kw,Son:2012vs}.} The black curves represent constraints coming from EWPT, {\it{i.e.}}~from the $\hat{S}$ parameter, which we computed in Appendix~\ref{AppB}. The black solid curve corresponds to the strict $95\%$ C.L. bound on $\hat{S}$ of Ref.~\cite{Ciuchini:2013vb}\footnote{The bound quoted in Ref.~\cite{Ciuchini:2013vb} is $S = 0.04 \pm 0.10$ obtained from an $STU$ fit. }, while the dashed line is obtained by artificially enlarging the latter bound by a factor of two. This second line is a more realistic quantification of the constraints than the strict limits because the EWPT observables are eminently off-shell observables and thus not calculable within the Simplified Model. Extra contributions, of the same order as the ones coming from the resonance exchange, can easily arise in the underlying complete model. By enlarging the bound on $\hat{S}$ we take these contributions into account and obtain a conservative exclusion limit. 

Any given explicit model corresponds to one point in the plots of Figure~\ref{Fig:Boundscfch}. The two points indicated by $A$ and $B$ correspond to the prediction of the two benchmarks models for the assumed values of $\gst$ and $M_{V}$. For small $\gst$ the lepton-neutrino search dominates the exclusion (first plot) and only a narrow band around $-1\lesssim c_{F}\lesssim 1$ remains allowed. Here EWPT are not competitive with direct searches and the di-boson searches are almost irrelevant due to the relatively small di-boson BR (see the discussion at the end of Section \ref{2.1}). Moreover, for small $\gst$ both our benchmark models are excluded. As $\gst$ increases we notice four main features: the constraints from EWPT become comparable to the direct searches, di-boson searches become more and more relevant due to the enhanced BRs, model B evades bounds from direct searches more and more compared to model A which remains close to the excluded region, and bounds from EWPT constrain model B more than model A. The last two features are due to the larger value of $c_{H}$ predicted by model B, corresponding to a region which is very difficult to access with direct searches.

\begin{figure}[t!]
\begin{tikzpicture}
    \begin{scope}
    \node {\includegraphics[scale=0.246]{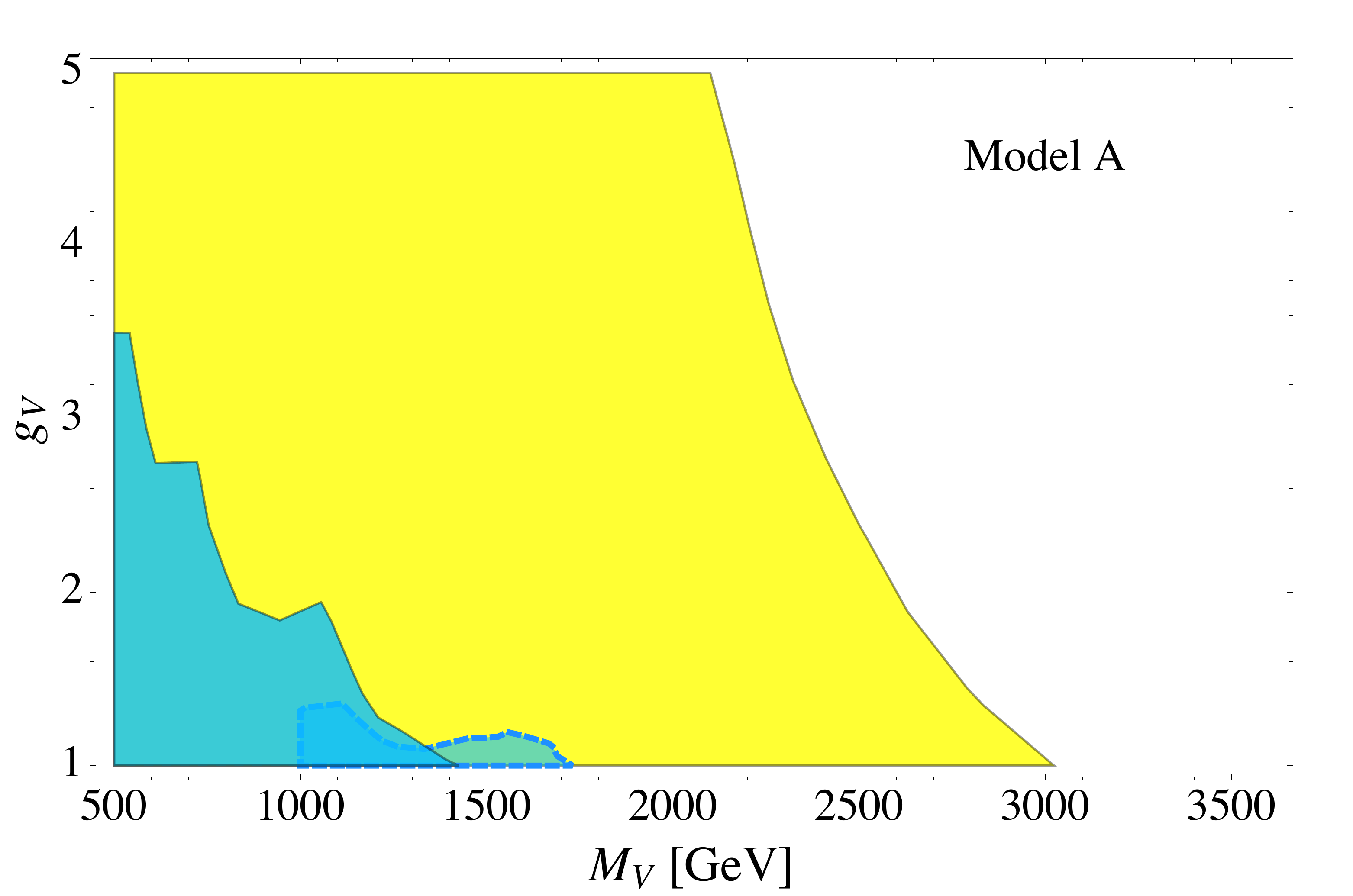}};
    \end{scope}
    \begin{scope}[xshift=8cm]
    \node {\includegraphics[scale=0.246]{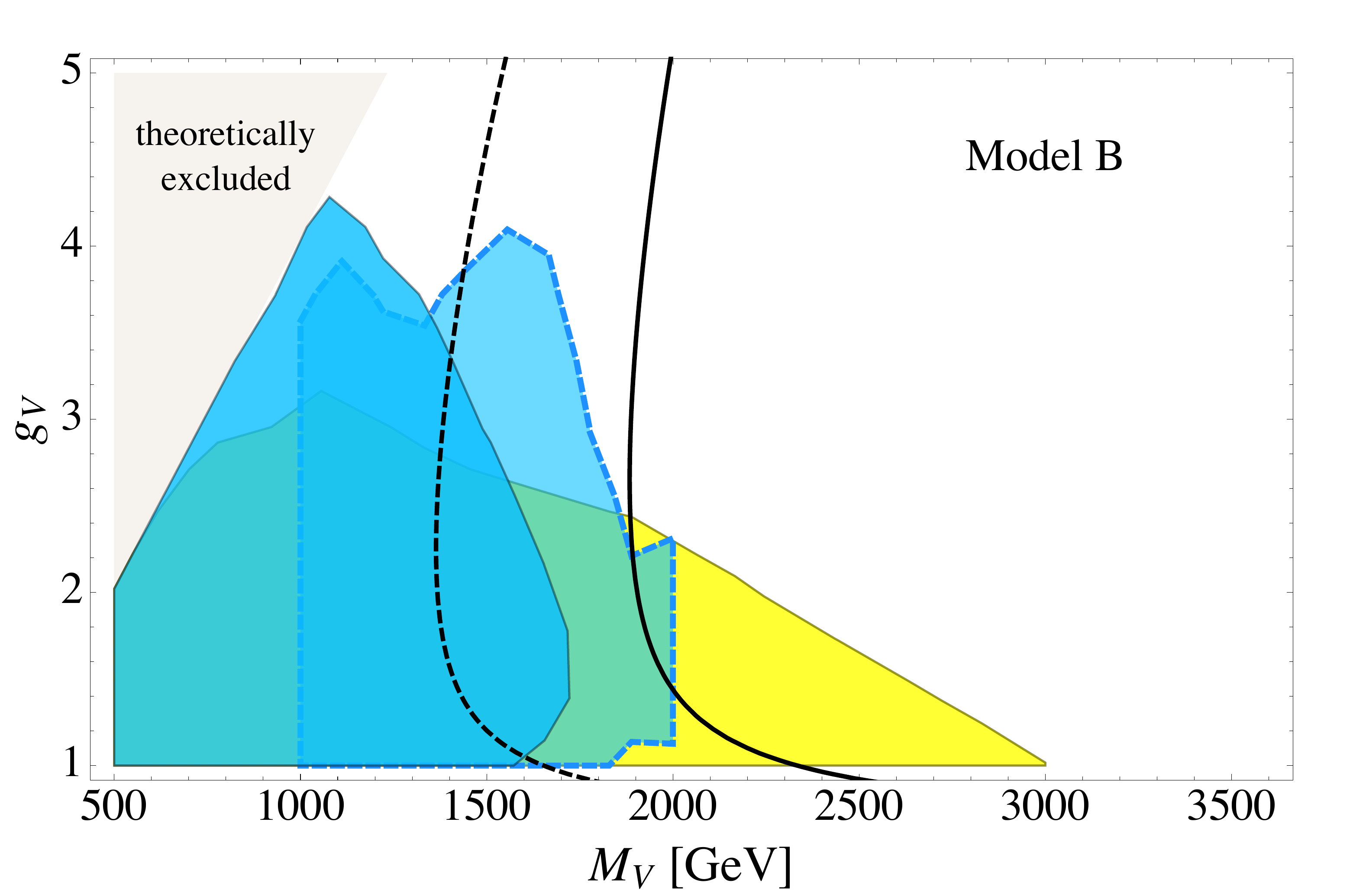}};
    \end{scope}
\end{tikzpicture}
\caption{\small Current experimental constraints in the $(M_{V},\gst)$ plane in models A and B. The notation is the same as in Figure~\ref{Fig:Boundscfch}.}\label{Fig:Boundscfch3}
\label{fmg}
\end{figure}

A second interesting way to present the experimental limits is to focus on explicit models and draw exclusion curves in the plane of their input parameters. In both models A and B we have two parameters, the coupling and the mass of the new vector. The limits in the $(M_{V}, \gst)$ plane are reported in Figure~\ref{fmg}. We find similar exclusions in the two models at low $\gst$, where the limit is dominated by leptonic final state searches, but the situation changes radically for large coupling. In particular the limit in model B is rather weak and barely competitive with EWPT already for intermediate couplings $\gst \sim3$ and it disappears when the coupling is large.

\begin{figure}[t!]
\begin{center}
\includegraphics[scale=0.169]{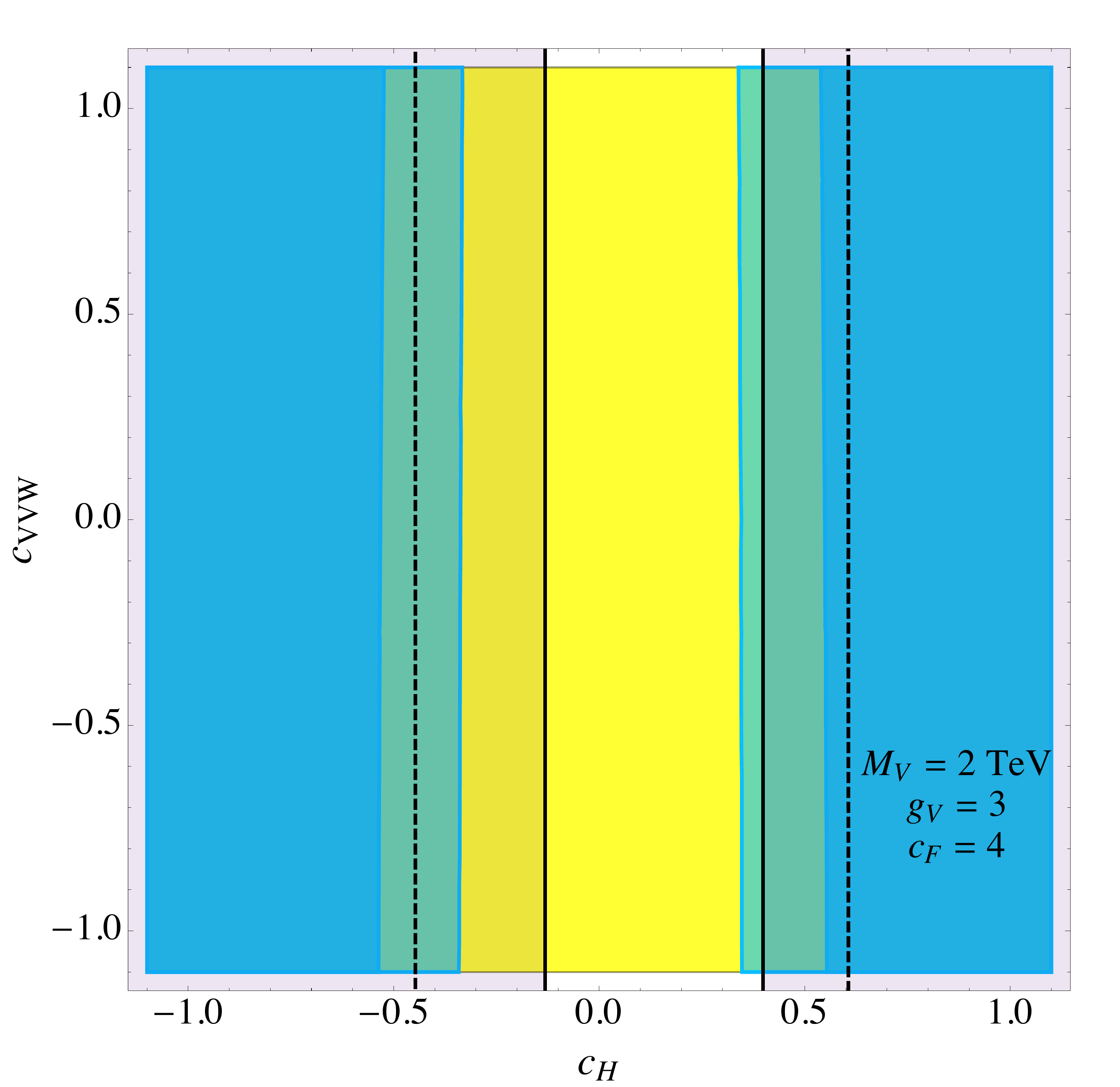}\hspace{-3mm}
\includegraphics[scale=0.169]{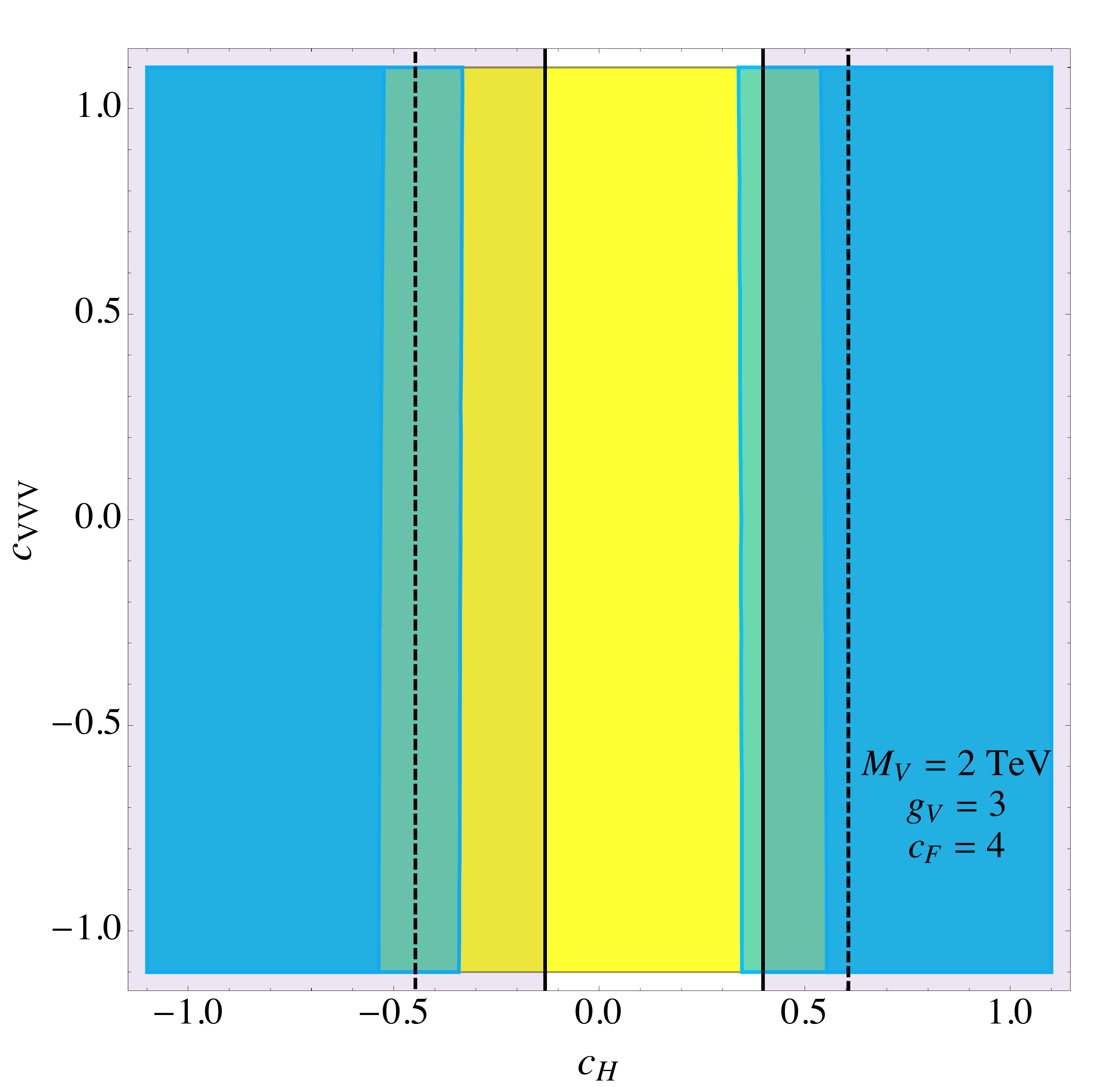}\hspace{-3mm}
\includegraphics[scale=0.169]{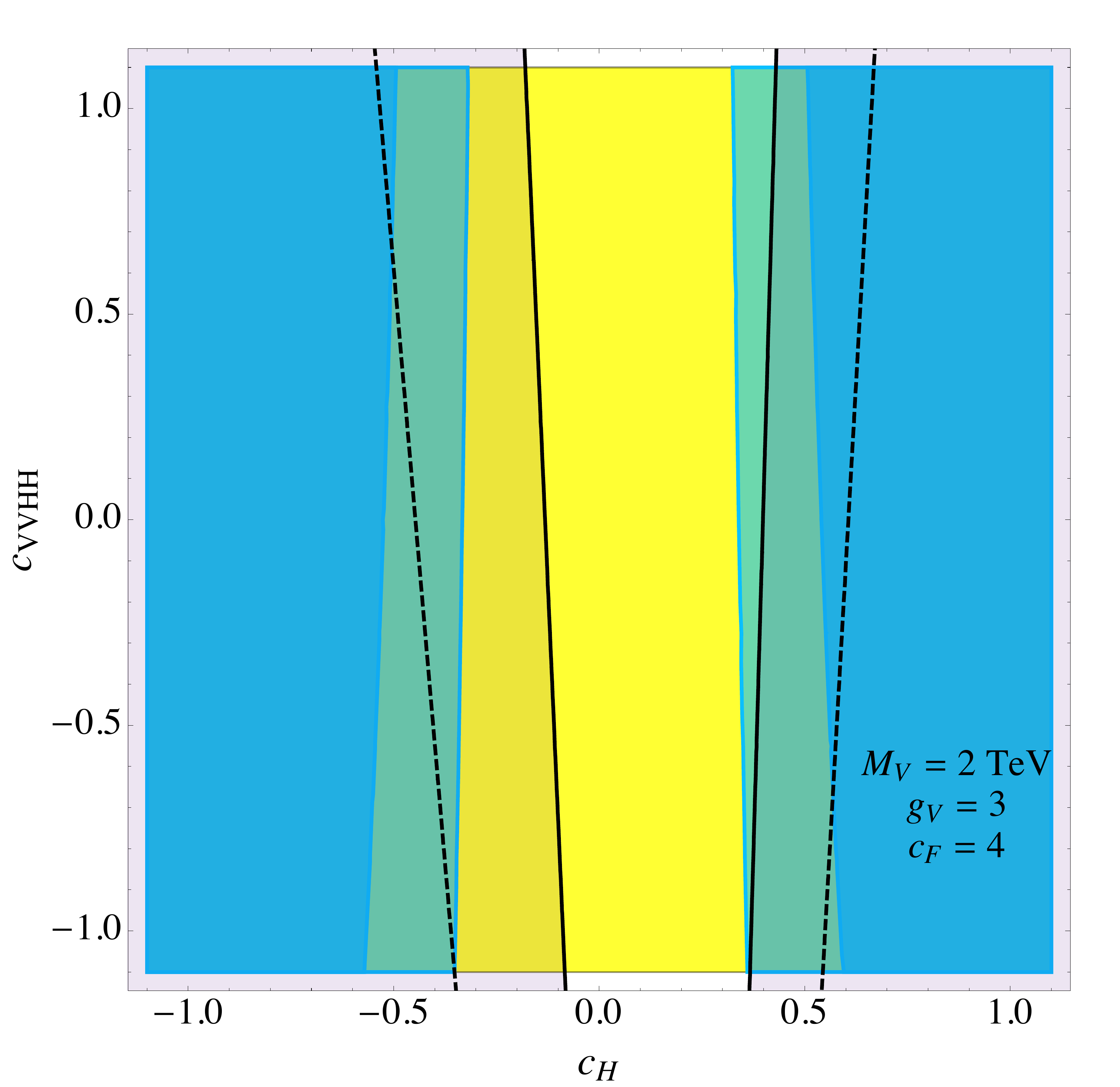}
\caption{\small Current experimental constrains in the $(c_{H}, c_{VVW})$, $(c_{H}, c_{VVV})$ and $(c_{H}, c_{VVHH})$ planes for $\gst=3$, $M_{V}=2$ TeV and $c_{F}=4$ (all the other parameters are fixed to their value in model A). The notation is the same as in Figure~\ref{Fig:Boundscfch}.} \label{Fig:Boundscfch2}
\end{center}
\end{figure}

Finally we want to check that, as expected from the discussion of Section \ref{2.1}, the parameters $c_{VVW}$, $c_{VVV}$ and $c_{VVHH}$ affect the exclusion only marginally. We thus plot the same constraints shown in Figure \ref{Fig:Boundscfch}, in the $(c_{H}, c_{VVW})$, $(c_{H}, c_{VVV})$ and $(c_{H}, c_{VVHH})$ planes in Figure \ref{Fig:Boundscfch2} for the benchmark models A and B at $\gst = 3$. The plots represent a horizontal slice at $c_{F}=4$ in the second plot of Figure \ref{Fig:Boundscfch} using the same coloring as previously. 
We find $c_{VVW}$ and $c_{VVV}$ indeed to be sub-leading with no variation in their direction. A slight tilt can be observed in the direction of $c_{VVHH}$, on the other hand. This is due to the enhanced sensitivity on $c_{VVHH}$ induced by the term $(1-4c_{VVHH}\zeta^{2})^{2}$ in the width in \eq{bbw} for relatively large $\zeta$. The correction induced by this term can be of the order of $20\%$ for $c_{H}\sim -0.5$ ($\zeta\approx0.4)$. One could expect the same enhancement in the central plot, due to the term $(1+c_{H}c_{VVV}\zeta^{2})^{2}$ in the width in \eq{bbw}. However, the absence of the factor of four only gives an effect of the order of the percent for $c_{H}\sim-0.5$, not clearly observable in the central plot.

\subsection{Limit setting for finite widths}

\begin{figure}[ht!]
\begin{tikzpicture}
    \begin{scope}
    \node {\includegraphics[scale=0.355]{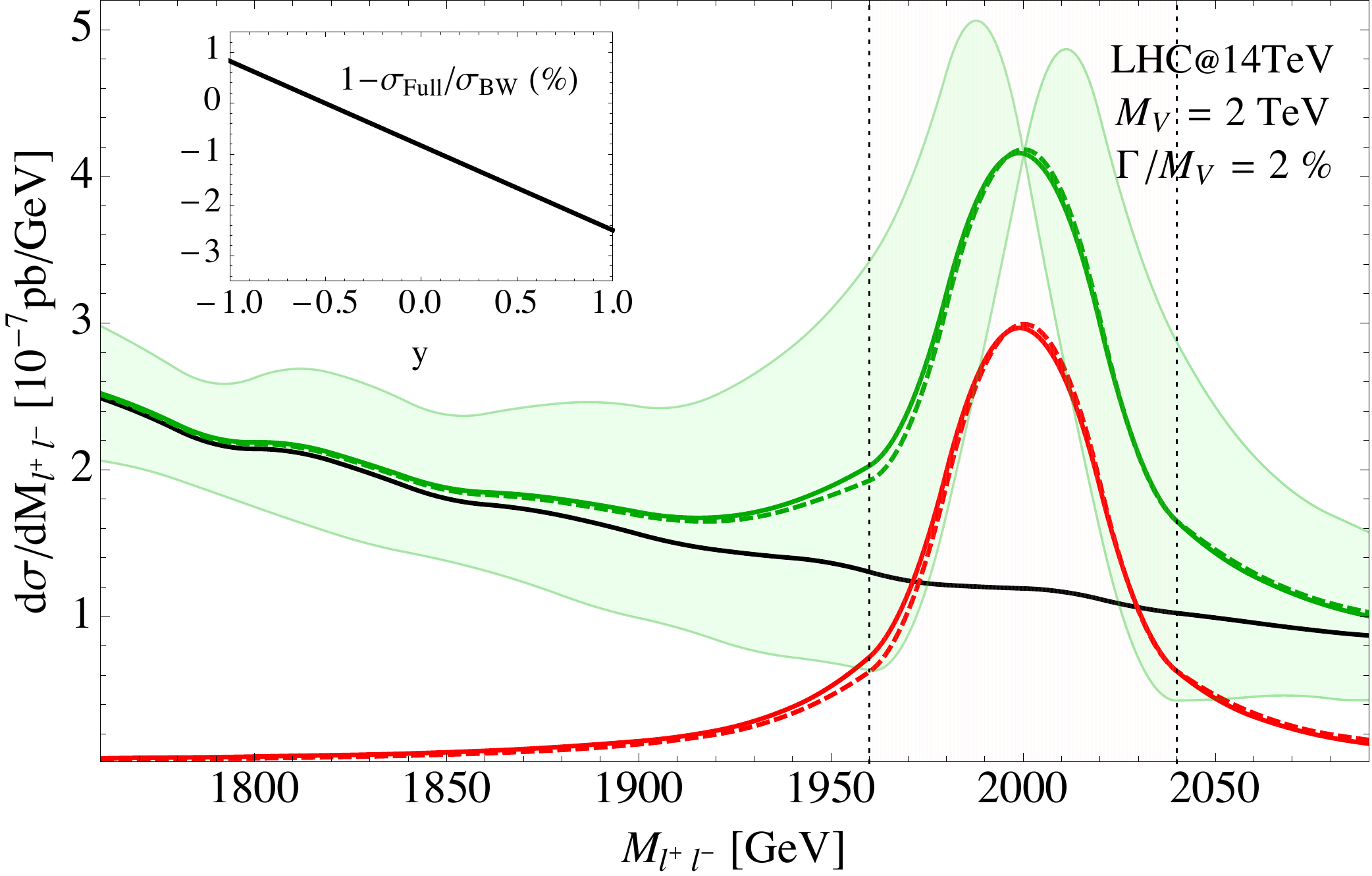}};
    \end{scope}
    \begin{scope}[xshift=8cm]
    \node {\includegraphics[scale=0.355]{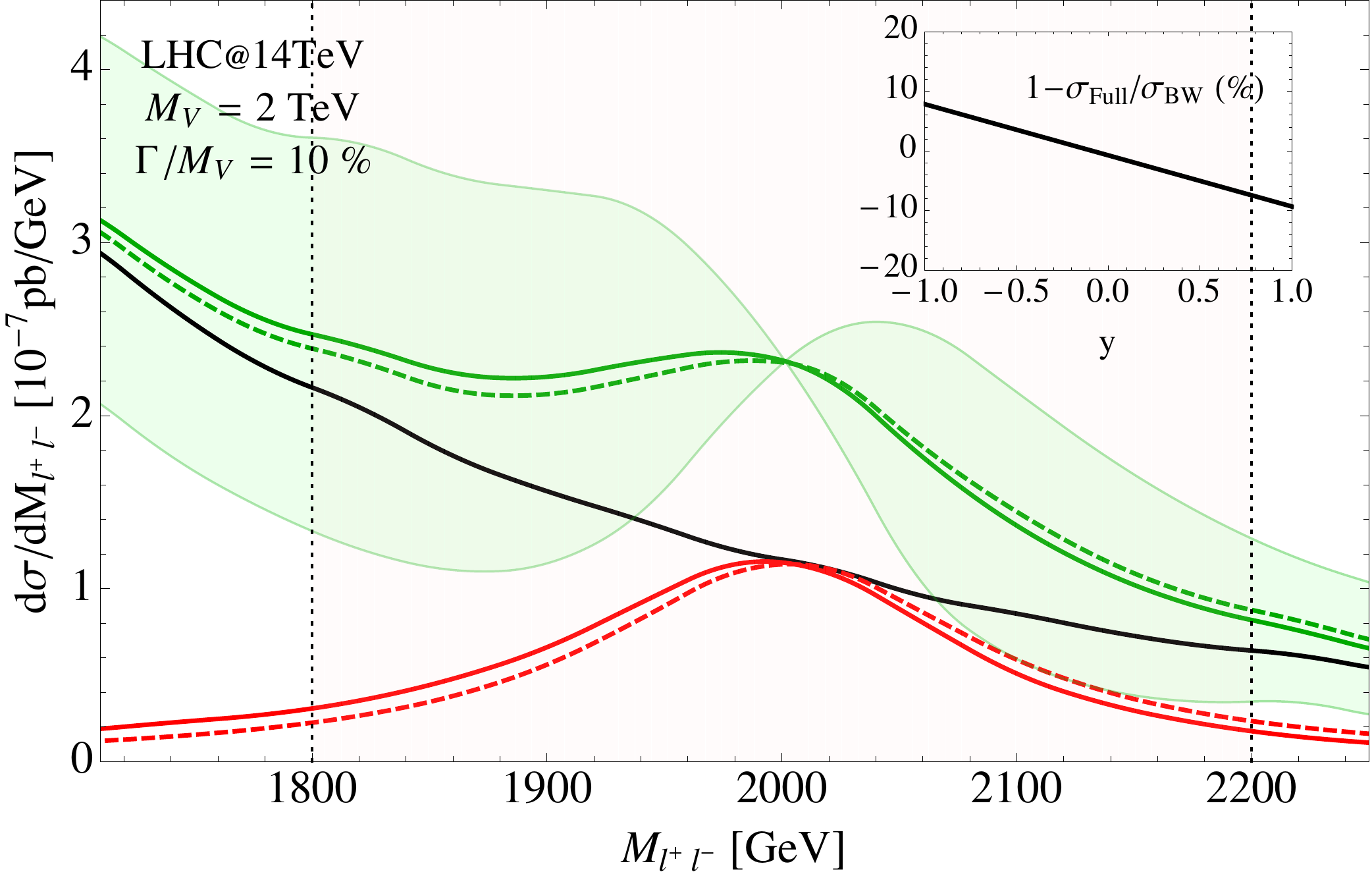}};
    \end{scope}
    \begin{scope}[xshift=8.255cm,yshift=5.2cm]
    \node {\includegraphics[scale=0.265]{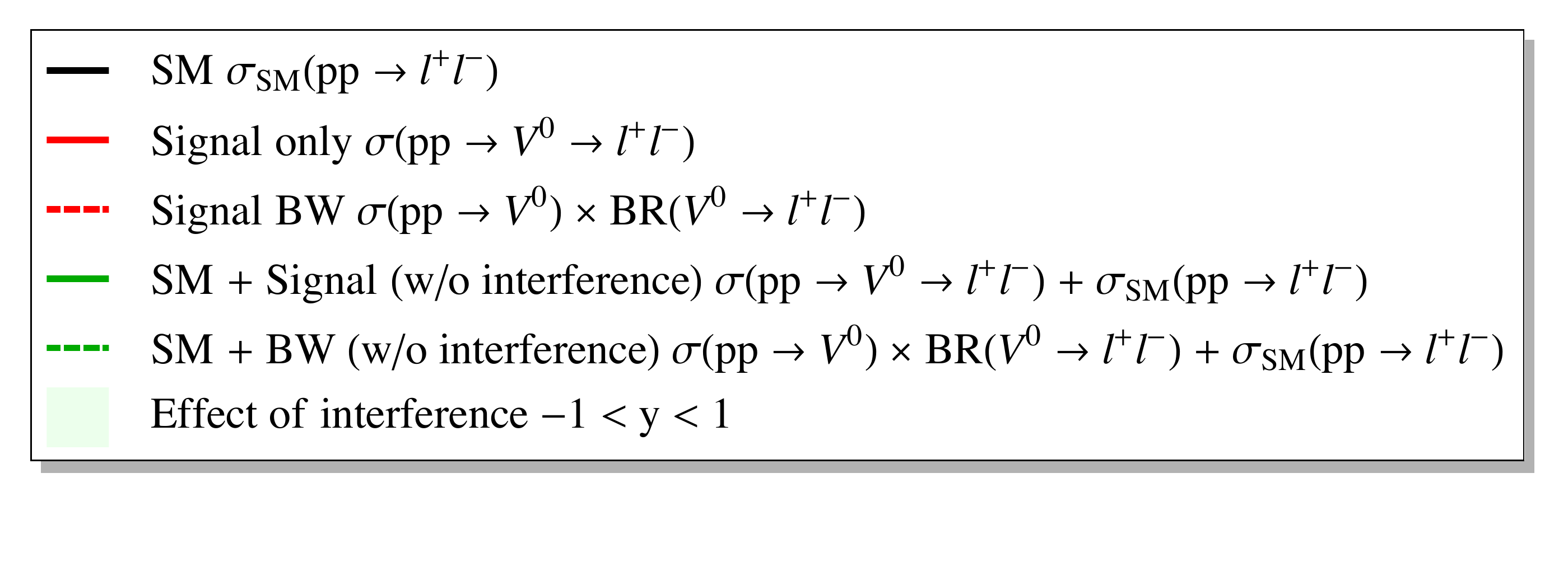}};
    \end{scope}
    \begin{scope}[yshift=5.2cm]
    \node {\includegraphics[scale=0.355]{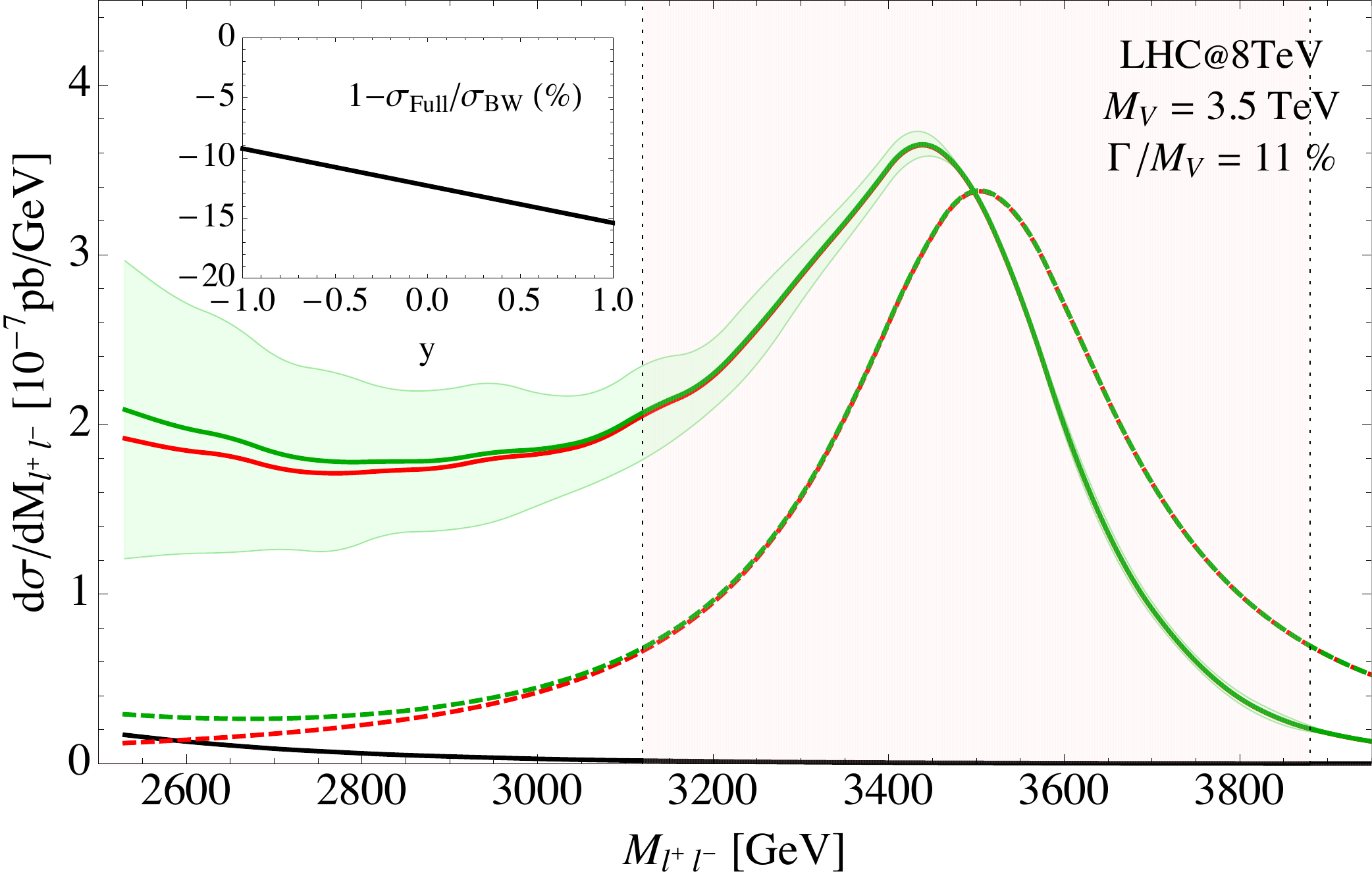}};
    \end{scope}
    \begin{scope}[yshift=10.4cm]
    \node {\includegraphics[scale=0.355]{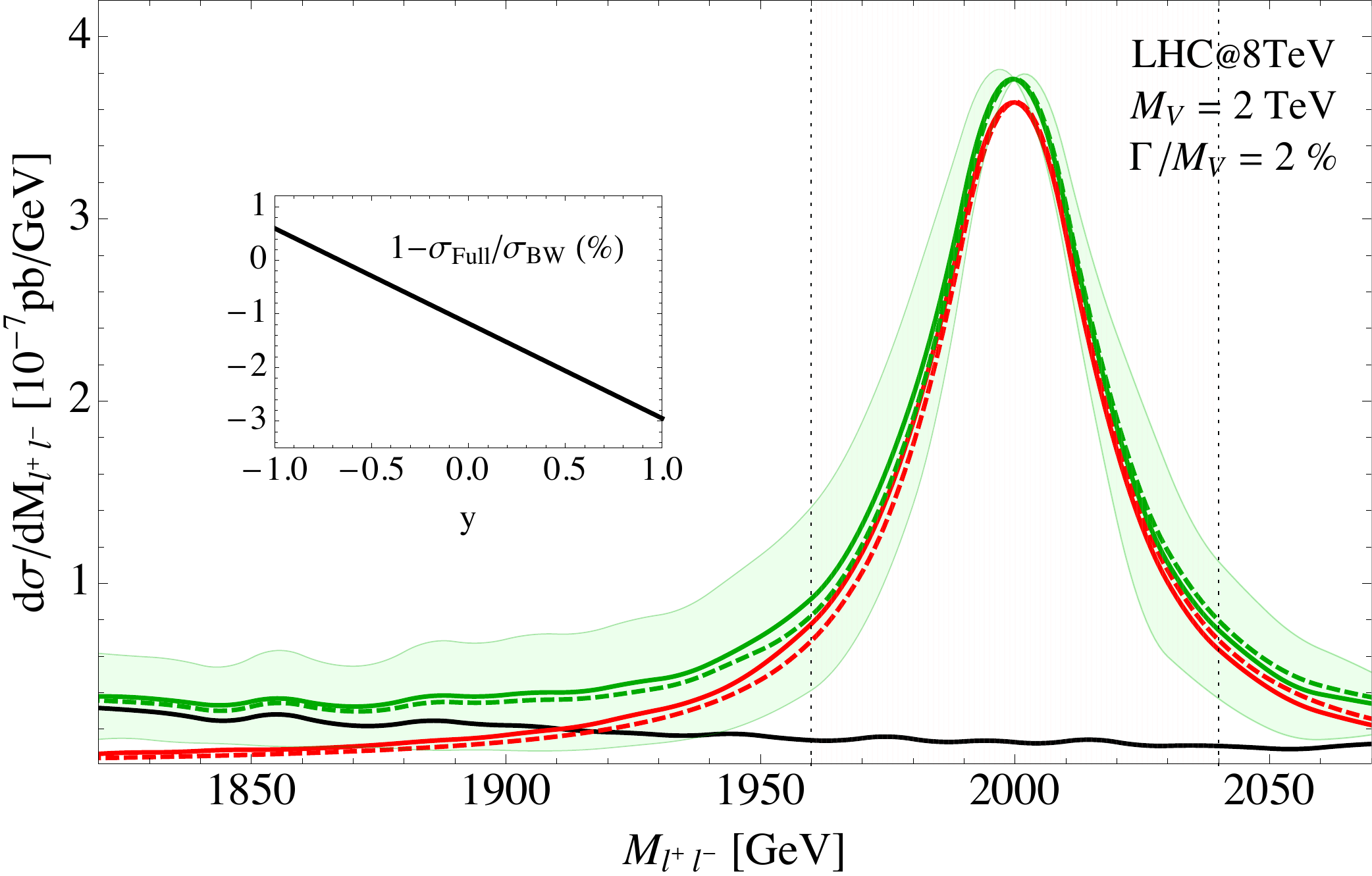}};
    \end{scope}
    \begin{scope}[yshift=10.4cm,xshift=8cm]
    \node {\includegraphics[scale=0.355]{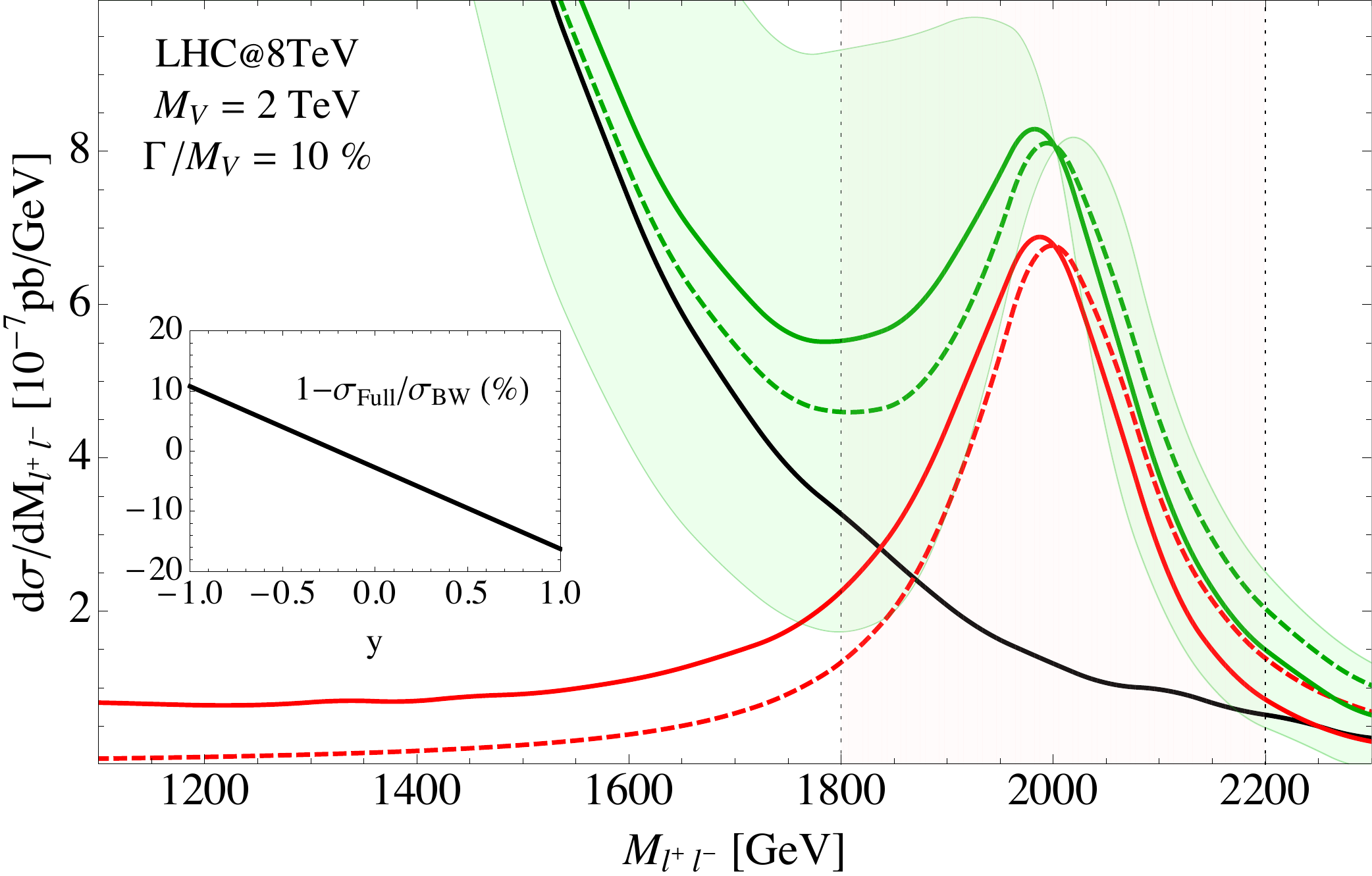}};
    \end{scope}
\end{tikzpicture}
\caption{\small Di-letpon invariant mass distribution for different choices of $M_{0}$ and $\Gamma_{0}/M_{0}$ at the LHC at 8 TeV (first three plots) and 14 TeV (last two plots) c.o.m energy. The Figures show the dependence of the difference between the full $2\to2$ calculation and a simple BW distribution normalized to the on-shell production cross-section and multiplied by the corresponding BR into di-leptons on $\Gamma_{0}/M_{0}$. The ``inset'' plots show the percentage agreement between the cross-sections obtained by integrating the full simulation with a $y$-dependent interference in the shaded ``bin'' (varying continuously from fully constructive ($y=1$) to fully deconstructive ($y=-1$)) and the simple sum of the BW plus the background.} \label{Figure:Breit-Wigner}
\end{figure}

The final goal of a resonance search is to set experimental limits, for either exclusion or discovery, on the resonance production cross-section times the BR into the relevant final states for different mass hypotheses. This way of presenting the experimental results is obviously the simplest and most convenient, as it is completely model-independent and can be very easily interpreted in any given model as we did above. However whether this goal can be really achieved or not, and with which accuracy, can depend crucially on the details of the analysis and on the assumed total width of the resonance. The aim of this Section is to illustrate two kinds of effects associated with the finite particle width that can make the extraction of \mbox{$\sigma\times$BR} limits from an experimental search rather involved. Both effects are well-known. Recent discussions can be found in Refs.~\cite{2011arXiv1111.1054C,Accomando:2011up,Accomando:2013ve}. Here we will quantify their importance for heavy vector searches at the LHC and propose some strategies to minimize their impact. The results of this Section are obviously not conclusive. A detailed analysis of these issues and their impact on the searches should be performed by the experimental collaborations. The final goal should be to quantify, and minimize, the systematic uncertainties associated with the determination of \mbox{$\sigma\times$BR}.

\subsubsection*{The example of the di-lepton invariant mass}

Let us study the width effects in detail by focusing on the simple case of di-lepton searches for the neutral vector. The relevant observable is the di-lepton invariant mass distribution which we show in Figure~\ref{Figure:Breit-Wigner} for different $V^{0}$ masses, widths and collider energies. We took a vector resonance with a mass of $2$ TeV, both narrow ($\Gamma/M_V=2\%$) and broad ($\Gamma/M_V=10\%$) at the LHC at $8$ TeV (first row of plots) and $14$ TeV (last row of plots) as reference. Finally, in the central plot we show the example of a resonance at $3.5$ TeV with $\Gamma/M_V\sim11\%$ at the $8$~TeV LHC.  

The first effect to be discussed is the distortion of the signal shape which can depart significantly from the prediction of the BW formula. This can be seen in the Figure by comparing the dashed red curves, which are obtained by the BW distribution normalized to \mbox{$\sigma\times$BR}, where $\sigma$ is defined by \eq{CS} with the red solid lines obtained by {\sc{MadGraph5}} simulations of the $2\to2$ process $pp\to V_0^*\to l^+ l^-$. We see that in the peak region the distortion is rather mild when the resonance is light ($2$~TeV) at both the $8$ and $14$~TeV LHC. The effect is barely visible for $\Gamma/M=2\%$ and more pronounced for a broad resonance $\Gamma/M=10\%$. The distortion is more significant for a $3.5$~TeV mass but the deviation is still under control. This can be seen by comparing the area of the two curves in the interval $[M-\Gamma,M+\Gamma]$ depicted as a shadowed region in the plots. The relative deviation is depicted in the inset plots for $y=0$. We see that it is of around $10\%$ in the worst case of $M_V=3.5$~TeV and $\Gamma/M_V\sim11\%$. Outside the peak, on the contrary, the situation is worrisome already for $M_V=2$~TeV. The simulated signal has a long tail extending towards small invariant masses which is due to the steep fall of the parton luminosities.

By focusing on the peak, where the signal is well approximated by the BW prediction, extracting the limit on \mbox{$\sigma\times$BR} is straightforward. For instance one could simply measure the cross-section of the $2\to2$ process integrated in a window around the resonance mass and convert it into a bound on \mbox{$\sigma\times$BR} by rescaling for the fraction of events that, according to the BW distribution, are expected to fall in the selected window. Alternatively, one could perform a shape analysis by assuming a BW signal shape and extract a limit on its normalization. Also in this second case the analysis should be restricted to the peak region because the tail is not well-described by the BW formula. Notice in particular that the total area of the simulated signal, that gives the total $2\to2$ cross-section, can differ considerably from \mbox{$\sigma\times$BR}. The low-mass tail, which extends in a wide range of masses, can indeed give a sizable contribution to the total integral.

In order to understand the effect in more detail, let us briefly remind the reader of the assumption under which the BW formula is derived. The measured signal is $pp\to l^+ l^-$ whose partonic cross-section is
\beq
\hat{\sigma}_{\text{S}}(\hat{s})=\frac{4\pi\hat{s}}{3 M_V^2}\frac{ \Gamma_{V\to q_{i}q_{j}} \Gamma_{V\to l^+l^-}}{(\hat{s}-M_{V}^{2})^{2}+M_{V}^{2}\Gamma^{2}}\,.
\eeq
After convoluting with the PDFs, and using $\hat{s}=M_{l^+l^-}^2$, the total differential cross-section reads
\beq
\frac{d\sigma_{\text{S}}}{dM_{l^+l^-}^2}=\sum_{i,j} \frac{4\pi}{3 }\frac{ \Gamma_{V\to q_{i}q_{j}} \Gamma_{V\to l^+l^-}}{(M_{l^+l^-}^2-M_{V}^{2})^{2}+M_{V}^{2}\Gamma^{2}}\frac{M_{l^+l^-}^2}{M_V^2\;\;}\frac{dL_{ij}}{d\hat{s}}\Bigg|_{\hat{s}=M_{l^+l^-}^2}\,.
\label{222}
\eeq
In the peak region, namely for $M_{l^+l^-}-M_V\sim\Gamma$, and only in that region, it is reasonable to approximate
\beq
\frac{M_{l^+l^-}^2}{M_V^2\;\;}\frac{dL_{ij}}{d\hat{s}}\Bigg|_{\hat{s}=M_{l^+l^-}^2}\simeq \frac{dL_{ij}}{d\hat{s}}\Bigg|_{\hat{s}=M_{V}^{2}}\,,
\label{as}
\eeq
from which, using \eq{CS}, the differential cross-section can be written in terms of the on-shell \mbox{$\sigma\times$BR}, times a universal function
\beq
\frac{d\sigma_{\text{S}}}{dM_{l^+l^-}^2}=\sigma\times\textrm{BR}_{V\to l^+l^-} {\textrm{BW}}(M_{l^+l^-}^2;M_V,\Gamma)\,,
\label{bw}
\eeq
where BW denotes the standard relativistic BW distribution
\beq
{\textrm{BW}}(\hat{s};M_V,\Gamma)=\frac{1}\pi \f{\Gamma M_V}{(\hat{s}-M_V^{2})^{2}+M_V^{2}\Gamma^{2}}\,.
\eeq
Whether \eq{bw} is a good description of the $2\to2$ shape or not depends on how accurately the assumption \eqref{as} holds, namely it depends on how fast the parton luminosities vary in the peak region. Therefore the agreement is better for smaller widths when the peak is narrower. But the level of agreement also depends on the resonance mass and decreases when the resonance approaches the kinematical production threshold of the collider. This is because after a certain threshold the parton luminosities start to decrease extremely fast, more than exponentials, so that regarding them as constants is less and less justified even for a narrow width. This threshold corresponds, in Figure~\ref{fig:partonlumi}, to the point where the luminosities become concave functions in logarithmic scale around $3$ or $3.5$~TeV at the $8$~TeV LHC. This explains why the peak distortion is so pronounced at the $3.5$~TeV mass point. Notice that the peak distortion could be modelled, starting from \eq{222}, and the agreement with the simulated signal improved by employing a ``distorted'' BW shape. We will not discuss this possibility because we consider the BW description to be sufficiently accurate in the cases at hand. However such an improvement could be helpful in order to deal with more problematic situations.

The second important effect to be taken into account originates from the quantum mechanical interference of the resonance production diagrams with those of the SM background. Differently from before the strength of this second effect crucially depends on the amount of background which is present in the peak region or, more precisely, on the signal to background ratio. Notice that only the strictly irreducible backgrounds matter, because interference can only occur among processes with the exact same initial and final states at the partonic level. In Figure \ref{Figure:Breit-Wigner}, the upper and lower boundaries of the green shaded region are the result of two complete simulations, including interference, of the $pp\to l^+ l^-$ process as obtained at two different points of the parameter space of our model. For each mass and collider energy the two points are chosen to have identical production rates and partial widths but, respectively, constructive and destructive interference. The two points are simply related by flipping the relative sign of $c_q$ and $c_l$, which leads to identical rates and widths but opposite interference. The green solid lines correspond instead to the ``signal plus background'' prediction, obtained by ignoring the interference and summing the background, reported in black, with the ``signal only'' line in red. In dashed green we show the signal plus background curves obtained by the BW prediction of the signal. Notice that the interference never vanishes in any model so that the signal-plus-background shape does not represent any point of the parameter space. However the interference could be reduced, and most of the shaded region could be populated by some explicit model. Therefore, imagining for simplicity that the interference can be continuously varied from constructive to destructive we define 
\beq
\label{sigmay}
\frac{d\sigma_{\text{Full}}}{dM_{l^+l^-}}(y)=\frac{d\sigma_{\text{B}}}{dM_{l^+l^-}}+\frac{d\sigma_{\text{S}}}{{dM_{l^+l^-}}}+y\, \frac{d\sigma_{\text{I}}}{dM_{l^+l^-}}\,.
\eeq
By varying $y$ among $-1$ and $1$ we can get a rough idea of how much the interference effect can change the shape in different regions of the parameter space.

We see in the Figure that the shape distortion due to the interference is considerable, and in most cases more significant than the one due to the PDFs. Notice however that we are voluntarily considering pessimistic cases where the interference distortion is enhanced. The idea is that if we manage to deal with these situations we will have no problems in covering more favorable cases. As mentioned above, the interference distortion depends on the signal to background ratio, therefore for a given mass and collider energy, where the background is fixed, the effect is maximal for the smallest possible signal cross-section. For the plots in Figure \ref{Figure:Breit-Wigner} we thus selected the minimal cross-sections that can be excluded at the $8$~TeV LHC with $20\;\textrm{fb}^{-1}$ and at $14$~TeV with  $100\;\textrm{fb}^{-1}$. Stated in a different way, we placed ourselves at the boundary of the excluded \mbox{$\sigma\times$BR} region for each mass hypothesis. In the bulk of the excluded region, where \mbox{$\sigma\times$BR} is well above the one assumed in the plots, the signal shape would grow and the interference effect would become relatively less important. With this choice, the interference  is more important at $14$ than at $8$~TeV because with the assumed luminosity the exclusion will be set in a region where the background is larger. For the $3.5$~TeV mass point the interference is negligible because the background is very small and the distortion is mainly due to the PDF effect as described above.

Notice that, differently from the PDF effect, the distortion due to the interference can not be modeled in any simple way. Namely, it is impossible to cast it in a way that only depends on the production rate and on the widths, indeed we saw above that it depends on other parameter combinations. Obviously it could be computed by a simulation, but the resulting shape would depend in a complicated way on all the model parameters and could not be taken as a universal template. Therefore by proceeding in this way it would not be possible to set model-independent limits on  \mbox{$\sigma\times$BR} and the comparison of the model with the data should be performed by scanning the parameter space with simulations on a grid of points.

Two different attitudes could be taken towards this problem. One could insist with a shape analysis, assuming a BW signal, and accept the intrinsic systematic uncertainty associated with this assumption. Of course the uncertainty should be quantified by comparing with the limits obtained with the ``true'' shape, taken for instance from \eq{sigmay} for different values of $y$. Alternatively, one could turn to a simpler cut-and-count experiment and try to reduce the impact of the interference by exploiting the following observation. In general, the interference contribution to the partonic cross-section has the functional form
\beq\label{interf}
{\hat{\sigma}}_{\text{I}}(\hat{s})\propto\f{(\hat{s}-M_V^{2})}{(\hat{s}-M_V^{2})^{2}+M_V^{2}\Gamma^{2}}\,,
\eeq
so that it vanishes exactly at $\hat{s}=M_V^2$ as is odd around this point. This explains why the shaded green region shrinks to a point for an invariant mass equal to the resonance mass. After PDF convolution one can show that, provided the approximation of constant parton luminosities in \eq{as}  holds accurately enough, the interference contribution to the signal shape is also an odd function around $M_{l^+l^-}=M_V$ and thus it cancels when integrated over a symmetric interval. The signal in the $[M_V-\Gamma, M_V+\Gamma]$ region is thus much less sensitive to the interference than the shape itself.\footnote{For a complete cancellation one should consider a domain which is symmetric in the squared invariant mass variable. The cancellation is only approximate in the window we have chosen. However we prefer to stick to this simpler prescription of a symmetric domain in $M_{l^+l^-}$, because the cancellation would not be exact anyhow due to the PDF variation in the peak region.} This is confirmed by the inset plots, where we report the relative deviation of the total signal in the window, compared to the BW signal plus background prediction as a function of the parameter $y$. We see that the deviation is typically below $10\%$ even in cases where the shape distortion due to the interference is considerable.

In view of the above results, let us briefly discuss the limit setting procedure employed by CMS \cite{CMS-PAS-EXO-12-061} and ATLAS \cite{Aad:2014cka} in the di-lepton searches. After suitable selection and identification cuts, both analyses perform a shape analysis on the di-lepton invariant mass distribution based on an un-binned (CMS) or binned (ATLAS) likelihood. The only relevant difference among the two methods is the choice of the assumed signal distribution and the mass-range where the analysis is performed. CMS employs a gaussian shape obtained by convoluting a narrow resonance peak with the detector resolution function and the analysis is performed in an invariant mass window around the resonance mass. If the resonance is assumed to be extremely narrow, the CMS strategy is definitely correct and leads to an accurate determination of \mbox{$\sigma\times$BR}. However, no finite width effect is taken into account with this method. It is not even clear, and we plan to study this and related aspects in a future publication, how narrow the resonance must be in order to make this method reliable. Notice that asking for a width below the experimental resolution might not be sufficient as the distortion effects outlined above take place already in the theoretical distribution and are completely unrelated with the detector resolution. Furthermore, assuming a too small width might be inconsistent with the amount of signal needed for exclusion. A given DY cross-section requires, at fixed mass, a given $q\overline{q}$ partial width and thus a minimal total width. Moreover, a non-vanishing BR into di-leptons is needed, leading to a larger minimal width. By exploiting this observation it is possible to compute the minimal width needed, at a given mass, for $3$ or more signal events at the $8$~TeV LHC. For a mass of $3.5$~TeV, for instance, the minimal width is $\Gamma/M_V\gtrsim10\%$ and therefore it would be inconsistent to set an exclusion limit for a very narrow resonance at this mass. The existence of a minimal width is the reason why we have not considered the case of a narrow $3.5$~TeV resonance in Figure~\ref{Figure:Breit-Wigner}. The ATLAS strategy is different from CMS in two respects. First, it performs the shape analysis in the full mass range rather than around the peak. In light of the above discussion, this is definitely a limitation. Second, it employs a template signal shape obtained by a sequential $Z'$ model \cite{Langacker:2008yv}. In this manner ATLAS somehow takes the effect of the width into account, but in a rather incomplete way because at each mass point the width is the one predicted by the sequential $Z'$ model. In other scenarios, with larger $\gst$, the width could be larger and the limit could change significantly. One might argue that at least the ATLAS limit, differently from the CMS one, is strictly correct within the specific model assumed in the simulation. However this is questionable as the interference effects, which are relevant close to the exclusion limit as shown above, are not included in the simulations.

\subsubsection*{The case of lepton-neutrino}

In the discussion above we focused on the simple example of the di-lepton final state, however our considerations are more general and apply to all those searches where the resonance invariant mass distribution can be reconstructed. This clearly includes di-jets \cite{Aad:2014aqa, CMS-PAS-EXO-12-059} and di-bosons in the hadronic channels \cite{CMS-PAS-EXO-12-024}, but also searches with one leptonic $W$ and reconstructed neutrino momentum \cite{Aad:2014pha, Khachatryan:2014xja}. In all these cases, the invariant mass distribution approximately follows the BW formula and the distortions due to the PDF and interference effects could be analyzed along the lines described above. Of course we expect that in these more complicated examples the experimental resolution, which we could safely ignore for di-leptons, could play an important role and should be taken into account. However, no qualitative difference is expected.

A radically different situation is instead encountered when the invariant mass can not be reconstructed, as in the CMS and ATLAS searches \cite{CMS-PAS-EXO-12-060,ATLAS-CONF-2014-017} in the lepton-neutrino final state. Setting a model-independent limit on $\sigma\times \text{BR}$ might seem hopeless in this case, because one can not rely on the BW formula which of course only describes the invariant mass distribution while the relevant observable is now the transverse mass $M_T$. This problem has been studied in detail in Ref.~\cite{Accomando:2011up} with the conclusion that indeed a model-independent limit can not be set and that the search must be reinterpreted in each given model separately. However, there could be a way out. Any pair of massless leptons, of any chirality, produced by the DY mechanism through the $s$--channel exchange of one vector, are characterized by a universal angular distribution relative to the beam direction in  the center of mass frame. Namely, the angular dependence of the partonic cross-section is effectively $1+\cos^2\theta$ because the term linear in $\cos\theta$ cancels out for a symmetric proton--proton collider such as the LHC.\footnote{Of course it does not cancel for asymmetric beams and this is why $Z$ boson asymmetries could be studied at the Tevatron.} Given that the angular dependence is fixed, the $p_{T}$ distribution of the final states can be uniquely computed and expressed, as usual in the limit \eqref{as} of slowly varying PDFs, in terms of $\sigma\times \text{BR}$. If the resonance is produced at rest in the transverse plane, which we expect to be a good approximation when it is sufficiently heavy, we have $M_T=2p_{T}$ and we predict
\beq
\displaystyle
\frac{d\sigma}{dM_T^2}=\sigma\times \text{BR}_{V\to l\nu} \textrm{TBW}(M_T; M_V,\Gamma)\,
\label{BWtransmass}
\eeq
where we denote as TBW a ``transverse BW'' distribution defined by the following integral
\beq
\displaystyle
\textrm{TBW}(M_T; M_V,\Gamma)= \f{3 \Gamma}{8 \pi M_{V}} \int_{M_{T}^{2}}^{s}\f{d\hat{s}}{\sqrt{\hat{s}-M_{T}^{2}}}\f{2\hat{s}-M_{T}^{2}}{(\hat{s}-M_V^{2})^{2}+\Gamma^{2}M_V^{2}}\f{1}{\sqrt{\hat{s}}}\,.
\eeq

\begin{figure}[t!]
\begin{tikzpicture}
    \begin{scope}
    \node {\includegraphics[scale=0.355]{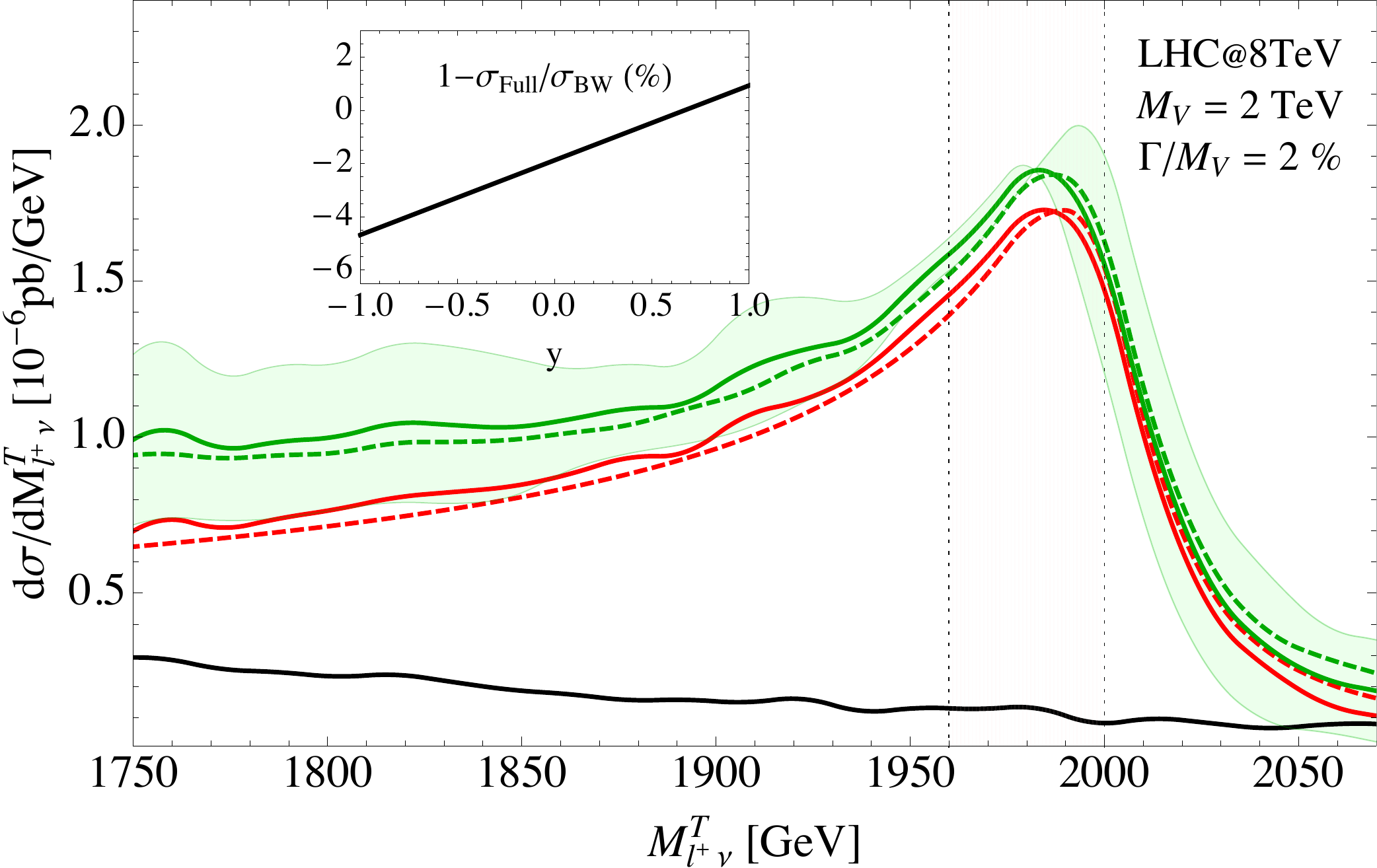}};
    \end{scope}
    \begin{scope}[xshift=8cm]
    \node {\includegraphics[scale=0.355]{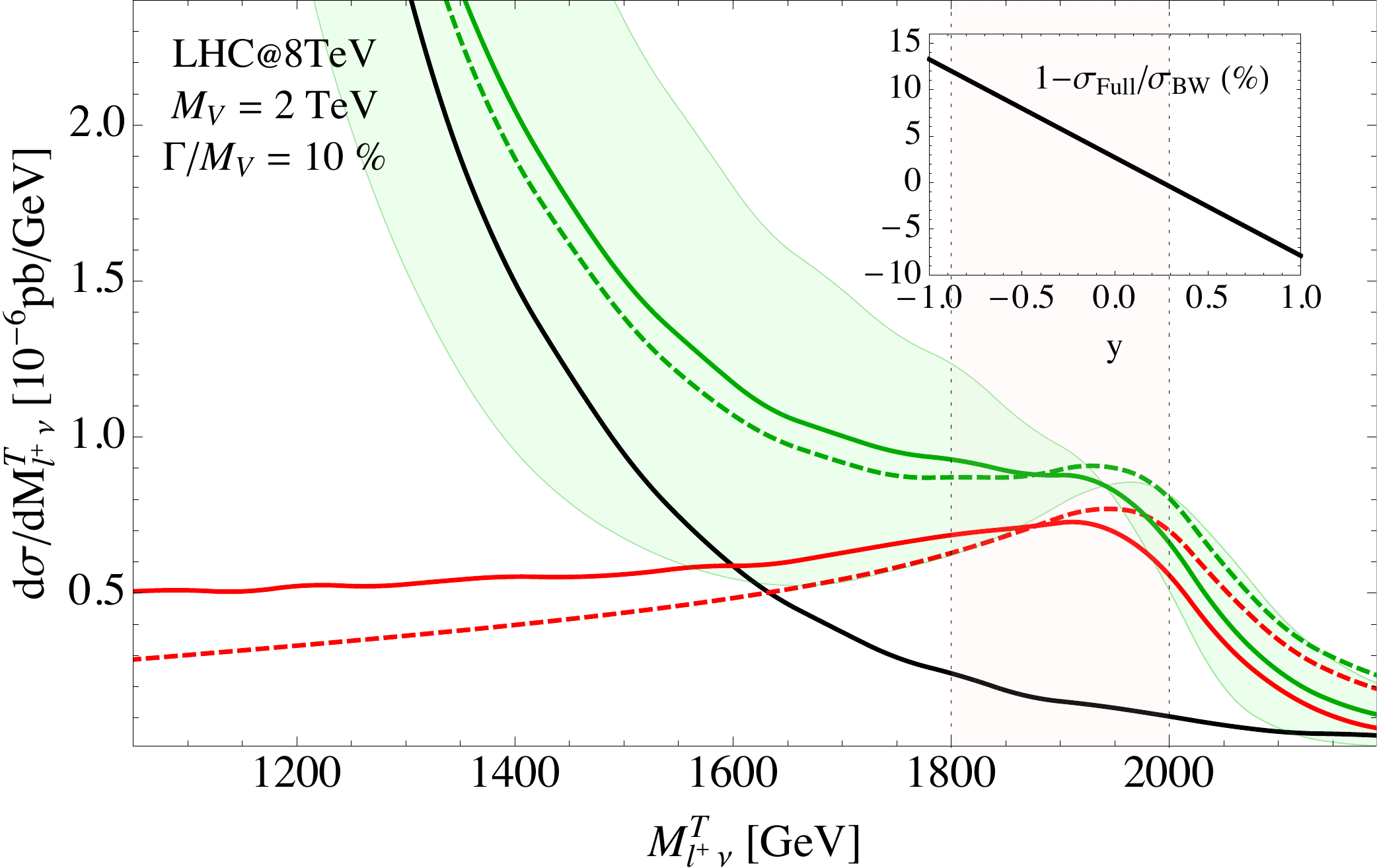}};
    \end{scope}
\end{tikzpicture}
\caption{\small Lepton-neutrino transverse mass distribution for the choices of $M_{+}$ and $\Gamma_{+}/M_{+}$ analogous to the ones of the first two plots of Figure \ref{Figure:Breit-Wigner} at the LHC at 8 TeV. The Figures show the dependence on $\Gamma_{+}/M_{+}$ of the difference between the full $2\to2$ calculation and a simple TBW distribution normalized to the on-shell production cross-section and multiplied by the corresponding BR into lepton-neutrino. The notation (dashing, coloring, ``inset'' plots) is identical to Figure \ref{Figure:Breit-Wigner}.}\label{Figure:Breit-WignerT}
\end{figure}

Needless to say, \eq{BWtransmass} is obtained by neglecting the interference, and in the approximation of slowly-varying PDF. The level of agreement with the ``true'' signal is illustrated by Figure~\ref{Figure:Breit-WignerT}. We considered the same points of the parameter space that were used in Figure~\ref{Figure:Breit-Wigner} for the $2$~TeV neutral resonances at the $8$~TeV LHC and we show the $M_T$ shape of the associated charged state. We see that the signal, defined as the cross-section in the window $M_T\in [M_V-\Gamma,M_V]$, is described by \eq{BWtransmass} at the $10\%$ level. However in this case, differently from the previous one, most of the signal is lost when restricting to the window we have selected, decreasing the sensitivity of the analysis. One should probably try to enlarge the window, accepting a larger error. Notice however that the interference, which is the dominant distortion effect, has been maximized in Figure~\ref{Figure:Breit-WignerT} by choosing the smallest possible rate as described above. The accuracy of the method would improve for higher rates, allowing at least to set a more conservative, but robust and model-independent limit.

\section{Explicit Models}
\label{4}

In this Section we present two examples of explicit models to populate the parameter space of the Simplified Model. The first one, called model A, describes the vector triplet emerging from the symmetry breaking pattern $SU(2)_1\times SU(2)_2\times U(1)_Y\to SU(2)_L\times U(1)_Y$ achieved through a linear $\sigma$-model \cite{1980PhRvD..22..727B}. The second model, B, describes the vector triplet considered in Ref.~\cite{Contino:2011np} based on a non-linearly realized $SO(5)/SO(4)$ global symmetry. 

\subsection{Model A: extended gauge symmetry}\label{4.1}

We consider the gauge theory $SU(2)_1\times SU(2)_2\times U(1)_Y$ \cite{1980PhRvD..22..727B}. The SM fermions are assumed to be charged under $SU(2)_1$ and $U(1)_Y$ with their usual quantum numbers. The SM Higgs doublet transforms as a $(\mathbf 2, \mathbf 1)_{1/2}$ under the enlarged gauge group. We also introduce an additional scalar field $\Phi$ transforming as a real bidoublet $(\mathbf 2, \mathbf 2)_{0}$. The bosonic part of the Lagrangian is
\begin{equation}\label{lagwc}
\mathcal L=-\frac{1}{4 g_1^2} W^a_{1\mu\nu} W_1^{a\mu\nu}-\frac{1}{4 g_2^2} W^a_{2\mu\nu} W_2^{a\mu\nu}+D_\mu H^\dagger D^\mu H+{\textrm{Tr}}(D_\mu \Phi^\dagger D^\mu \Phi)- \mathcal V(H,\Phi) \, .
\end{equation}
In order to obtain the SM at low energies we assume the potential $\mathcal V$ in \eq{lagwc} to be such that $\Phi$ obtains a vacuum expectation value
\begin{equation}\label{vevphi}
\langle\Phi\rangle=\begin{pmatrix}f&0\\0&f
\end{pmatrix}.
\end{equation}
This VEV breaks the $SU(2)_1\times SU(2)_2$ gauge symmetry to its vectorial subgroup which is identified with the SM $SU(2)_L$ gauge group. By going to the unitary gauge for the heavy vector triplet one obtains the following mass term from the kinetic term of $\Phi$
\begin{equation}
\textrm{Tr}(D_\mu \Phi^\dagger D^\mu \Phi)\supset \frac{f^2}{2}(W^a_{1\mu}-W^a_{2\mu})^2.
\end{equation}
A single gauge invariance under which both $W_1$ and $W_2$ shift in the same way is preserved. It is thus useful to perform the following field redefinition
\begin{equation}\label{FRtriplet0}
W^a_{2\mu}=V^a_{\mu}+W^a_{1\mu}.
\end{equation}
In this way $V$ transforms as the triplet of Section \ref{2} and $W_1$ is just the SM $W$ boson field (the index ``$1$'' will be dropped from now on).

The only part of the Lagrangian that transforms non trivially under the field redefinition in \eq{FRtriplet0} is the kinetic term of $W_2$. One has
\begin{equation}
W_{2\mu\nu}^a=
 D_{[\mu}V^a_{\nu ]}+\epsilon^{abc}V^b_{\mu}V^c_{\nu}+W^a_{\mu\nu}\,,
\end{equation}
which leads to
\begin{equation}
\bry{lll}
W^a_{2\mu\nu}W^{a\mu\nu}_{2}&=&\dst W^a_{\mu\nu}W^{a\mu\nu}+D_{[\mu}V^a_{\nu ]} D^{[\mu}V^{a\nu ]}+2W^a_{\mu\nu}D^{[\mu}V^{a\nu ]}\vspace{2mm}\\ 
&&+ 2 \epsilon^{abc} W^a_{\mu\nu}V^{b\mu}V^{c\nu}+O(V^4)\,.
\ery
\end{equation}
After the redefinition, the Lagrangian develops a kinetic mixing between $V$ and $W$ and thus it can be matched with \eq{sml} only after the mixing is removed by one further redefinition. This is performed in Appendix \ref{AppA}, starting from a``tilded'' field basis in which the kinetic mixing term is present. By identifying
\begin{equation}
\gst \equiv g_2\quad {\textrm{and}}\quad  \frac{1}{g^2}\equiv\frac{1}{g_1^2}+\frac{1}{g_2^2} \,,
\end{equation}
we have
\begin{equation}\label{matctriplet0}
\tilde m_{V} =\gst  f, \quad \tilde c_{V W}=-\tilde c_{VV W}=\tilde c_{VVV}=-1,\quad \tilde c_{H}=\tilde c_{VV HH}=\tilde c_{F}=0\,,
\end{equation}
from which we obtain the parameters in \eq{sml} by the relations in \eq{cnontilde}. In particular, we see that in all cases, $g_\ast\sim g$ or $g_\ast\gg g$
\begin{equation}
c_H\sim -g^{2}/\gst^{2}~~~{\textrm{and}}~~~ c_F\sim 1\,.
\end{equation}

Depending on the precise form of the potential $\mathcal V(H,\Phi)$ and in particular depending on the presence of a $\lambda |H|^2{\textrm{Tr}\,\Phi^\dagger \Phi}$ term, an additional contribution to $c_{VVHH}$ proportional to $\lambda g_2^2 f^2/m_\Phi^2$ is generated by integrating out the physical mode of $\Phi$. We define our benchmark model setting $\lambda=0$. However, finite $\lambda$ effects can be easily accounted for by modifying \eq{matctriplet0}. We will come back to this point in Section~4.3.  Notice that integrating out $\Phi$ also generates (irrelevant) contributions to the quartic interaction of $V$.

\subsection{Model B: Minimal Composite Higgs Model}\label{4.2}

Models in which the Higgs boson emerges as a light state (a pseudo Nambu-Goldstone boson) from an underlying strong dynamics predict the existence of heavy vector resonances with electroweak quantum numbers. In the case of the Minimal Composite Higgs Model (MCHM), where the Higgs doublet emerges from the spontaneous breaking of a global $SO(5)$ symmetry to an $SO(4)$ subgroup, these resonances have been discussed in Refs.~\cite{Contino:2011np,Contino:2013un}. Here we want to show how the lightest vector resonance in these models can be described by our Simplified Model.
In order to enforce the constraints imposed by the underlying symmetry structure a minimal amount of technical complications is required. Here we follow Ref.~\cite{Contino:2011np} which uses the Callan-Coleman-Wess-Zumino (CCWZ) formalism reviewed, for instance, in Appendix A of Ref.~\cite{DeSimone:2012ul}. The matching with the Lagrangian of the Simplified Model can be found at the end of this Section and the reader who is not interested in the derivation can jump there directly.

We introduce a spin 1 field ${\rho}_{\mu}$ transforming under the unbroken $SO(4)$ subgroup as a $({\bf{3}},{\bf{1}})$ irreducible representation
\begin{equation}\label{rhotrans}
{\rho}_\mu\equiv {\rho}_\mu^{a} t^{a}\;\rightarrow\;h_4{\rho}_\mu h_4^T-ih_4\partial_\mu h_4^T~~~{\textrm{for}}~~~a=1,2,3\,, 
\end{equation}
where $t^a$ are generators of the $SU(2)_L$ subgroup of $SO(4)$ in the vector representation and $h_4$ is a non-linear $SO(4)$ transformation whose construction is described in Appendix A of Ref.~\cite{DeSimone:2012ul}. 
We consider the following Lagrangian
\begin{equation}\label{rholag}
\mathcal L_{\rho}=-\frac{1}{4\hat g'^2}(B_{\mu\nu})^2-\frac{1}{4 \hat g^2}(W^a_{\mu\nu})^2+\frac{f^2}{4} d_\mu^{i}d^{\mu i}-\frac{1}{4 g_\rho^2}({\rho}_{\mu\nu}^{a})^2+\frac{m_{\rho}^2}{2g_\rho^2}\left({\rho}_\mu^{a}-e_\mu^{a}\right)^2.
\end{equation}
The $\rho$ field strength is given by $\dst {\rho}_{\mu\nu}^{a}= \partial_\mu {\rho}_{\nu}^{a}- \partial_\nu {\rho}_{\mu}^{a}-\epsilon^{abc}{\rho}_{\mu}^{b}{\rho}_{\nu}^{c}$. The full expressions for the $d$ and $e$ symbols for $SO(5)/SO(4)$ are given in Appendix A of Ref.~\cite{DeSimone:2012ul}. Here we will only need approximate formulas in the large $f$ limit

\begin{equation}\label{dmudmu}
d_\mu^i d^{\mu\, i}= \frac{4}{f^2}|D_\mu H|^2+\frac{2}{3 f^{4}}\left[(\partial_\mu |H|^2)^2-4 |H|^2 |D_\mu H|^2\right]+O(1/f^6) \, ,
\end{equation}
and
\begin{equation}\label{emu}
\rho^a_\mu- e_\mu^a=\rho_{\mu}^{a}+W_\mu^a-\frac{i}{f^{2}}H^\dagger \tau^a \overset{\longleftrightarrow}{D_{\mu}} H+\frac{i}{f^{4}}|H|^2H^\dagger \tau^a \overset{\longleftrightarrow}{D_{\mu}} H+O(1/f^6).
\end{equation}
We can thus define the triplet $V$, which does not shift under the SM gauge group, as 
\begin{equation}
V_\mu^a\equiv\rho_{\mu}^{a}+W_\mu^a.
\end{equation}
Under this field redefinition the $\rho$ kinetic term transforms as
\begin{equation}\label{rhohatrho}
\rho_{\mu\nu}^a=D_{[\mu}V^a_{\nu ]}-\epsilon^{abc}V_\mu^bV_\nu^c-W^a_{\mu\nu}\,,
\end{equation}
and using the large $f$ expressions in \eq{dmudmu} and (\ref{emu}) it is now straightforward to match $\mathcal L_\rho$ with the ``tilded'' basis of Appendix \ref{AppA}. By identifying
\begin{equation}
\gst=g_\rho,~~~\frac{1}{g^2}=\frac{1}{\hat g^2}+\frac{1}{g_\rho^2}~~~{\textrm{and}}~~~g'=\hat g' \,,
\end{equation}
and after normalizing the kinetic term of $V$ we obtain
\begin{equation}\label{mathchingNSM}
\tilde m_{V} =m_\rho, \quad  \tilde c_{V W}=\tilde c_{VV W}=\tilde c_{VVV}=1,\quad \tilde c_{H}=-\frac{m_\rho^2}{g_\rho^2 f^2}\equiv - a_\rho^2,\quad\tilde c_{VV HH}=\tilde c_{F}=0,
\end{equation}
where $a_\rho$ is an $O(1)$ free parameter as defined in Ref.~\cite{Contino:2011np}.
Using \eq{cnontilde} we see that
\begin{equation}
c_H\sim ~ c_F\sim 1\,.
\end{equation}
The difference with the linear model of the previous Section arises from the fact that $\tilde c_H$ is now non vanishing.

In order to perform the matching we ignored both higher dimension operators coming from subleading corrections to \eq{emu} and higher derivative terms which could be added to the Lagrangian in \eq{rholag}. We will discuss their effects in the next Section.

\subsection{The role of higher dimensional operators}
The simple phenomenological Lagrangian in \eq{sml} has been the starting point of our discussion. Its usefulness stems from the fact that it contains just a handful of parameters due to neglecting all the higher dimensional operators. 

As already stressed throughout the paper \eq{sml} has to be understood as an intermediate step to compare a more or less complete model of New Physics with the experimental data. That is, not as the leading subset of terms of the effective field theory describing the interactions of $V$ with the SM. From this point of view the fact that \eq{sml} is all that is needed has to be guaranteed by the underlying theory. We will now check this assumption for the two models we presented in the last two Sections.

This discussion is almost straightforward in the context of the linear model. Since the model is renormalizable higher dimensional operators can only be generated by integrating out the heavy physical fluctuations of the scalar field $\Phi$. A hierarchy of masses $m_\Phi\gg m_V$ is understood in order to be allowed to study the vector in isolation. The real bidoublet $\Phi$ can be written as
\beq
\Phi=\frac{\phi_0}{2}+i\tau^a\phi_a.
\eeq
In this way \eq{vevphi} can be rephrased as $\langle\phi_0\rangle=2 f$. The three scalar fields $\phi_a$ are unphysical and only $\phi_0$ remains in the spectrum with a mass $m_\Phi$ which depends on the parameters in the potential $\mathcal V(H,\Phi)$. In the unitary gauge for $V_\mu$, the only relevant $\phi_0$ interactions come from the kinetic term of $\Phi$ and from a mixed $\Phi$-$H$ quartic coupling which can be present in $\mathcal V$
\beq
\textrm{Tr}(D_\mu \Phi^\dagger D^\mu \Phi)-\lambda H^\dagger H\textrm{Tr}(\Phi^\dagger\Phi)=\frac{1}{2}(\partial_\mu \phi_0)^2+\frac{g_V^2}{8} \phi_0^2 V_\mu^aV^{\mu\,a}-\frac{\lambda}{2}\phi_0^2 H^\dagger H.
\eeq
By integrating out the heavy $\phi_0$ field we obtain the following Lagrangian containing operators up to dimension 6
\beq\label{deltalagwc1}
\Delta \mathcal L_A=\frac{2\lambda^2 f^2}{m_\Phi^2}|H|^4-\frac{\lambda g_2^2 f^2}{m_\Phi^2}V_\mu^aV^{\mu\,a}|H|^2+\frac{3}{2}\frac{\lambda^2 g_2^2 f^2}{m_\Phi^4}V_\mu^aV^{\mu\,a}|H|^4+O(V^4, |H|^6,\ldots) \, .
\eeq
The first term is an unobservable modification of the Higgs quartic coupling while the second is the contribution to $c_{VVHH}$ that was already anticipated in Section 4.1. 
It modifies the matching in Eq.~(\ref{matctriplet0}) by
\beq\label{deltacvvhh}
\tilde c_{VVHH}=\frac{\lambda f^2}{m_\Phi^2}.
\eeq
The operator
\beq
\mathcal O'_{VVHH}\equiv V_\mu^aV^{\mu\,a}|H|^4 \, ,
\eeq
is new and not included in the phenomenological model. It is easy to verify that all its effects, both in the mass matrix and in the couplings of $V$ to $W_L$, $Z_L$ and $h$, are suppressed by a factor $\lambda v^2/ m_\Phi^2$ with respect to those emerging from $\tilde c_{VVHH}$ in Eq.~(\ref{deltacvvhh}). In a reasonably weakly coupled theory these effects are small and can be safely neglected.

A similar discussion for the non-linear model described in Section 4.2 is necessarily more involved. This is due to its intrinsically finite energy range of validity. In order to have any predictive power the theory has to be endowed with a criterion to estimate the size of the coefficients of the higher dimensional operators. Using this criterion one must be able to show that only a finite number of operators is relevant to achieve a given precision. Here we adopt a slight modification of the \emph{partial UV completion} criterion used in Ref.~\cite{Contino:2011np}. We assume that a New Physics mass scale $m_*$ is defined (which could for instance characterize the mass scale of other resonances) such that $m_V\ll m_*$. We furthermore assume that all ``composite'' states in the theory, which include $V$, $H$ and the other resonances at $m_*$, interact with a strength of order $g_*$ when probed at energies of order $m_*$. More in detail we require that for $E\sim m_*$, amplitudes involving ``composite'' fields have size $g_* m_*$ and $g_*^2$ for three and four point functions respectively. Applying this to the scattering amplitude of four Goldstone bosons, it implies in particular that $m_*\sim g_* f$. This criterion has to be extended to estimate the size of those amplitudes involving weakly coupled fields, for instance insertions of the SM gauge bosons. These amplitudes originate from the EW force and not from the strong sector interactions. We thus require them to be suppressed by an additional factor $(g/g_*)^n$ where $n$ is the number of weakly coupled field insertions. In this last point we depart from the prescription of Ref.~\cite{Contino:2011np}.
The first intuitive consequence of this criterion is the fact that, in order for the model defined by \eq{rholag} to be consistent, it is not only necessary to have $m_V\ll m_*$, but also to have the vector $\rho$ weakly coupled, $g_\rho \ll g_*$.

Before discussing the role of higher derivative terms in the model of Section 4.2, it is worth noticing that the Lagrangian in \eq{rholag} already contains dimension-6 and higher operators which have not been included in the matching with the Simplified Model. The existence of these operators, even in the absence of heavy matter fields to be integrated out, is due to Higgs non-linearities emerging from the $\sigma$-model structure. Using Eqs.~(\ref{dmudmu}) and (\ref{emu}) one finds at the dimension-6 level
\beq\label{deltalagsc}
\Delta \mathcal L_B=\frac{1}{6f^2}\left(1- \f{3m_\rho^2}{4 g_\rho^2 f^2}\right)\left[(\partial_\mu |H|^2)^2-4 |H|^2 |D_\mu H|^2\right]+\frac{m_\rho^2}{g_\rho f^4}|H|^2 iV_\mu^a H^\dagger \tau^a {\overset{{}_{\leftrightarrow}}{D}}^\mu H+\ldots \, .
\eeq
The first term renormalizes the Higgs and Goldstones kinetic terms and through this affects all their interactions. However its contribution is suppressed by $\xi=v^2/f^2$ which is necessarily small in this scenario as mentioned already in Section 2.
The second term is a new dimension-6 operator
\beq
\mathcal O _{H}'=|H|^2 iV_\mu^a H^\dagger \tau^a {\overset{{}_{\leftrightarrow}}{D}}^\mu H \, ,
\eeq
with a coefficient of order $g_V/f^2$ (one should recall that $g_V\equiv g_\rho$ and $m_\rho=a_\rho g_\rho f\sim g_\rho f$). Qualitatively $\mathcal O'_H$ has the same effect as the $c_H$ operator both in the mass matrix and in the coupling of $V$ to $W_L$, $Z_L$ and $h$ and it induces small $O(\xi)$ corrections relative to the latter.
Additional operators collectively denoted by ``...'' in \eq{deltalagsc} and containing extra insertions of $|H|^2$ are always accompanied by more powers of $1/f^2$ so their contribution to the mass matrix and to the decay widths are further suppressed by powers of $\xi$.

The addition of higher derivative terms to the leading order Lagrangian of the non-linear model is thoroughly discussed in Ref.~\cite{Contino:2011np}. The analysis shows that only two CP even operators can give a relative contribution to physical processes which is larger than the typical size of a higher derivative correction, $m_V^2/m_*^2$. The two operators are
\begin{equation}
\mathcal O_1={\textrm{Tr}}(\rho_{\mu\nu}i[d_\mu,d_\nu])~~~{\textrm{and}}~~~\mathcal O_2={\textrm{Tr}}(\rho_{\mu\nu}A_{\mu\nu}).
\end{equation}
Here $A_{\mu\nu}$ is defined by
\beq
A_{\mu\nu}=U^\dagger \left(T^a_LW_{\mu\nu}+T^3_R B_{\mu\nu}\right) U \,,
\eeq
in terms of the Goldstone matrix $U$ and the SO(4) generators $T^a_{L,R}$,  which are defined in Appendix A of Ref.~\cite{DeSimone:2012ul}.
The two operators can be expanded at order $1/f^2$
\begin{eqnarray}
{\textrm{Tr}} (\rho^{\mu\nu} i[d_\mu, d_\nu])&=&-\frac{4i}{f^2}\rho^{\mu\nu\,a}D_\mu H^\dagger\tau^a D_\nu H+O(1/f^4)\,,\\
{\textrm{Tr}} (\rho_{\mu\nu} A_{\mu\nu})&=&-W_{\mu\nu}^{a}\rho_{\mu\nu}^{a}  \left(1-\frac{|H|^2}{2f^2}\right)+\frac{1}{f^2} B_{\mu\nu} \rho_{\mu\nu}^{a} H^\dagger \tau^a H+O(1/f^4).
\end{eqnarray}
According to our refined partial UV completion they should appear in the Lagrangian as 
\beq
\Delta \mathcal L_B=c_1 \frac{1}{g_\rho g_*} \mathcal O_1+c_2 \frac{1}{g_\rho^2} \mathcal O_2 \, ,
\eeq
with $c_{1,2}\sim 1$. Let us start focusing on $\mathcal O_1$.
Applying the field redefinition in \eq{rhohatrho} and normalizing the kinetic term of $V$, we obtain
\beq
\Delta \mathcal L_B=-\frac{4c_1}{g_*f^2}iD_{[\mu}V^a_{\nu ]} D_\mu H^\dagger \tau^a D_\nu H.
\eeq
We did not write operators contributing only to 4-point functions involving $V$, nor  operators involving only the SM fields.
The main effect of $\mathcal O_1$ is to modify the width of $V$ to longitudinal gauge bosons.
Using the equivalence theorem and the field redefinitions in Eqs.~(\ref{fr1}) and (\ref{fr2}) we obtain the corrections to \eq{egV1}
\bea\label{egV2}
\displaystyle
&&-\frac{g_V }{4(1-c_H^2\zeta^2)}\left[4c_{1}\frac{M_V^2}{g_\rho^2 f^2}\frac{M_V}{m_*}\right]\epsilon^{abc}V_\mu^a\pi^b\partial^\mu\pi^c\nn\\
&&+\frac{g_V }{2\sqrt{1-c_H^2\zeta^2}}\left[4c_{1}\frac{M_V^2}{g_\rho^2 f^2}\frac{M_V}{m_*}\right]\,h\,V_\mu^a\partial_\mu\pi^a.
\eea
Relative to the leading term which is proportional to $c_H\sim m_\rho^2/g_\rho^2 f^2$ the above contributions are suppressed even though only by a single power of $M_V/m_*$, the parameter controlling the derivative expansion.
 
To discuss the effects of $\mathcal O_2$, we apply the shift of \eq{rhohatrho} and obtain
\bea\label{deltalagsc1}
\Delta \mathcal L_B=&-&\frac{c_2}{g_V} D_{[\mu}V^a_{\nu ]} W^{\mu\nu\,a}+c_2\epsilon_{abc}W^{\mu\nu\,a} V_\mu^b V_\nu^c\\\nn
&-&\frac{c_2}{2f^2}|H|^2\epsilon_{abc}W^{\mu\nu\,a} V_\mu^b V_\nu^c+\frac{c_2}{2g_Vf^2}|H|^2 D_{[\mu}V^a_{\nu ]} W^{\mu\nu\,a}\\\nn
&-&\frac{c_2}{f^2}B^{\mu\nu}\epsilon_{abc}H^\dagger \tau^a H\, V_\mu^bV_\nu^c+\frac{c_2}{g_V f^2} B^{\mu\nu}D_{[\mu}V^a_{\nu ]}H^\dagger\tau^a  H+\ldots\,.
\eea
The first line of Eq.~(\ref{deltalagsc1}) contains $O(1)$ corrections to the matching conditions in \eq{mathchingNSM} which now become
\beq\label{mathchingNSM2}
\tilde c_{VW}=\tilde c_{VVW}=1-2 c_2.
\eeq
All the other operators except for the last one in \eq{deltalagsc1} induce negligible $O(\xi)$ corrections to the spectrum and to the width of $V$ into transverse gauge bosons. Finally the effect of 
\beq
\mathcal O _{VB}=B^{\mu\nu}D_{[\mu}V^a_{\nu ]}  H^\dagger\tau^a  H \, ,
\eeq
is qualitatively new. After EWSB it generates a kinetic mixing between the hypercharge gauge boson and $V^3$
\begin{equation}
\Delta\mathcal L_B\supset c_{2}\tan\theta_W\zeta\frac{\hat m_W}{m_V}\left(\frac{m_V}{g_\rho f}\right)^2B^{\mu\nu} V_{\mu\nu}^3.
\end{equation}
Such a mixing can be eliminated by a field redefinition of the form given in \eq{fieldred} but involving $B_\mu$
\beq\label{fieldredB}
\left\{
\bry{l}
B_\mu\rightarrow B_\mu + \alpha V_\mu^3\\
V_\mu^3\rightarrow \beta V_\mu^3
\ery
\right. \, ,
\eeq
with $\alpha \sim m_W/M_V$ and $\beta\sim1$. It is simple to show that after this field redefinition the spectrum is only modified by corrections of order $m_W^2/M_V^2$. The shift also affects the couplings $g_{L,R}^N$ of $V^3$ to fermions. The corrections are at most of order $m_W/M_V$, hence safely negligible.

To summarize, our study of higher derivative terms in the context of the non-linear model shows that the only additional structure to consider is the operator $\mathcal O_2$. Its effects can be included in the dimension four phenomenological Lagrangian by the modified matching conditions in Eq.~(\ref{mathchingNSM2}). Notice that among those dimension-6 operators that have not been listed in Eq.~(\ref{deltalagsc1}) because they only involve SM fields one operator is particularly relevant as it contributes to the $\hat S$ parameter
\beq\label{mathchingNSM1}
\Delta \mathcal L_B\supset-\frac{c_2}{g_V^2 f^2} B^{\mu\nu}W^{a\,\mu\nu}H^\dagger\tau^a  H, \quad \Delta\hat S=c_2\frac{\hat m_W^2}{g_V^2 f^2} \, .
\eeq
If $c_2\sim 1$ this correction is of the same size as those calculated in Appendix B and can have both signs.

\section{Conclusions}
\label{5}

We described a model-independent strategy to study heavy spin one particles in the triplet of the SM gauge group. Our method, depicted as a bridge in Figure~\ref{fig:uno}, is based on a Simplified Model Lagrangian, introduced in Section~\ref{2}, designed to reproduce a large class of explicit descriptions of the heavy vector in different regions of its parameter space. Two explicit examples, describing vectors with rather different properties and physical origin, are discussed in Section~\ref{4}. Those are denoted as model A and model B and correspond, respectively, to heavy vectors emerging from an underlying weakly-coupled extensions of the SM gauge group \cite{1980PhRvD..22..727B} and a strongly coupled Composite Higgs scenario \cite{Contino:2011np}. 

By studying the Simplified Model we derived a set of generic phenomenological features of the heavy vectors in Section~\ref{2}. In particular, we have seen that the charged and neutral states are essentially degenerate in mass and thus have comparable production rates. As discussed in Ref.~\cite{deBlas:2012tc}, this fact is a strong motivation for combining the searches of the two charge states. We have also seen that the heavy vector always has a negligible coupling with the transversely polarized EW bosons and the only relevant interactions are with the longitudinals. The longitudinal coupling is generically comparable with the one to the SM fermions in the region of the parameter space that corresponds to weakly coupled models and it becomes dominant in the strongly-coupled case. This is the main phenomenological difference between the two scenarios. Finally, we showed that not all the parameters of the model are equally relevant. The partial decay widths, and in turn the single production rate, are to a good approximation completely determined by the parameter combinations $\gst c_H$ and $g^2 c_F/\gst$. If we assume, for simplicity, a universal coupling to fermions, then the experimental limits on the heavy vector can be conveniently represented, for a given mass, on a two-dimensional plane as we did in Figure~\ref{Fig:Boundscfch}. The dependence on the other parameters is extremely mild and can be safely ignored. Moreover, the phenomenology being controlled by a few parameters implies tight model-independent correlations among different observables. For instance, the relative BRs of the charged and neutral states in different bosonic decay channels, including the ones with the Higgs boson in the final state, are basically fixed. This would make the combination of different experimental searches extremely easy.

In Section~\ref{3} we quantified the impact of the present experimental searches. Following the Bridge method we firstly translated the experimental results into limits on the Simplified Model parameters (Figure~\ref{fig:uno}) and afterwards converted them into the ``fundamental'' parameters of the explicit models A and B. The results are shown in Figure~\ref{Fig:Boundscfch3} in a mass-coupling two-dimensional plane. We see that model A is excluded for masses below around $2$ or $3$~TeV, depending on the coupling, while the limit is weaker in model B. For large coupling, which is expected in model B as this is supposed to represent a strongly-coupled scenario, the exclusion never exceeds $2$~TeV and is still comparable with the indirect limits from EWPT.

For our analysis we took all the experimental results at face value, and used the exclusions on \mbox{$\sigma\times$BR} at each mass point. However we pointed out in Section~3.3 that this might not be completely correct because of the effects associated with the finite resonance width, which might affect the limit setting procedures adopted by the experimental collaborations. We illustrated the expected impact of these effects on the invariant mass and transverse mass distributions that are employed in the di-lepton and in the lepton-neutrino searches respectively. Our conclusion is that finite-width effects can be considerable and can distort the signal shape in a significant way. In spite of this, we identified some strategies by which their impact could be reduced and a robust model-independent limit on  \mbox{$\sigma\times$BR} could be extracted. We plan to elaborate more on these aspects in a forthcoming publication.

Our work could be extended in at least three directions. First, one could easily consider other representations of the SM group. Aside from the triplet which we studied in the present paper, another relevant representation is the singlet, either neutral like a $Z'$ \cite{Langacker:2008yv,Salvioni:2009mt,Salvioni:2010p2769} or charged like a $W'$ \cite{Grojean:2011vu}. These particles emerge together in strongly-coupled models where they arise from a $(\mathbf{1},\mathbf{3})$ representation of the custodial group. Another interesting representation, which is present in models with a Composite pseudo-Nambu-Goldstone boson Higgs, is the doublet with $1/2$ hypercharge \cite{Agashe:2009ve}. A second limitation of our approach, which could be easily overcome, is the assumption of a linearly realised EW group, broken by the VEV of the Higgs doublet like in the SM. This is clearly a well-motivated assumption, but it might be worth studying also technicolor-like theories where the strong sector condensate breaks the EW symmetry directly. For this purpose our parametrization is insufficient because some higher dimensional operators involving extra powers of the Higgs field would be unsuppressed and should be included in the Simplified Model Lagrangian. Finally, in this paper we did not discuss the possibility of non-universal fermion couplings $c_F=\{c_l,c_q,c_3\}$ in detail. In particular, $c_3$ being different from the light fermions couplings $c_{l,q}$ is well-motivated in strongly coupled scenarios with partial fermion compositeness \cite{Giudice:1024017}. In this case, the large compositeness of the top quark induces a potentially large coupling to the third family quarks. Its effect on the searches with third family final states should be investigated.

\section*{Note added}
During the publication process of this paper new experimental searches and updates to previous ones have been published. We updated the searches in the paper as follows: Ref.~\cite{Aad:2014cka} supersedes Ref.~\cite{ATLAS-CONF-2013-017}, Ref.~\cite{ATLAS-CONF-2014-017} added, Ref.~\cite{Aad:2014aqa} supersedes Ref.~\cite{ATLAS-CONF-2012-148}, Ref.~\cite{Aad:2014pha} supersedes Ref.~\cite{ATLAS-CONF-2013-015} and Ref.~\cite{Khachatryan:2014xja} supersedes Ref.~\cite{CMS-PAS-EXO-12-025}.

\section*{Acknowledgements}
%
We would like to thank Brando Bellazzini, Roberto Contino, Davide Greco, Da Liu, Hui Luo and Manuel Perez-Victoria for useful conversations. We are grateful to the Galileo Galilei Institute for Theoretical Physics in Florence for hospitality during the initial stage of this project.
D.P. has been supported by the NSF Grant PHY-0855653. A.T.~acknowledges support from the Swiss National Science Foundation under contract no.~200020-138131. The work of R.T. was supported by the ERC Advanced Grant no.~267985 {\it DaMeSyFla} and by the Research Executive Agency (REA) of the European Union under the Grant Agreement number PITN-GA- 2010-264564 {\it LHCPhenoNet}. A.W.~acknowledges the MIUR-FIRB grant RBFR12H1MW. We finally thank Heidi for computing resources, the grant SNF Sinergia no.~CRSII2-141847 and the computing support of INFN Genova and INFN Padova.

\appendix


\section{The tilded basis}\label{AppA}

The field redefinition in \eq{fieldred} allows many equivalent Lagrangian description of the Simplified Model. In all but one of them a kinetic mixing between $V$ and $W$ is present. We define each of these bases by the same Lagrangian in \eq{sml} with all the couplings replaced by ``tilded'' ones
\begin{equation}
c\to \tilde c, \quad m_V\to \tilde m_V \, ,
\end{equation}
and with the addition of the kinetic mixing term
\begin{equation}
\tilde c_{VW} \frac{g}{2\gst }D_{[\mu}V^a_{\nu ]}W^{\mu\nu a}.
\end{equation}
Using the field redefinition of \eq{fieldred} with
\begin{equation}
\dst \alpha = \frac{g~\tilde{c}_{VW}}{\sqrt{\gst ^{2}  -\tilde{c}_{VW}^{2}g^{2}}}~~~{\textrm{and}}~~~
\dst \beta=\frac{\gst }{\sqrt{\gst ^{2}  -\tilde{c}_{VW}^{2}g^{2}}},
\end{equation}
we get the following relations between the parameters in the two bases
\begin{equation}\label{cnontilde}
\bry{lll}
\dst m_{V}^{2}=\frac{\gst ^{2}}{\gst ^{2}  -\tilde{c}_{VW}^{2}g^{2}} \tilde m_{V}^{2}\,,\vspace{2mm}\\
\dst c_{VV W}=\frac{\gst ^{2}}{\gst ^{2}  -\tilde{c}_{W \rho }^{2}g^{2}}\bigg[\tilde{c}_{VV  W}-\frac{g^{2}}{\gst ^{2}}\tilde{c}_{VW }^{2}\bigg]\,,\vspace{2mm}\\
\dst c_{VVV}=\frac{\gst ^{3}}{\left(\gst ^{2}  -\tilde{c}_{VW }^{2}g^{2}\right)^{3/2}}\bigg[\tilde{c}_{VVV }- \frac{g^{2}}{\gst ^{2}}\tilde{c}_{VW }\left(\tilde{c}_{VV  W}+2\right)+2\frac{g^{4}}{\gst ^{4}}\tilde{c}_{VW }^{3} \bigg] \,,\vspace{2mm}\\
\dst c_{H}=\frac{\gst }{\sqrt{\gst ^{2}  -\tilde{c}_{VW }^{2}g^{2}}}\left[\tilde{c}_{H}+\frac{g^2}{\gst ^2}\tilde{c}_{VW }\right]\,,\vspace{2mm}\\
\dst c_{VV  HH}=\frac{\gst ^{2}}{\gst ^{2}  -\tilde{c}_{VW }^{2}g^{2}}\bigg[\tilde{c}_{VV  H H} + \frac{g^{2}}{2\gst ^{2}}\tilde{c}_{VW } \tilde{c}_{ H}+ \frac{g^4}{4 \gst ^4}\tilde{c}_{VW }^{2}\bigg]\,,\vspace{2mm}\\
\dst c_{F}=\frac{\gst }{\sqrt{\gst ^{2}  -\tilde{c}_{VW }^{2}g^{2}}}\left[\tilde c_F+\tilde c_{VW}\right].
\ery
\end{equation}

\section{Electroweak precision tests}\label{AppB}
In this Appendix we discuss the constraints of EWPT on the Simplified Model parameter space. In order to do this we integrate out the vector triplet and describe the resulting theory as the SM supplemented by higher dimensional operators. We expect all the relevant corrections to be oblique, that is encoded in corrections to the vacuum polarization of the SM gauge bosons. This is not immediate to see in the basis of \eq{sml} as $V$ couples, though universally, to the light fermions. It is then useful to remove this coupling through the field redefinition\footnote{For convenience we work in the basis in which the gauge coupling appears only in front of the gauge kinetic term.}
\begin{equation}
W_\mu^a\to W_\mu^a-c_F \frac{g^2}{\gst } V_\mu^a.
\end{equation}
The resulting Lagrangian reads
\beq\label{smlEWPT1}
\bry{lll}
\dst {\mathcal{L}}_V&=&\dst -\frac{1}{4} \left(1+c_F^2\frac{g^2}{\gst ^2}\right)D_{[\mu}V_{\nu ]}^a D^{[\mu}V^{\nu ]\;a}+\frac{m_V^{2}}2V_\mu^a V^{\mu\;a}\vspace{2mm}\\
&&\dst+\, i\,\gst  \left(c_H-c_F\frac{g^2}{\gst ^2}\right) V_\mu^a H^\dagger \tau^a {\overset{{}_{\leftrightarrow}}{D}}^\mu H+\frac{c_F}{2 \gst }D_{[\mu}V^a_{\nu ]}W^{\mu\nu a}\vspace{2mm}\\
&&\dst +\,\gst ^{2} \left(c_{VVHH}+\frac{c_F^2}{4}\frac{g^4}{\gst ^4}-\frac{c_F c_H}{2}\frac{g^2}{\gst ^2}\right) V_\mu^aV^{\mu\;a} H^\dagger H+\ldots\,,
\ery
\eeq
while the coupling of $V$ to the light fermions is removed. Notice that a kinetic mixing between the $W$ and $V$ is reintroduced. The dots include terms of order $WV^2$, $V^3$, $V^4$. These are not relevant in the discussion of the EWPT. Normalizing the kinetic term gives the leading order equation of motion of $V$ 

\begin{equation}\label{VEOM1}
\left[(\square+ \mu_V^2) g_{\mu\nu}-\partial_\mu\partial_\nu\right]V_\nu^a=-i \gst   \gamma_{H}H^\dagger\tau^a \overset{\longleftrightarrow}{{D}_{\mu}} H+\gamma_F\frac{1}{\gst } D_\nu W^a_{\nu\mu}\equiv\mathcal J_\mu^a \, ,
\end{equation}
where
\begin{eqnarray}
 \left(1+c_F^2\frac{g^2}{\gst ^2}\right)\mu_V^2&=&m_V^2+2\left(c_{VVHH}+\frac{c_F^2}{4}\frac{g^4}{\gst ^4}-\frac{c_F c_H}{2}\frac{g^2}{\gst ^2}\right)\gst ^2 |H|^2,\\
 \left(1+c_F^2\frac{g^2}{\gst ^2}\right)^{1/2}\gamma_H&=& c_H-c_F\frac{g^2}{\gst ^2},\\
 \left(1+c_F^2\frac{g^2}{\gst ^2}\right)^{1/2} \gamma_F&=&c_F.
\end{eqnarray}
The solution of \eq{VEOM1} is
\begin{equation}
V_\mu^a=D_{\mu\nu} \mathcal J_\nu^a,\qquad D_{\mu\nu}=\frac{g_{\mu\nu}+\partial_\mu\partial_\nu/\mu_V^2}{\square+\mu_V^2}.
\end{equation}
Plugging this solution into \eq{smlEWPT1} (with normalised kinetic terms) and expanding in derivatives we get the leading terms contributing to the EWPT
\beq\label{dim6ewpt}
\bry{lll}
\dst \mathcal L_V&=&\dst -\frac{1}{2 \mu_V^2}\left(-i \gst  \gamma_{H}H^\dagger\tau^a \overset{\longleftrightarrow}{{D}_{\mu}} H+\gamma_F\frac{1}{\gst } D_\nu W^a_{\nu\mu}\right)^2\vspace{2mm}\\
&&\dst +\frac{1}{2 \mu_V^2}\left(-i \gst  \gamma_{H}H^\dagger\tau^a \overset{\longleftrightarrow}{{D}_{\mu}} H\right)\frac{\square^T_{\mu \nu}}{\mu_V^2}\left(-i \gst  \gamma_{H}H^\dagger\tau^a \overset{\longleftrightarrow}{{D}_{\nu}} H\right)+\ldots\,,\vspace{2mm}
\ery
\eeq
where we defined $\square^T_{\mu\nu}=g_{\mu\nu}\square-\partial_\mu\partial_\nu$. All other terms in the expansion, represented by the dots, give subleading contributions to the EWPT in a $\hat{m}_W^2/\mu_V^2$ expansion. Following Ref.~\cite{Cacciapaglia:2006pk} we rewrite the quadratic part of $\mathcal L_V$ as
\begin{equation}
\mathcal L=-\frac{1}{2}W_\mu^3\Pi_{33}(p^2)W^{\mu3}-\frac{1}{2}B_\mu\Pi_{00}(p^2)B^{\mu}-W_\mu^3\Pi_{30}(p^2)B^{\mu}-W_\mu^+\Pi_{\pm}(p^2)W^{\mu-} \, .
\end{equation}
The various form factors are then expanded in powers of $p^2$
\begin{equation}
\Pi(p^2)=\Pi(0)+p^2\Pi'(0)+\frac{p^4}{2}\Pi''(0)+\ldots.
\end{equation}
Starting from \eq{dim6ewpt} and following the procedure we outlined above we get the leading order contributions (as in the text we define $z\equiv \gst \hat v/2\mu_V$, $\hat m_W=g \hat v/2$ and $t_W\equiv \tan\theta_W=g'/g$)
\begin{eqnarray}
\Pi_{00}(0)&=&\Pi_{33}(0)=\Pi_{\pm}(0)=-\Pi_{30}(0)=-\frac{\hat v^2}{4}\left(1-z^2\gamma_H^2\right),\\\nn
\Pi'_{00}(0)&=&\frac{1}{g'^{2}}\left(1+  t_W^2 \gamma_H^2 z^2\frac{\hat m_W^2}{\mu_V^2}\right),\\\nn
\Pi'_{30}(0)&=&\frac{1}{g^{2}}\left(\gamma_H^2z^2 \frac{\hat m_W^2}{\mu_V^2}-\gamma_H \gamma_F \frac{\hat m_W^2}{\mu_V^2}\right),\\\nn
\Pi'_\pm(0)&=&\Pi'_{33}(0)=\frac{1}{g^{2}}\left(1+\gamma_H^2 z^2 \frac{\hat m_W^2}{\mu_V^2}+2 \gamma_H \gamma_F \frac{\hat m_W^2}{\mu_V^2}\right),\\\nn
\Pi''_\pm(0)&=&\Pi''_{33}(0)=\frac{1}{g^2\hat m_W^2}\left(2\gamma_F^2\frac{g^2}{\gst ^2}\frac{\hat m_W^2}{\mu_V^2}\right).
\end{eqnarray}
We thus obtain the following relations
\begin{eqnarray}
v^2|_{\textrm{exp}}&\equiv&-4\Pi_\pm(0)=\hat v^2\left(1-z^2\gamma_H^2\right),\\\nn
\frac{1}{g^2}\bigg|_{\textrm{exp}}&\equiv&\Pi'_\pm(0)=\frac{1}{g^{2}}\left(1+\gamma_H^2 z^2\frac{\hat m_W^2}{\mu_V^2}+2 \gamma_H \gamma_F \frac{\hat m_W^2}{\mu_V^2}\right),\\\nn
\frac{1}{g'^2}\bigg|_{\textrm{exp}}&\equiv&\Pi'_{00}(0)=\frac{1}{g'^{2}}\left(1+  t_W^2 \gamma_H^2z^2\frac{\hat m_W^2}{\mu_V^2}\right).
\end{eqnarray}
The relevant custodial invariant oblique parameters are defined by
\begin{equation}
\hat S=g^2 \Pi'_{30}(0),\quad W=\frac{g^2m_W^2}{2}\Pi''_{33}(0).
\end{equation}
The natural size of the coefficients $\gamma_H$ and $\gamma_F$ is $\gamma_H\sim\gamma_F\sim 1$. This implies that the oblique parameters will be at most of order $\hat m_W^2/\mu_V^2$, while
\begin{equation}
g|_{\textrm{exp}}=g+O(\hat m_W^2/\mu_V^2),\quad g'|_{\textrm{exp}}=g'+O(\hat m_W^2/\mu_V^2) \, ,
\end{equation}
so that the corrections to $g$ and $g'$ can be neglected in the calculation of the oblique parameters. Notice on the other hand that $\hat v$ can depart from its measured value $246$\,GeV by $O(1)$ corrections 
\begin{equation}
v^2|_{\textrm{exp}}=\hat v^2(1-\gamma_H^2z^2).
\end{equation}
One thus finds
\begin{equation}\label{sparam}
\hat S=\gamma_H^2 z^2 \frac{\hat m_W^2}{\mu_V^2}-\gamma_H \gamma_F \frac{\hat m_W^2}{\mu_V^2},\qquad W=\gamma_F^2\frac{g^2}{\gst ^2}\frac{m_W^2}{\mu_V^2} .
\end{equation}
where one has still to express $\hat v$ in terms of the physical  $v\simeq 246$\,GeV. Notice that under the assumption that $\gamma_F\sim 1$ the correction to the $V$ kinetic term which is present in \eq{smlEWPT1} is always subleading in a $\hat{m}_W^2/\mu_V^2$ expansion and can be neglected.

\section{Tools provided with this paper}\label{AppC}
In addition to the present paper we provide a set of tools useful to perform analyses using the Simplified Model. We make them available on the webpage of this project \cite{projectwebpage}.

The Simplified Model Lagrangian in \eq{sml} in the mass eigenstate basis and in the unitary gauge was implemented into different Matrix Element Generators (MEG) using the FeynRules \cite{Alloul:2013vc,Duhr:tm} {\sc Mathematica} package. Model files for the CalcHEP \cite{Belyaev:2012tt,Belyaev:Df-Kj9yc} and {\sc{MadGraph5}} \cite{Alwall:2011fk} MEG and the FeynRules source model are registered in the HEPMDB model database \cite{hepmdb} with the unique number hepmdb:0214.0151 and are available at the link \cite{hepmdbmodel}.

The model was implemented into FeynRules  taking $\alpha_{EW}$, $G_{F}$ and $M_{Z}$ as SM electroweak input parameters and the mass of the neutral heavy vector $M_{0}$, the overall coupling $\gst$ and all the parameters $c_{i}$'s as described in the paper as the new vector input parameters. The Higgs mass is also an input parameter, that we fix to a default value of $125.5$ GeV. All the other parameters appearing in this paper are dependent parameters, defined as functions of the aforementioned inputs. Free parameters $a,b,c,d_{3},d_{4}$ for the Higgs sector are also implemented with the notation of Ref.~\cite{Contino:2010vc}. For $a=b=c=d_{3}=d_{4}=1$ the Higgs sector is exactly SM like. 

In addition to the MEG model files, we make available different Computable Document Format CDF\copyright\,\cite{cdfwebpage} files on the webpage \cite{projectwebpage}. For each CDF we make available both a web interface and a downloadable file which can be opened with {\sc Mathematica} (version 9 or later). The web version is intended for simple studies, while for more intensive tasks we recommend the use of the local versions. The first CDF file allows the user to compute the dependent parameters, the widths and the BRs in the model and to plot the relevant cross-sections at $8$, $14$ and $100$ TeV by simply inputing the independent parameters. It also automatically generates the {\sc{MadGraph5}} ``param\_card.dat'' for the chosen point of the parameter space. The second CDF file allows the used to simply scan cross-sections, widths and BRs over all the independent parameters by simply setting initial and final values and number of points. Further information and possibly additional tools can be found directly on the webpage of this project \cite{projectwebpage}.

\bibliographystyle{mine}
\bibliography{bridge}

\providecommand{\href}[2]{#2}\begingroup\raggedright\begin{thebibliography}{100}

\bibitem{projectwebpage}
D.~Pappadopulo, A.~Thamm, R.~Torre, and A.~Wulzer, ``Tools for the study of
  heavy vector triplets'' \href{http://heidi.pd.infn.it/html/vector/index.html}{Webpage}.

\bibitem{Alves:2011dz}
{\bfseries LHC New Physics Working Group} Collaboration, D.~Alves {\em et al.},
  ``{Simplified Models for LHC New Physics Searches}'',
  \href{http://dx.doi.org/10.1088/0954-3899/39/10/105005}{{\em J. Phys.}
  {\bfseries G 39} (2012) 105005},
  \href{http://xxx.lanl.gov/abs/1105.2838}{{\tt arXiv:1105.2838}} [\href{http://inspirehep.net/record/900212}{Inspire}].

\bibitem{Bauer:2009p1295}
C.~W. Bauer, Z.~Ligeti, M.~Schmaltz, J.~Thaler, and D.~G.~E. Walker,
  ``{S}upermodels for early {LHC}'',
  \href{http://dx.doi.org/10.1016/j.physletb.2010.05.032}{{\em Phys. Lett.}
  {\bfseries B 690} (2010) 280--288},
  \href{http://xxx.lanl.gov/abs/0909.5213}{{\tt arXiv:0909.5213}} [\href{http://inspirebeta.net/record/832461}{Inspire}].

\bibitem{Barbieri:2011p2759}
R.~Barbieri and R.~Torre, ``Signals of single particle production at the
  earliest {LHC}'',
  \href{http://dx.doi.org/10.1016/j.physletb.2010.11.037}{{\em Phys. Lett.}
  {\bfseries B 695} (2011) 259--263},
  \href{http://xxx.lanl.gov/abs/1008.5302}{{\tt arXiv:1008.5302}} [\href{http://inspirebeta.net/record/866732}{Inspire}].

\bibitem{Han:2010p2696}
T.~Han, I.~Lewis, and Z.~Liu, ``Colored {R}esonant {S}ignals at the {LHC}:
  {L}argest {R}ate and {S}implest {T}opology'',
  \href{http://dx.doi.org/10.1007/JHEP12(2010)085}{{\em JHEP} {\bfseries 12}
  (2010) 085}, \href{http://xxx.lanl.gov/abs/1010.4309}{{\tt arXiv:1010.4309}} [\href{http://inspirebeta.net/record/873674}{Inspire}].

\bibitem{Accomando:2010p2249}
E.~Accomando, A.~S. Belyaev, L.~Fedeli, S.~F. King, and C.~H.
  Shepherd-Themistocleous, ``{$Z'$ physics with early LHC data}'',
  \href{http://dx.doi.org/10.1103/PhysRevD.83.075012}{{\em Phys. Rev.}
  {\bfseries D 83} (2012) 075012},
  \href{http://xxx.lanl.gov/abs/1010.6058}{{\tt arXiv:1010.6058}} [\href{http://inspirehep.net/record/874714}{Inspire}].

\bibitem{Schmaltz:2010p2610}
M.~Schmaltz and C.~Spethmann, ``Two {S}imple ${W}^\prime$ {M}odels for the
  {E}arly {LHC}'', \href{http://dx.doi.org/10.1007/JHEP07(2011)046}{{\em JHEP}
  {\bfseries 07} (2011) 046}, \href{http://xxx.lanl.gov/abs/1011.5918}{{\tt
  arXiv:1011.5918}} [\href{http://inspirehep.net/record/878831}{Inspire}].

\bibitem{Grojean:2011vu}
C.~Grojean, E.~Salvioni, and R.~Torre, ``A weakly constrained ${W}'$ at the
  early {LHC}'', \href{http://dx.doi.org/10.1007/JHEP07(2011)002}{{\em JHEP}
  {\bfseries 07} (2011) 002}, \href{http://xxx.lanl.gov/abs/1103.2761}{{\tt
  arXiv:1103.2761}} [\href{http://inspirebeta.net/record/892770}{Inspire}].

\bibitem{Chiang:2011kq}
C.-W. Chiang, N.~D. Christensen, G.-J. Ding, and T.~Han, ``{Discovery in
  Drell-Yan Processes at the LHC}'',
  \href{http://dx.doi.org/10.1103/PhysRevD.85.015023}{{\em Phys. Rev.}
  {\bfseries D 85} (2012) 015023},
  \href{http://xxx.lanl.gov/abs/1107.5830}{{\tt arXiv:1107.5830}} [\href{http://inspirehep.net/record/921469}{Inspire}].

\bibitem{Torre:2011vn}
R.~Torre, ``Limits on leptophobic ${W}'$ after 1 fb$^{-1}$ of {LHC} data: a
  lesson on parton level simulations'',
  \href{http://xxx.lanl.gov/abs/1109.0890}{{\tt arXiv:1109.0890}} [\href{http://inspirehep.net/record/926294}{Inspire}].

\bibitem{Torre:2012ub}
R.~Torre, ``{An isosinglet $W'$ at the LHC: updated bounds from direct
  searches}'', {\em PoS CORFU2011} (2011) 036,
  \href{http://xxx.lanl.gov/abs/1204.4364}{{\tt arXiv:1204.4364}} [\href{http://inspirebeta.net/record/1111661}{Inspire}].

\bibitem{deBlas:2012tc}
J.~{de Blas}, J.~M. Lizana, and M.~Perez-Victoria, ``{Combining searches of
  $Z'$ and $W'$ bosons}'',
  \href{http://dx.doi.org/10.1007/JHEP01(2013)166}{{\em JHEP} {\bfseries 01}
  (2013) 166}, \href{http://xxx.lanl.gov/abs/1211.2229}{{\tt arXiv:1211.2229}} [\href{http://inspirehep.net/record/1201947}{Inspire}].

\bibitem{DeSimone:2012ul}
A.~{De Simone}, O.~Matsedonskyi, R.~Rattazzi, and A.~Wulzer, ``{A First Top
  Partner's Hunter Guide}'',
  \href{http://dx.doi.org/10.1007/JHEP04(2013)004}{{\em JHEP} {\bfseries 04}
  (2013) 004}, \href{http://xxx.lanl.gov/abs/1211.5663}{{\tt arXiv:1211.5663}} [\href{http://inspirehep.net/record/1203860}{Inspire}].

\bibitem{Buchkremer:2013uj}
M.~Buchkremer, G.~Cacciapaglia, A.~Deandrea, and L.~Panizzi, ``{Model
  Independent Framework for Searches of Top Partners}'',
  \href{http://dx.doi.org/10.1016/j.nuclphysb.2013.08.010}{{\em Nucl. Phys.}
  {\bfseries B 876} (2013) 376--417},
  \href{http://xxx.lanl.gov/abs/1305.4172}{{\tt arXiv:1305.4172}} [\href{http://inspirehep.net/record/1233883}{Inspire}].

\bibitem{AguilarSaavedra:2013hg}
J.~A. Aguilar-Saavedra, R.~Benbrik, S.~Heinemeyer, and M.~Perez-Victoria, ``{A
  handbook of vector-like quarks: mixing and single production}'',
  \href{http://dx.doi.org/10.1103/PhysRevD.88.094010}{{\em Phys. Rev.}
  {\bfseries D 88} (2013) 094010},
  \href{http://xxx.lanl.gov/abs/1306.0572}{{\tt arXiv:1306.0572}} [\href{http://inspirehep.net/record/1236810}{Inspire}].

\bibitem{Lizana:2013vz}
J.~M. Lizana and M.~Perez-Victoria, ``{Vector triplets at the LHC}'',
  \href{http://dx.doi.org/10.1051/epjconf/20136017008}{{\em EPJ Web Conf.}
  {\bfseries 60} (2013) 17008}, \href{http://xxx.lanl.gov/abs/1307.2589}{{\tt
  arXiv:1307.2589}} [\href{http://inspirehep.net/record/1242121}{Inspire}].

\bibitem{1980PhRvD..22..727B}
V.~Barger, W.-Y. Keung, and E.~Ma, ``{Gauge model with light W and Z bosons}'',
  \href{http://dx.doi.org/10.1103/PhysRevD.22.727}{{\em Phys. Rev.} {\bfseries
  D 22} (1980) 727} [\href{http://inspirehep.net/record/152417}{Inspire}].

\bibitem{Hewett:1989dr}
J.~L. Hewett and T.~G. Rizzo, ``{Low-energy phenomenology of
  superstring-inspired $E_{6}$ models}'',
  \href{http://dx.doi.org/10.1016/0370-1573(89)90071-9}{{\em Phys. Rept.}
  {\bfseries 183} (1989) 193--381} [\href{http://inspirehep.net/record/268529}{Inspire}].

\bibitem{Cvetic:1995vc}
M.~Cvetic and S.~Godfrey, ``{Discovery and Identification of Extra Gauge
  Bosons}'', {\em In *Barklow, T.L. (ed.) et al.: Electroweak symmetry breaking
  and new physics at the TeV scale*} (1995) 383--415,
  \href{http://xxx.lanl.gov/abs/hep-ph/9504216}{{\tt hep-ph/9504216}} [\href{http://inspirehep.net/record/393993}{Inspire}].

\bibitem{Rizzo:2006wq}
T.~G. Rizzo, ``{$Z'$ Phenomenology and the LHC}'',
  \href{http://xxx.lanl.gov/abs/hep-ph/0610104}{{\tt hep-ph/0610104}} [\href{http://inspirehep.net/record/728548}{Inspire}].

\bibitem{Langacker:2008yv}
P.~Langacker, ``The {P}hysics of {H}eavy ${Z}^\prime$ {G}auge {B}osons'',
  \href{http://dx.doi.org/10.1103/RevModPhys.81.1199}{{\em Rev. Mod. Phys.}
  {\bfseries 81} (2009) 119--1228},
  \href{http://xxx.lanl.gov/abs/0801.1345}{{\tt arXiv:0801.1345}} [\href{http://inspirebeta.net/record/777086}{Inspire}].

\bibitem{Salvioni:2009mt}
E.~Salvioni, G.~Villadoro, and F.~Zwirner, ``Minimal {Z}' models: present
  bounds and early {LHC} reach'',
  \href{http://dx.doi.org/10.1088/1126-6708/2009/11/068}{{\em JHEP} {\bfseries
  11} (2009) 068}, \href{http://xxx.lanl.gov/abs/0909.1320}{{\tt
  arXiv:0909.1320}} [\href{http://inspirebeta.net/record/830532}{Inspire}].

\bibitem{Salvioni:2010p2769}
E.~Salvioni, A.~Strumia, G.~Villadoro, and F.~Zwirner, ``Non-universal minimal
  {Z}' models: present bounds and early {LHC} reach'',
  \href{http://dx.doi.org/10.1007/JHEP03(2010)010}{{\em JHEP} {\bfseries 03}
  (2010) 010}, \href{http://xxx.lanl.gov/abs/0911.1450}{{\tt arXiv:0911.1450}} [\href{http://inspirebeta.net/record/836375}{Inspire}].

\bibitem{Salvioni:2010p1209}
E.~Salvioni, ``{Minimal $Z'$ models and the early LHC}'', {\em Frascati
  Phys.Ser.} {\bfseries 51} (2010) ,
  \href{http://xxx.lanl.gov/abs/1007.0490}{{\tt arXiv:1007.0490}} [\href{http://inspirebeta.net/record/860483}{Inspire}].

\bibitem{Chanowitz:2011ew}
M.~S. Chanowitz, ``{A heavy little Higgs and a light $Z'$ under the radar}'',
  \href{http://dx.doi.org/10.1103/PhysRevD.84.035014}{{\em Phys. Rev.}
  {\bfseries D 84} (2011) 035014},
  \href{http://xxx.lanl.gov/abs/1102.3672}{{\tt arXiv:1102.3672}} [\href{http://inspirehep.net/record/889847}{Inspire}].

\bibitem{Salvioni:2012gya}
E.~Salvioni, ``{Some Z-prime and W-prime models facing current LHC searches}'',
  \href{http://dx.doi.org/10.3204/DESY-PROC-2012-02/177}{{\em 20th
  International Workshop on Deep-Inelastic Scattering and Related Subjects (DIS
  2012) Proceeding} (2012)} [\href{http://inspirebeta.net/record/1229598}{Inspire}].

\bibitem{Accomando:2013ve}
E.~Accomando, D.~Becciolini, A.~S. Belyaev, S.~Moretti, and C.~H.
  Shepherd-Themistocleous, ``{$Z'$ at the LHC: Interference and Finite Width
  Effects in Drell-Yan}'',
  \href{http://dx.doi.org/10.1007/JHEP10(2013)153}{{\em JHEP} {\bfseries 10}
  (2013) 153}, \href{http://xxx.lanl.gov/abs/1304.6700}{{\tt arXiv:1304.6700}} [\href{http://inspirehep.net/record/1229792}{Inspire}].

\bibitem{Langacker:1989p2578}
P.~Langacker and S.~U. Sankar, ``Bounds on the mass of ${W}_{R}$ and the
  ${W}_{L}-{W}_{R}$ mixing angle $\zeta$ in general ${SU}(2)_{L} \times
  {SU}(2)_{R} \times {U}(1)$ models'',
  \href{http://dx.doi.org/10.1103/PhysRevD.40.1569}{{\em Phys. Rev.} {\bfseries
  D 40} (1989) 1569--1585} [\href{http://inspirebeta.net/record/277249}{Inspire}].

\bibitem{Sullivan:2002p2617}
Z.~Sullivan, ``Fully differential ${W}^\prime$ production and decay at
  next-to-leading order in {QCD}'',
  \href{http://dx.doi.org/10.1103/PhysRevD.66.075011}{{\em Phys. Rev.}
  {\bfseries D 66} (2002) 075011},
  \href{http://xxx.lanl.gov/abs/hep-ph/0207290}{{\tt hep-ph/0207290}} [\href{http://inspirebeta.net/record/591241}{Inspire}].

\bibitem{Rizzo:2007bk}
T.~G. Rizzo, ``{The Determination of the Helicity of $W'$ Boson Couplings at
  the LHC}'', \href{http://dx.doi.org/10.1088/1126-6708/2007/05/037}{{\em JHEP}
  {\bfseries 05} (2007) 037}, \href{http://xxx.lanl.gov/abs/0704.0235}{{\tt
  arXiv:0704.0235}} [\href{http://inspirehep.net/record/747947}{Inspire}].

\bibitem{Frank:2010p2250}
M.~Frank, A.~Hayreter, and I.~Turan, ``Production and {D}ecays of ${W}_{R}$
  bosons at the {LHC}'',
  \href{http://dx.doi.org/10.1103/PhysRevD.83.035001}{{\em Phys. Rev.}
  {\bfseries D 83} (2011) 035001},
  \href{http://xxx.lanl.gov/abs/1010.5809}{{\tt arXiv:1010.5809}} [\href{http://inspirebeta.net/record/874800}{Inspire}].

\bibitem{Accomando:2011up}
E.~Accomando, D.~Becciolini, S.~de~Curtis, D.~Dominici, L.~Fedeli, and C.~H.
  Shepherd-Themistocleous, ``{Interference effects in heavy $W'$-boson searches
  at the LHC}'', \href{http://dx.doi.org/10.1103/PhysRevD.85.115017}{{\em Phys.
  Rev.} {\bfseries D 85} (2011) 115017},
  \href{http://xxx.lanl.gov/abs/1110.0713}{{\tt arXiv:1110.0713}} [\href{http://inspirehep.net/record/930438}{Inspire}].

\bibitem{Agashe:2007hh}
K.~Agashe, H.~Davoudiasl, S.~Gopalakrishna, T.~Han, G.-Y. Huang, G.~Perez,
  Z.-G. Si, and A.~Soni, ``{LHC Signals for Warped Electroweak Neutral Gauge
  Bosons}'', \href{http://dx.doi.org/10.1103/PhysRevD.76.115015}{{\em Phys.
  Rev.} {\bfseries D 76} (2007) 115015},
  \href{http://xxx.lanl.gov/abs/0709.0007}{{\tt arXiv:0709.0007}} [\href{http://inspirehep.net/record/759584}{Inspire}].

\bibitem{Agashe:2009bj}
K.~Agashe, S.~Gopalakrishna, T.~Han, G.-Y. Huang, and A.~Soni, ``{LHC signals
  for warped electroweak charged gauge bosons}'',
  \href{http://dx.doi.org/10.1103/PhysRevD.80.075007}{{\em Phys. Rev.}
  {\bfseries D 80} (2009) 075007},
  \href{http://xxx.lanl.gov/abs/0810.1497}{{\tt arXiv:0810.1497}} [\href{http://inspirehep.net/record/798910}{Inspire}].

\bibitem{Agashe:2009ve}
K.~Agashe, A.~Azatov, T.~Han, Y.~Li, Z.-G. Si, and L.~Zhu, ``{LHC Signals for
  Coset Electroweak Gauge Bosons in Warped/Composite PGB Higgs Models}'',
  \href{http://dx.doi.org/10.1103/PhysRevD.81.096002}{{\em Phys. Rev.}
  {\bfseries D 81} (2010) 096002},
  \href{http://xxx.lanl.gov/abs/0911.0059}{{\tt arXiv:0911.0059}} [\href{http://inspirehep.net/record/835687}{Inspire}].

\bibitem{Contino:2011np}
R.~Contino, D.~Marzocca, D.~Pappadopulo, and R.~Rattazzi, ``On the effect of
  resonances in composite {H}iggs phenomenology'',
  \href{http://dx.doi.org/10.1007/JHEP10(2011)081}{{\em JHEP} {\bfseries 10}
  (2011) 081}, \href{http://xxx.lanl.gov/abs/1109.1570}{{\tt arXiv:1109.1570}} [\href{http://inspirehep.net/record/926810}{Inspire}].

\bibitem{Bellazzini:2012tv}
B.~Bellazzini, C.~Csaki, J.~Hubisz, J.~Serra, and J.~Terning, ``{Composite
  Higgs Sketch}'', \href{http://dx.doi.org/10.1007/JHEP11(2012)003}{{\em JHEP}
  {\bfseries 11} (2012) 003}, \href{http://xxx.lanl.gov/abs/1205.4032}{{\tt
  arXiv:1205.4032}} [\href{http://inspirehep.net/record/1115304}{Inspire}].

\bibitem{Accomando:2012us}
E.~Accomando, L.~Fedeli, S.~Moretti, S.~D. Curtis, and D.~Dominici, ``{Charged
  di-boson production at the LHC in a 4-site model with a composite Higgs
  boson}'', \href{http://dx.doi.org/10.1103/PhysRevD.86.115006}{{\em Phys.Rev.}
  {\bfseries D86} (2012) 115006}, \href{http://xxx.lanl.gov/abs/1208.0268}{{\tt
  arXiv:1208.0268}} [\href{http://inspirehep.net/record/1124599}{Inspire}].

\bibitem{Hernandez:2013wd}
A.~E. {C\'arcamo Hern\'andez}, C.~O. Dib, and A.~R. Zerwekh, ``{The Effect of
  Composite Resonances on Higgs decay into two photons}'',
  \href{http://xxx.lanl.gov/abs/1304.0286}{{\tt arXiv:1304.0286}} [\href{http://inspirehep.net/record/1226002}{Inspire}].

\bibitem{Chanowitz:1993fc}
M.~S. Chanowitz and W.~Kilgore, ``{Complementarity of Resonant and Nonresonant
  Strong $WW$ Scattering at the LHC}'',
  \href{http://dx.doi.org/10.1016/0370-2693(94)90503-7}{{\em Phys. Lett.}
  {\bfseries B 322} (1993) 147--153},
  \href{http://xxx.lanl.gov/abs/hep-ph/9311336}{{\tt hep-ph/9311336}} [\href{http://inspirehep.net/record/360435}{Inspire}].

\bibitem{Barbieri:2008p1580}
R.~Barbieri, G.~Isidori, V.~S. Rychkov, and E.~Trincherini, ``Heavy {V}ectors
  in {H}iggs-less models'',
  \href{http://dx.doi.org/10.1103/PhysRevD.78.036012}{{\em Phys. Rev.}
  {\bfseries D 78} (2008) 036012},
  \href{http://xxx.lanl.gov/abs/0806.1624}{{\tt arXiv:0806.1624}} [\href{http://inspirebeta.net/record/787754}{Inspire}].

\bibitem{Barbieri:2010p144}
R.~Barbieri, A.~E. {C{\'a}rcamo Hern{\'a}ndez}, G.~Corcella, R.~Torre, and
  E.~Trincherini, ``Composite vectors at the {L}arge {H}adron {C}ollider'',
  \href{http://dx.doi.org/10.1007/JHEP03(2010)068}{{\em JHEP} {\bfseries 03}
  (2010) 068}, \href{http://xxx.lanl.gov/abs/0911.1942}{{\tt arXiv:0911.1942}} [\href{http://inspirebeta.net/record/836568}{Inspire}].

\bibitem{Barbieri:2010p1577}
R.~Barbieri, V.~S. Rychkov, and R.~Torre, ``Signals of composite
  electroweak-neutral {D}ark {M}atter: {LHC}/{D}irect {D}etection interplay'',
  \href{http://dx.doi.org/10.1016/j.physletb.2010.04.010}{{\em Phys. Lett.}
  {\bfseries B 688} (2010) 212--215},
  \href{http://xxx.lanl.gov/abs/1001.3149}{{\tt arXiv:1001.3149}} [\href{http://inspirebeta.net/record/843274}{Inspire}].

\bibitem{CarcamoHernandez:2010p1578}
A.~E. {C\'arcamo Hern\'andez} and R.~Torre, ``A `composite' scalar-vector
  system at the {LHC}'',
  \href{http://dx.doi.org/10.1016/j.nuclphysb.2010.08.004}{{\em Nucl. Phys.}
  {\bfseries B 841} (2010) 188}, \href{http://xxx.lanl.gov/abs/1005.3809}{{\tt
  arXiv:1005.3809}} [\href{http://inspirebeta.net/record/856011}{Inspire}].

\bibitem{Hernandez:2010qp}
A.~E. {C\'arcamo Hern\'andez}, ``{Top quark effects in composite vector pair
  production at the LHC}'',
  \href{http://dx.doi.org/10.1140/epjc/s10052-012-2154-3}{{\em Eur. Phys. J.}
  {\bfseries C 72} (2012) 2154}, \href{http://xxx.lanl.gov/abs/1008.1039}{{\tt
  arXiv:1008.1039}} [\href{http://inspirehep.net/record/864538}{Inspire}].

\bibitem{Cata:2009ka}
O.~Cat{\`a}, G.~Isidori, and J.~F. Kamenik, ``Drell-{Y}an production of {H}eavy
  {V}ectors in {H}iggsless models'',
  \href{http://dx.doi.org/10.1016/j.nuclphysb.2009.07.015}{{\em Nucl. Phys.}
  {\bfseries B 822} (2009) 230}, \href{http://xxx.lanl.gov/abs/0905.0490}{{\tt
  arXiv:0905.0490}} [\href{http://inspirebeta.net/record/819405}{Inspire}].

\bibitem{Accomando:2011gt}
E.~Accomando, D.~Becciolini, S.~D. Curtis, D.~Dominici, and L.~Fedeli, ``{$W'$
  production at the LHC in the 4-site Higgsless model}'',
  \href{http://dx.doi.org/10.1103/PhysRevD.84.115014}{{\em Phys. Rev.}
  {\bfseries D 84} (2011) 115014},
  \href{http://xxx.lanl.gov/abs/1107.4087}{{\tt arXiv:1107.4087}} [\href{http://inspirehep.net/record/919245}{Inspire}].

\bibitem{Falkowski:2011ua}
A.~Falkowski, C.~Grojean, A.~Kaminska, S.~Pokorski, and A.~Weiler, ``{If no
  Higgs then what?}'', \href{http://dx.doi.org/10.1007/JHEP11(2011)028}{{\em
  JHEP} {\bfseries 11} (2011) 028},
  \href{http://xxx.lanl.gov/abs/1108.1183}{{\tt arXiv:1108.1183}} [\href{http://inspirehep.net/record/922186}{Inspire}].

\bibitem{ATLASCollaboration:2012ex}
{\bfseries ATLAS} Collaboration, G.~Aad {\em et al.}, ``{Measurement of
  $W\gamma$ and $Z\gamma$ production cross sections in $pp$ collisions at
  $\sqrt{s} = 7$ TeV and limits on anomalous triple gauge couplings with the
  ATLAS detector}'',
  \href{http://dx.doi.org/10.1016/j.physletb.2012.09.017}{{\em Phys. Lett.}
  {\bfseries B 717} (2012) 49--69},
  \href{http://xxx.lanl.gov/abs/1205.2531}{{\tt arXiv:1205.2531}} [\href{http://inspirehep.net/record/1114319}{Inspire}].

\bibitem{ATLASCollaboration:2012gi}
{\bfseries ATLAS} Collaboration, G.~Aad {\em et al.}, ``{Measurement of $WZ$
  production in proton-proton collisions at $\sqrt{s} = 7$ TeV with the ATLAS
  detector}'', \href{http://dx.doi.org/10.1140/epjc/s10052-012-2173-0}{{\em
  Eur. Phys. J.} {\bfseries C 72} (2012) 2173},
  \href{http://xxx.lanl.gov/abs/1208.1390}{{\tt arXiv:1208.1390}} [\href{http://inspirehep.net/record/1126131}{Inspire}].

\bibitem{ATLAS-CONF-2012-130}
{\bfseries ATLAS} Collaboration, ``{Search for exotic same-sign dilepton
  signatures (b' quark, $T_{5/3}$ and four top quarks production) in 4.7/fb of
  pp collisions at $\sqrt{s}=7$ TeV with the ATLAS detector}'' [\href{http://cds.cern.ch/record/1478217}{ATLAS-CONF-2012-130}].

\bibitem{ATLAS-CONF-2012-150}
{\bfseries ATLAS} Collaboration, ``{Search for resonant $ZZ$ production in the
  $ZZ \to llqq$ channel with the ATLAS detector using 7.2 fb$^{-1}$ of
  $\sqrt{s} = 8$ TeV $pp$ collision data}'' [\href{http://cds.cern.ch/record/1493489?ln=en}{ATLAS-CONF-2012-150}].

\bibitem{ATLAS-CONF-2013-050}
{\bfseries ATLAS} Collaboration, ``{Search for $W' \to t\bar{b}$ in
  proton-proton collisions at a centre-of-mass energy of $\sqrt{s} = 8$ TeV
  with the ATLAS detector}'' [\href{http://cds.cern.ch/record/1547566?ln=en}{ATLAS-CONF-2013-050}].

\bibitem{ATLAS-CONF-2013-052}
{\bfseries ATLAS} Collaboration, ``{A search for $t\bar{t}$ resonances in the
  lepton plus jets final state with ATLAS using 14 fb$^{−1}$ of $pp$
  collisions at $\sqrt{s}=8$ TeV}'' [\href{http://cds.cern.ch/record/1547568?ln=en}{ATLAS-CONF-2013-052}].

\bibitem{ATLAS-CONF-2013-066}
{\bfseries ATLAS} Collaboration, ``{A search for high-mass ditau resonances
  decaying in the fully hadronic final state in $pp$ collisions at $\sqrt{s}=8$
  TeV with the ATLAS detector}'' [\href{http://cds.cern.ch/record/1562841?ln=en}{ATLAS-CONF-2013-066}].

\bibitem{ATLAS-CONF-2014-017}
{\bfseries ATLAS} Collaboration, ``{Search for high-mass states with one lepton
  plus missing transverse momentum in pp collisions at $\sqrt{s}=8\,$TeV with
  the ATLAS detector}'' [\href{http://cds.cern.ch/record/1692660?ln=en}{ATLAS-CONF-2014-017}].

\bibitem{Aad:2014cka}
{\bfseries ATLAS} Collaboration, G.~Aad {\em et al.}, ``{Search for high-mass
  dilepton resonances in pp collisions at $\sqrt{s}$ = 8 TeV with the ATLAS
  detector}'', \href{http://xxx.lanl.gov/abs/1405.4123}{{\tt arXiv:1405.4123}} [\href{http://inspirehep.net/record/1296830}{Inspire}].

\bibitem{Aad:2014pha}
{\bfseries ATLAS} Collaboration, G.~Aad {\em et al.}, ``{Search for $WZ$
  resonances in the fully leptonic channel using pp collisions at $\sqrt{s} =
  8\,$TeV with the ATLAS detector}'',
  \href{http://xxx.lanl.gov/abs/1406.4456}{{\tt arXiv:1406.4456}} [\href{http://inspirehep.net/record/1300821}{Inspire}].

\bibitem{Aad:2014aqa}
{\bfseries ATLAS} Collaboration, G.~Aad {\em et al.}, ``{Search for new
  phenomena in the dijet mass distribution using $pp$ collision data at
  $\sqrt{s}=8\,$TeV with the ATLAS detector}'',
  \href{http://xxx.lanl.gov/abs/1407.1376}{{\tt arXiv:1407.1376}} [\href{http://inspirehep.net/record/1305096}{Inspire}].

\bibitem{CMS-PAS-EXO-12-021}
{\bfseries CMS} Collaboration, ``{Search for new resonances decaying to $WW \to
  l \nu q \bar{q}$ in the final state with a lepton, missing transverse energy,
  and single reconstructed jet}'' [\href{http://cds.cern.ch/record/1590301?ln=en}{CMS-PAS-EXO-12-021}].

\bibitem{CMS-PAS-EXO-12-022}
{\bfseries CMS} Collaboration, ``{Search for a narrow spin-2 resonance decaying
  to Z bosons in the semileptonic final state}'' [\href{http://cds.cern.ch/record/1596494?ln=en}{CMS-PAS-EXO-12-022}].

\bibitem{CMS-PAS-EXO-12-023}
{\bfseries CMS} Collaboration, ``{Search for Heavy Resonances Decaying into
  $bb$ and $bg$ Final States in $pp$ Collisions at $\sqrt{s} = 8$ TeV}'' [\href{http://cds.cern.ch/record/1542405?ln=en}{CMS-PAS-EXO-12-023}].

\bibitem{CMS-PAS-EXO-12-024}
{\bfseries CMS} Collaboration, ``{Search for heavy resonances in the
  $W/Z$-tagged dijet mass spectrum in $pp$ collisions at 8 TeV}'' [\href{http://cds.cern.ch/record/1563153?ln=en}{CMS-PAS-EXO-12-024}].

\bibitem{CMS-PAS-EXO-12-059}
{\bfseries CMS} Collaboration, ``{Search for Narrow Resonances using the Dijet
  Mass Spectrum with 19.6 fb$^{-1}$ of $pp$ Collisions at $\sqrt{s}=8$ TeV}'' [\href{http://cds.cern.ch/record/1519066?ln=en}{CMS-PAS-EXO-12-059}].

\bibitem{CMS-PAS-EXO-12-060}
{\bfseries CMS} Collaboration, ``{Search for leptonic decays of $W'$ bosons in
  $pp$ collisions at $\sqrt{s}=8$ TeV}'' [\href{http://cds.cern.ch/record/1522476?ln=en}{CMS-PAS-EXO-12-060}].

\bibitem{CMS-PAS-EXO-12-061}
{\bfseries CMS} Collaboration, ``{Search for Resonances in the Dilepton Mass
  Distribution in $pp$ Collisions at $\sqrt{s} = 8$ TeV}'' [\href{http://cds.cern.ch/record/1519132?ln=en}{CMS-PAS-EXO-12-061}].

\bibitem{CMS-PAS-B2G-12-005}
{\bfseries CMS} Collaboration, ``{Search for Anomalous Top Quark Pair
  Production in the Boosted All-Hadronic Final State using $pp$ Collisions at
  $\sqrt{s} = 8$ TeV}'' [\href{http://cds.cern.ch/record/1545285?ln=en}{CMS-PAS-B2G-12-005}].

\bibitem{CMS-PAS-B2G-12-010}
{\bfseries CMS} Collaboration, ``{Search for narrow $t + b$ resonances in the
  leptonic final state at $\sqrt{s} = 8$ TeV}'' [\href{http://cds.cern.ch/record/1525924?ln=en}{CMS-PAS-B2G-12-010}].

\bibitem{Khachatryan:2014xja}
{\bfseries CMS Collaboration} Collaboration, V.~Khachatryan {\em et al.},
  ``{Search for new resonances decaying via WZ to leptons in proton-proton
  collisions at $\sqrt{s}=8\,$TeV}'',
  \href{http://xxx.lanl.gov/abs/1407.3476}{{\tt arXiv:1407.3476}} [\href{http://inspirehep.net/record/1306289}{Inspire}].

\bibitem{Chanowitz:1985hj}
M.~S. Chanowitz and M.~K. Gaillard, ``{The TeV Physics of Strongly Interacting
  W's and Z's}'', \href{http://dx.doi.org/10.1016/0550-3213(85)90580-2}{{\em
  Nucl.Phys.} {\bfseries B261} (1985)} [\href{http://inspirebeta.net/record/216026}{Inspire}].

\bibitem{Giudice:1024017}
G.~F. Giudice, C.~Grojean, A.~Pomarol, and R.~Rattazzi, ``{The
  Strongly-Interacting Light Higgs. }'',
  \href{http://dx.doi.org/10.1088/1126-6708/2007/06/045}{{\em JHEP} {\bfseries
  06} (2007) 045}, \href{http://xxx.lanl.gov/abs/hep-ph/0703164}{{\tt
  hep-ph/0703164}} [\href{http://inspirehep.net/record/746568}{Inspire}].

\bibitem{Aguila:2010p1781}
F.~del Aguila, J.~{de Blas}, and M.~Perez-Victoria, ``{Electroweak Limits on
  General New Vector Bosons}'',
  \href{http://dx.doi.org/10.1007/JHEP09(2010)033}{{\em JHEP} {\bfseries 09}
  (2010) 033}, \href{http://xxx.lanl.gov/abs/1005.3998}{{\tt arXiv:1005.3998}} [\href{http://inspirehep.net/record/855936}{Inspire}].

\bibitem{Ciuchini:2013vb}
M.~Ciuchini, E.~Franco, S.~Mishima, and L.~Silvestrini, ``{Electroweak
  Precision Observables, New Physics and the Nature of a 126 GeV Higgs
  Boson}'', \href{http://dx.doi.org/10.1007/JHEP08(2013)106}{{\em JHEP}
  {\bfseries 08} (2013) 106}, \href{http://xxx.lanl.gov/abs/1306.4644}{{\tt
  arXiv:1306.4644}} [\href{http://inspirehep.net/record/1239175}{Inspire}].

\bibitem{Contino:2013un}
R.~Contino, C.~Grojean, D.~Pappadopulo, R.~Rattazzi, and A.~Thamm, ``{Strong
  Higgs Interactions at a Linear Collider}'',
  \href{http://dx.doi.org/10.1007/JHEP02(2014)006}{{\em JHEP} {\bfseries 02}
  (2014) 006}, \href{http://xxx.lanl.gov/abs/1309.7038}{{\tt arXiv:1309.7038}} [\href{http://inspirehep.net/record/1255676}{Inspire}].

\bibitem{Panico:1359049}
G.~Panico and A.~Wulzer, ``{The Discrete Composite Higgs Model}'',
  \href{http://dx.doi.org/10.1007/JHEP09(2011)135}{{\em JHEP} {\bfseries 1109}
  (2011) 135}, \href{http://xxx.lanl.gov/abs/1106.2719}{{\tt arXiv:1106.2719}} [\href{http://inspirehep.net/record/913588}{Inspire}].

\bibitem{Matsedonskyi:2012ws}
O.~Matsedonskyi, G.~Panico, and A.~Wulzer, ``{Light Top Partners for a Light
  Composite Higgs}'', \href{http://dx.doi.org/10.1007/JHEP01(2013)164}{{\em
  JHEP} {\bfseries 1301} (2013) 164},
  \href{http://xxx.lanl.gov/abs/1204.6333}{{\tt arXiv:1204.6333}} [\href{http://inspirehep.net/record/1112825}{Inspire}].

\bibitem{Chacko:2012wh}
Z.~Chacko, R.~Franceschini, and R.~K. Mishra, ``{Resonance at 125 GeV: Higgs or
  Dilaton/Radion?}'', \href{http://dx.doi.org/10.1007/JHEP04(2013)015}{{\em
  JHEP} {\bfseries 04} (2013) 015},
  \href{http://xxx.lanl.gov/abs/1209.3259}{{\tt arXiv:1209.3259}} [\href{http://inspirehep.net/record/1185589}{Inspire}].

\bibitem{Bellazzini:2012td}
B.~Bellazzini, C.~Csaki, J.~Hubisz, J.~Serra, and J.~Terning, ``{A Higgslike
  Dilaton}'', \href{http://dx.doi.org/10.1140/epjc/s10052-013-2333-x}{{\em Eur.
  Phys. J.} {\bfseries C 73} (2013) 2333},
  \href{http://xxx.lanl.gov/abs/1209.3299}{{\tt arXiv:1209.3299}} [\href{http://inspirehep.net/record/1185590}{Inspire}].

\bibitem{Peskin:1992p1662}
M.~E. Peskin and T.~Takeuchi, ``Estimation of oblique electroweak
  corrections'', \href{http://dx.doi.org/10.1103/PhysRevD.46.381}{{\em Phys.
  Rev.} {\bfseries D 46} (1992) 381--409} [\href{http://inspirebeta.net/record/321491}{Inspire}].

\bibitem{Barbieri:2004p1607}
R.~Barbieri, A.~Pomarol, R.~Rattazzi, and A.~Strumia, ``Electroweak symmetry
  breaking after {LEP1} and {LEP2}'',
  \href{http://dx.doi.org/10.1016/j.nuclphysb.2004.10.014}{{\em Nucl. Phys.}
  {\bfseries B 703} (2004) 127--146},
  \href{http://xxx.lanl.gov/abs/hep-ph/0405040}{{\tt hep-ph/0405040}} [\href{http://inspirebeta.net/record/649700}{Inspire}].

\bibitem{Cacciapaglia:2006pk}
G.~Cacciapaglia, C.~Csaki, G.~Marandella, and A.~Strumia, ``{The Minimal Set of
  Electroweak Precision Parameters}'',
  \href{http://dx.doi.org/10.1103/PhysRevD.74.033011}{{\em Phys. Rev.}
  {\bfseries D 74} (2006) 033011},
  \href{http://xxx.lanl.gov/abs/hep-ph/0604111}{{\tt hep-ph/0604111}} [\href{http://inspirehep.net/record/714334}{Inspire}].

\bibitem{Wulzer:2013tp}
A.~Wulzer, ``{An Equivalent Gauge and the Equivalence Theorem}'',
  \href{http://xxx.lanl.gov/abs/1309.6055}{{\tt arXiv:1309.6055}} [\href{http://inspirehep.net/record/1255318}{Inspire}].

\bibitem{Domenech:2012ai}
O.~Domenech, A.~Pomarol, and J.~Serra, ``{Probing the SM with Dijets at the
  LHC}'', \href{http://dx.doi.org/10.1103/PhysRevD.85.074030}{{\em Phys. Rev.}
  {\bfseries D 85} (2012) 074030},
  \href{http://xxx.lanl.gov/abs/1201.6510}{{\tt arXiv:1201.6510}} [\href{http://inspirehep.net/record/1086831}{Inspire}].

\bibitem{deBlas:2013qqa}
J.~de~Blas, M.~Chala, and J.~Santiago, ``{Global Constraints on Lepton-Quark
  Contact Interactions}'',
  \href{http://dx.doi.org/10.1103/PhysRevD.88.095011}{{\em Phys. Rev.}
  {\bfseries D 88} (2013) 095011},
  \href{http://xxx.lanl.gov/abs/1307.5068}{{\tt arXiv:1307.5068}} [\href{http://inspirehep.net/record/1243596}{Inspire}].

\bibitem{1985NuPhB.249...42D}
S.~Dawson, ``{The effective W approximation}'',
  \href{http://dx.doi.org/10.1016/0550-3213(85)90038-0}{{\em Nucl. Phys.}
  {\bfseries B 249} (1985) 42} [\href{http://inspirehep.net/record/200030}{Inspire}].

\bibitem{2011arXiv1110.3906T}
R.~Torre, {\em {Signals of composite particles at the LHC}}.
\newblock PhD thesis, Universit{\`a} di Pisa, 2011, \newblock \href{http://xxx.lanl.gov/abs/1110.3906}{{\tt arXiv:1110.3906}} \newblock [\href{http://inspirehep.net/record/940384}{Inspire}].

\bibitem{Borel:2012wg}
P.~Borel, R.~Franceschini, R.~Rattazzi, and A.~Wulzer, ``{Probing the
  Scattering of Equivalent Electroweak Bosons}'',
  \href{http://dx.doi.org/10.1007/JHEP06(2012)122}{{\em JHEP} {\bfseries 06}
  (2012) 122}, \href{http://xxx.lanl.gov/abs/1202.1904}{{\tt arXiv:1202.1904}} [\href{http://inspirehep.net/record/1088565}{Inspire}].

\bibitem{Han:2013wa}
T.~Han, P.~Langacker, Z.~Liu, and L.-T. Wang, ``{Diagnosis of a New Neutral
  Gauge Boson at the LHC and ILC for Snowmass 2013}'',
  \href{http://xxx.lanl.gov/abs/1308.2738}{{\tt arXiv:1308.2738}} [\href{http://inspirehep.net/record/1247674}{Inspire}].

\bibitem{Hayden:2013ve}
D.~Hayden, R.~Brock, and C.~Willis, ``{Z Prime: A Story}'',
  \href{http://xxx.lanl.gov/abs/1308.5874}{{\tt arXiv:1308.5874}} [\href{http://inspirehep.net/record/1251060}{Inspire}].

\bibitem{Godfrey:2013uu}
S.~Godfrey and T.~Martin, ``{Z' Discovery Reach at Future Hadron Colliders: A
  Snowmass White Paper}'', \href{http://xxx.lanl.gov/abs/1309.1688}{{\tt
  arXiv:1309.1688}} [\href{http://inspirehep.net/record/1253126}{Inspire}].

\bibitem{Katz:2010jx}
A.~Katz, M.~Son, and B.~Tweedie, ``{Jet Substructure and the Search for Neutral
  Spin-One Resonances in Electroweak Boson Channels}'',
  \href{http://dx.doi.org/10.1007/JHEP03(2011)011}{{\em JHEP} {\bfseries 03}
  (2011) 011}, \href{http://xxx.lanl.gov/abs/1010.5253}{{\tt arXiv:1010.5253}} [\href{http://inspirehep.net/record/874477}{Inspire}].

\bibitem{Katz:2010kw}
A.~Katz, M.~Son, and B.~Tweedie, ``{Ditau-Jet Tagging and Boosted Higgses from
  a Multi-TeV Resonance}'',
  \href{http://dx.doi.org/10.1103/PhysRevD.83.114033}{{\em Phys. Rev.}
  {\bfseries D 83} (2011) 114033},
  \href{http://xxx.lanl.gov/abs/1011.4523}{{\tt arXiv:1011.4523}} [\href{http://inspirehep.net/record/878411}{Inspire}].

\bibitem{Son:2012vs}
M.~Son, C.~Spethmann, and B.~Tweedie, ``{Diboson-Jets and the Search for
  Resonant $Zh$ Production}'',
  \href{http://dx.doi.org/10.1007/JHEP08(2012)160}{{\em JHEP} {\bfseries 08}
  (2012) 160}, \href{http://xxx.lanl.gov/abs/1204.0525}{{\tt arXiv:1204.0525}} [\href{http://inspirehep.net/record/1102907}{Inspire}].

\bibitem{2011arXiv1111.1054C}
D.~Choudhury, R.~M. Godbole, and P.~Saha, ``{Dijet resonances, widths and all
  that}'', \href{http://dx.doi.org/10.1007/JHEP01(2012)155}{{\em JHEP}
  {\bfseries 01} (2012) 155}, \href{http://xxx.lanl.gov/abs/1111.1054}{{\tt
  arXiv:1111.1054}} [\href{http://inspirehep.net/record/944599}{Inspire}].

\bibitem{ATLAS-CONF-2013-017}
{\bfseries ATLAS} Collaboration, ``{Search for high-mass dilepton resonances in
  20 fb$^{-1}$ of $pp$ collisions at $\sqrt{s} = 8$ TeV with the ATLAS
  experiment}'' [\href{http://cds.cern.ch/record/1525524?ln=en}{ATLAS-CONF-2013-017}].

\bibitem{ATLAS-CONF-2012-148}
{\bfseries ATLAS} Collaboration, ``{Search for New Phenomena in the Dijet Mass
  Distribution updated using 13.0 fb$^{-1}$ of $pp$ Collisions at $\sqrt{s}=8$
  TeV collected by the ATLAS Detector}'' [\href{http://cds.cern.ch/record/1493487?ln=en}{ATLAS-CONF-2012-148}].

\bibitem{ATLAS-CONF-2013-015}
{\bfseries ATLAS} Collaboration, ``{Search for resonant $WZ \to 3l\nu$
  production in $\sqrt{s} = 8$ TeV $pp$ collisions with 13 fb$^{−1}$ at
  ATLAS}'' [\href{http://cds.cern.ch/record/1525522?ln=en}{ATLAS-CONF-2013-015}].

\bibitem{CMS-PAS-EXO-12-025}
{\bfseries CMS} Collaboration, ``{Search for $W'$/technirho in $WZ$ using
  leptonic final states}'' [\href{http://cds.cern.ch/record/1558197?ln=en}{CMS-PAS-EXO-12-025}].

\bibitem{Alloul:2013vc}
A.~Alloul, N.~D. Christensen, C.~Degrande, C.~Duhr, and B.~Fuks, ``{FeynRules
  2.0 - A complete toolbox for tree-level phenomenology}'',
  \href{http://xxx.lanl.gov/abs/1310.1921}{{\tt arXiv:1310.1921}} [\href{http://inspirehep.net/record/1257621}{Inspire}].

\bibitem{Duhr:tm}
C.~Duhr, N.~D. Christensen, and B.~Fuks, ``Feyn{R}ules - {A} {M}athematica
  package to calculate {F}eynman rules'', \url{http://feynrules.irmp.ucl.ac.be}.

\bibitem{Belyaev:2012tt}
A.~Belyaev, N.~D. Christensen, and A.~Pukhov, ``{CalcHEP 3.4 for collider
  physics within and beyond the Standard Model}'',
  \href{http://dx.doi.org/10.1016/j.cpc.2013.01.014}{{\em Comput. Phys.
  Commun.} {\bfseries 184} (2013) 1729--1769},
  \href{http://xxx.lanl.gov/abs/1207.6082}{{\tt arXiv:1207.6082}} [\href{http://inspirehep.net/record/1123804}{Inspire}].

\bibitem{Belyaev:Df-Kj9yc}
A.~Belyaev, N.~Christensen, and A.~Pukhov, ``Calc{HEP} - a package for
  calculation of {F}eynman diagrams and integration over multi-particle phase
  space'', {\em \href{http://theory.sinp.msu.ru/~pukhov/calchep.html}{Web page
  of the package}} (2005) .

\bibitem{Alwall:2011fk}
J.~Alwall, M.~Herquet, F.~Maltoni, O.~Mattelaer, and T.~Stelzer, ``{MadGraph 5:
  going beyond}'', \href{http://dx.doi.org/10.1007/JHEP06(2011)128}{{\em JHEP}
  {\bfseries 06} (2011) 128}, \href{http://xxx.lanl.gov/abs/1106.0522}{{\tt
  arXiv:1106.0522}} [\href{http://inspirehep.net/record/912611}{Inspire}].

\bibitem{hepmdb}
``High energy physics models database'', \url{http://hepmdb.soton.ac.uk}.

\bibitem{hepmdbmodel}
D.~Pappadopulo, A.~Thamm, R.~Torre, and A.~Wulzer, ``Vector triplet (bridge
  model)'', \url{http://hepmdb.soton.ac.uk/hepmdb:0214.0151}.

\bibitem{Contino:2010vc}
R.~Contino, C.~Grojean, M.~Moretti, F.~Piccinini, and R.~Rattazzi, ``{S}trong
  {D}ouble {H}iggs {P}roduction at the {LHC}'',
  \href{http://dx.doi.org/10.1007/JHEP05(2010)089}{{\em JHEP} {\bfseries 05}
  (2010) 089}, \href{http://xxx.lanl.gov/abs/1002.1011}{{\tt arXiv:1002.1011}} [\href{http://inspirebeta.net/record/845195}{Inspire}].

\bibitem{cdfwebpage}
Wolfram, ``Computable document format'', \url{http://www.wolfram.com/cdf/}.

\end{thebibliography}\endgroup

\end{document}